\documentclass[a4paper,fleqn,sort&compress]{cas-sc}
\usepackage[numbers]{natbib}
\usepackage[nodots]{numcompress}
\usepackage{caption,subcaption}
\usepackage{lineno}

\begin{document}
\let\WriteBookmarks\relax
\def\floatpagepagefraction{1}
\def\textpagefraction{.001}


\shorttitle{Wave-energy-harvesting liquid tank}
\shortauthors{C Zhang et~al.}

\title [mode = title]{Coupled hydro-aero-turbo dynamics of liquid-tank system for wave energy harvesting: Numerical modellings and scaled prototype tests}                      

\author[1]{Chongwei Zhang}[
      orcid=0000-0002-0062-4548]
\fnmark[1]
\ead{chongweizhang@dlut.edu.cn}
\credit{Conceptualisation, Methodology,  Data curation, Investigation, Formal analysis, Visualisation, Validation, Software, Resources, Writing - Original Draft, Writing - Review \& Editing}
\affiliation[1]{organization={State Key Laboratory of Coastal and Offshore Engineering, Dalian University of Technology},
    city={Dalian},
    postcode={116024}, 
    country={P.R. China}}
\author[1]{Xunhao Zhu}
\credit{Methodology,  Data curation, Investigation, Formal analysis, Visualisation, Validation, Software, Writing - Original Draft, Writing - Review \& Editing}
\author[1]{Cheng Zhang}
\credit{Methodology,  Data curation, Investigation, Formal analysis, Visualisation, Validation, Software, Writing - Original Draft, Writing - Review \& Editing}

\author[2]{Luofeng Huang}
\credit{Methodology, Writing - Review $\&$ Editing}
\affiliation[2]{organization={School of Water, Energy and Environment, Cranfield University},
    city={Cranfield},
    postcode={MK43 0AL}, 
    country={United Kingdom}}

\author[1]{Dezhi Ning}
\credit{Methodology,  Resources, Supervision, Writing - Review \& Editing}
\cortext[cor1]{corresponding author. E-mail: dzning@dlut.edu.cn}
\cormark[1]
\begin{abstract}
An integrated numerical model is proposed for the first time to explore the coupled hydro- aero-turbo dynamics of wave-energy-harvesting (WEH) liquid tanks.
A scaled prototype of the WEH liquid tank with an impulse air turbine system is made to experimentally validate the numerical model.
Multi-layered impulse air turbine systems (MLATS) are creatively introduced into the liquid-tank system.
The inherent mechanisms of the coupled hydro-aero-turbo dynamics of the WEH liquid tank with different turbine properties are systematically investigated.
Compared with the experimental data, the numerical model can accurately reproduce the rotor speed, liquid motion, and air pressure of the WEH liquid tank.
Upon analysing mechanical parameters of the turbine rotor, it is found that the rotor's moment of inertia mainly affects the rotor speed's variation range, while the damping coefficient significantly influences the averaged rotor speed.
The optimal power take-off damping for the WEH liquid tank is identified. 
Considering the efficiency performances of three MLATSs, improving Turbine-L1 to Turbine-L2 or Turbine-L3 can increase the averaged power output by about $25\%$ or $40\%$, respectively.
Increasing the tank breadth can effectively boost the power output in a nonlinear way.
Under the considered excitation conditions, if the tank breadth is doubled, the maximum averaged power output can be increased by around four times. 
Through a series of failure tests, Turbine-L3 shows greater reliability in extreme conditions compared to a conventional single-rotor turbine. 
Even if the most important rotor of Turbine-L3 fails to work, the maximum loss of the averaged power output is only $44\%$.
The present WEH liquid with Turbine - L3 shows improved efficiency and reliability compared to the conventional liquid-tank system with a single - rotor turbine.

\end{abstract}

\begin{keywords}
\sep wave energy 
\sep liquid tank
\sep impulse air turbine  
\sep wave energy prototype
\sep power take-off
\end{keywords}

\maketitle

\section{Introduction}

Ocean wave energy has received extraordinary attention in recent years due to its great potential to reduce carbon footprint  \cite{Clement2002}.
A large variety of wave energy converter (WEC) concepts have been proposed to convert wave power into electricity.
To date, three dominant technical streams of WECs can be identified according to the power take-off (PTO) principle, \textit{i.e.} the pneumatic \cite{Liu2021}, the oscillating-buoy (OB) \cite{Zhang2024}, and the overtopping (OT) types \cite{Liu2024,Liu2018,Vicinanza2014}. 
The pneumatic type WECs include the oscillation water column (OWC)\cite{Zhang20231,Mayon2021,Carrelhas2022}, the backward bent duct buoy (BBDB)\cite{Wu2017,Kim2015,Liu20241}, and so on; and the OB type WECs can be further classified as point absorbers\cite{Chen2017,LaiI2021,Eng2011}, buoyancy pendulums\cite{Chen2024,Renzi2013}, and gravitational pendulums\cite{Yu2016}.
Over the past decade, massive efforts have been made to improve the energy conversion efficiency of some specific WEC concept bit by bit, and tens of prototypes have finally experienced short-term sea-trial tests in practice \cite{Wang2023}.

However, the area of wave power is still in a infancy stage from the commercial point of view\cite{Guanche2014,Guo2021}.
From laboratory to real sea, survivability is an important criterion of successfulness of a WEC product \cite{Cordonnier2015}. 
A practical WEC product, often with a designed lifetime of at least twenty years, must withstand extreme wave conditions with a return period of 100 years\cite{Tiron2015}. 
Especially in large storm waves or rogue waves, the wave height may reach 20 m\cite{Magnusson2013,McAllister2019}.
Such extreme waves can generate destructive slamming loads on the device, threatening the survivability of a WEC device\cite{Renzi2018,Medina2022}.

Unfortunately, for conventional WECs, a common inherent weakness in the design is that the moving parts, which are often the core power take-off (PTO) components of the device, are constantly exposed to the marine environment\cite{Gallutia2022}.
For example, the turbo-generator system of an OWC-type WEC is immersed in moisture and salt air near the sea surface\cite{Bruschi2019}; the hinges or hydraulic cylinders of the PTO system of an OB-type WEC normally exist in the seawater splash zone\cite{Ahamed2020}; and the hydro-turbine of OT-type WECs is situated in a tunnel filled with bubbly sea water.
These ``exposed'' moving parts have to interact with the active water mass, air and all its living forms\cite{Langhamer2009}.
Extreme wave loads and various types of corrosions that are often overlooked in the design phase have been identified as the main cause of most technical failures of WECs in real ocean conditions \cite{Cordonnier2015}.
Exposing the most essential but weakest working units to harsh marine environment is a significant threat to the survivability of a WEC device\cite{Wang2024}.

\begin{figure}[!htp]
\centering
\includegraphics[scale=0.42]{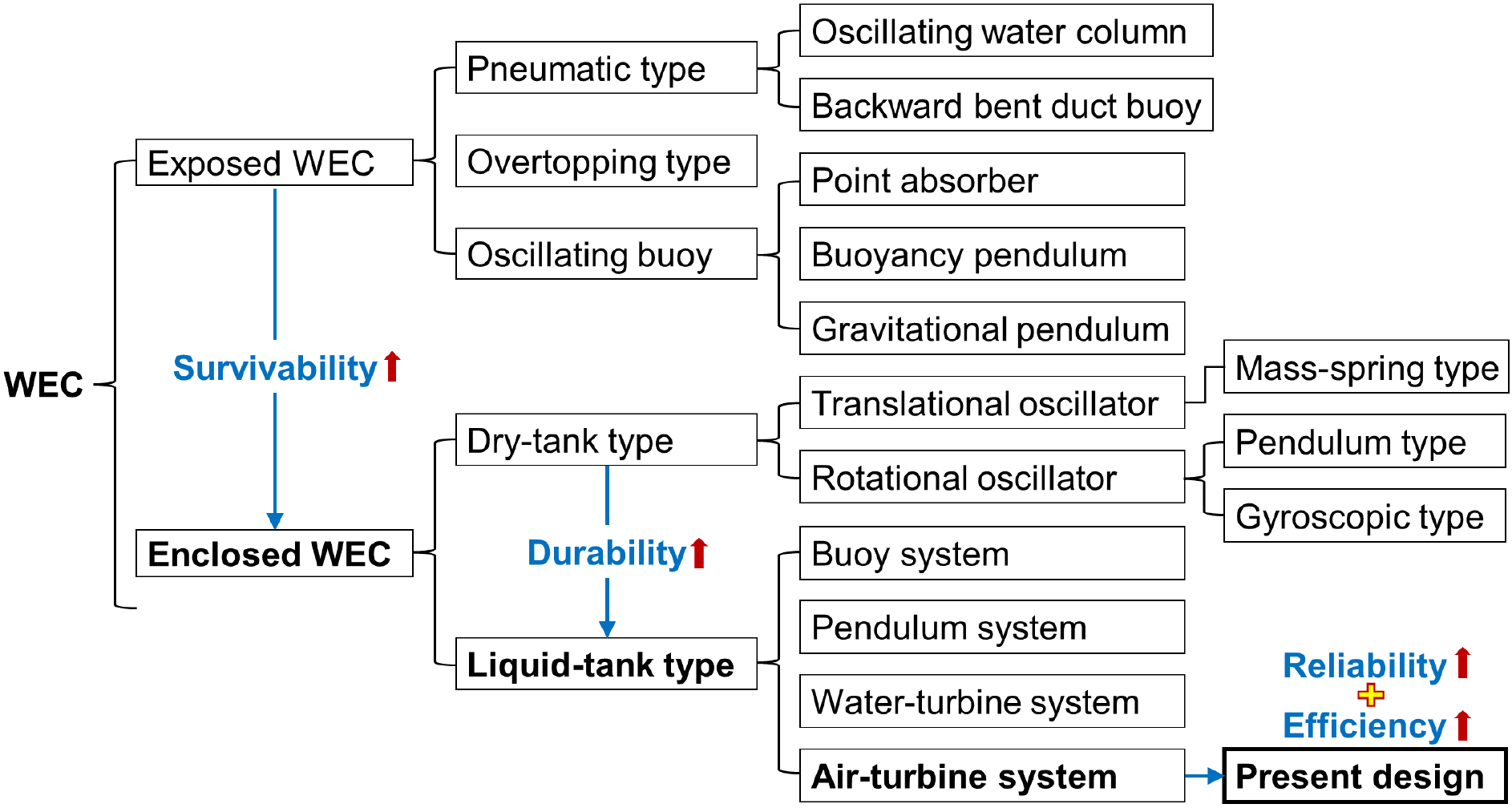}
\caption{Classification of working principles of wave energy converters}
\label{fig:category}
\end{figure}

To improve the survivability of WECs in real sea, some researchers turn to an alternative direction of concept design, i.e. the ``enclosed WEC'' (En-WEC for short).
The key idea is to shelter all mechanical and electrical components inside a closed hull, so that all moving parts are isolated from the aggressive sea environment.
For ease of distinction, the conventional pneumatic, OB-, or OT-type WECs are referred to by the term ``exposed WEC'' (Ex-WEC for short).
En-WECs can be further categorised as the ``dry-tank'' and ``liquid-tank'' type, based on whether the internal PTO system is directly driven by the hull or indirectly driven via oscillating liquids inside the hull.
A detailed classification of working principles of WECs is summarised in Fig. \ref{fig:category}.

\begin{figure}[!htp]
\centering
{
  \begin{minipage}{0.32\linewidth}
  \centering
  \includegraphics[width=1.0\linewidth]{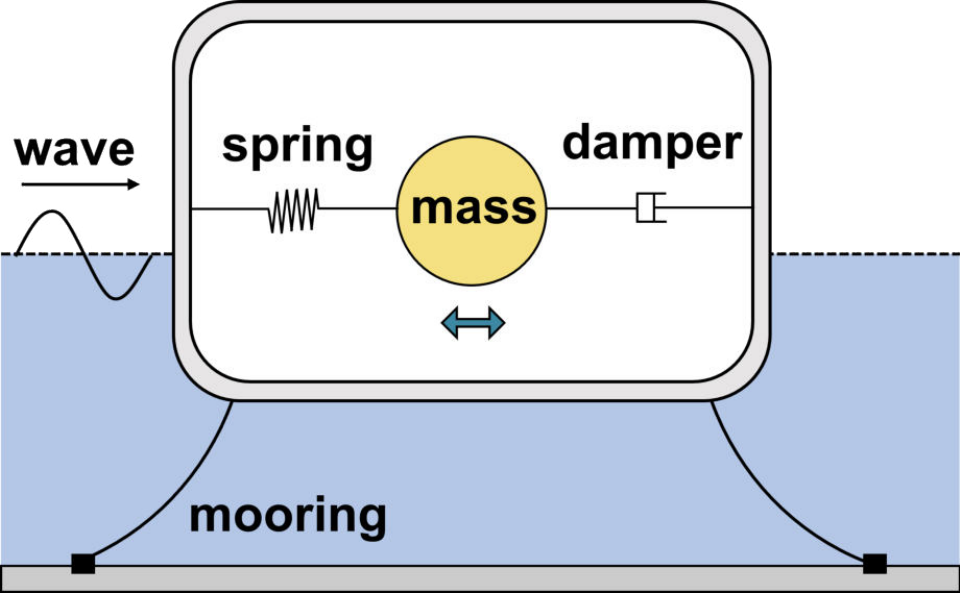}
  \subcaption{}
  \end{minipage}
}
{
  \begin{minipage}{0.32\linewidth}
  \centering
  \includegraphics[width=1.0\linewidth]{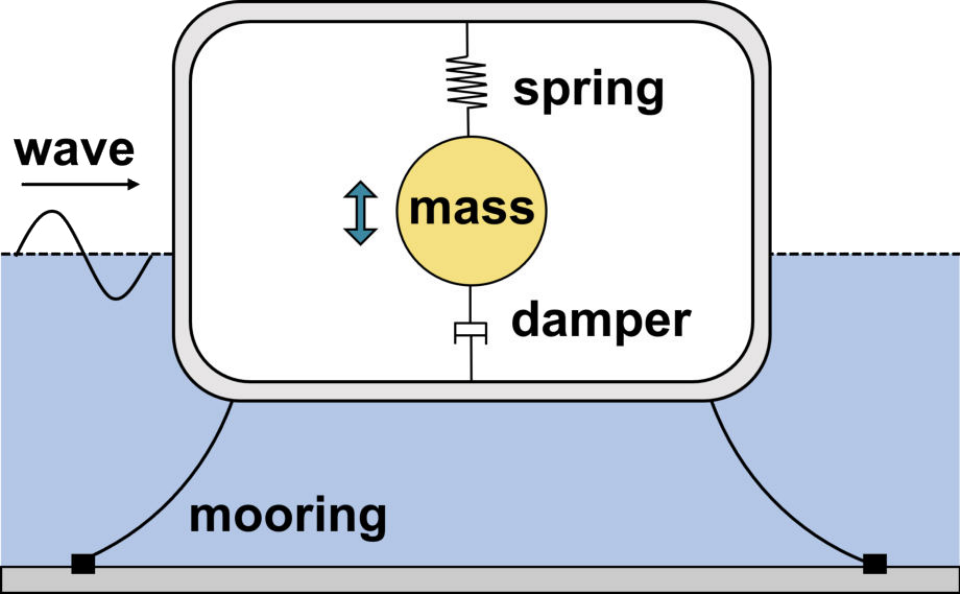}
  \subcaption{}
  \end{minipage}
}
{
  \begin{minipage}{0.32\linewidth}
  \centering
  \includegraphics[width=1.0\linewidth]{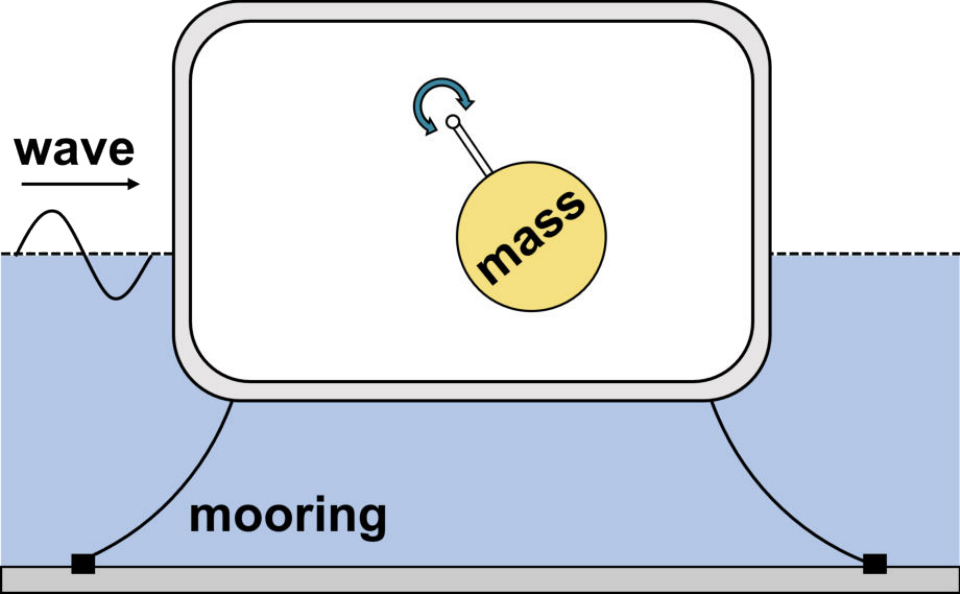}
  \subcaption{}
  \end{minipage}
}
{
  \begin{minipage}{0.32\linewidth}
  \centering
  \includegraphics[width=1.0\linewidth]{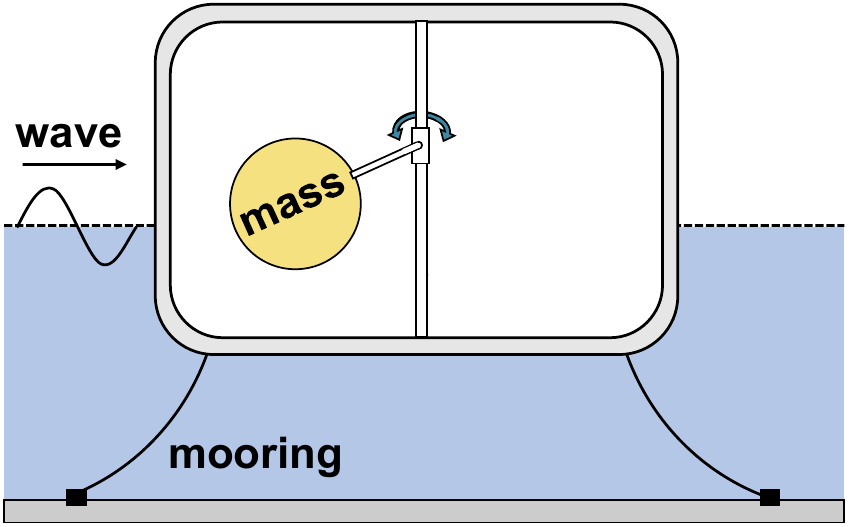}
  \subcaption{}
  \end{minipage}
}
{
  \begin{minipage}{0.32\linewidth}
  \centering
  \includegraphics[width=1.0\linewidth]{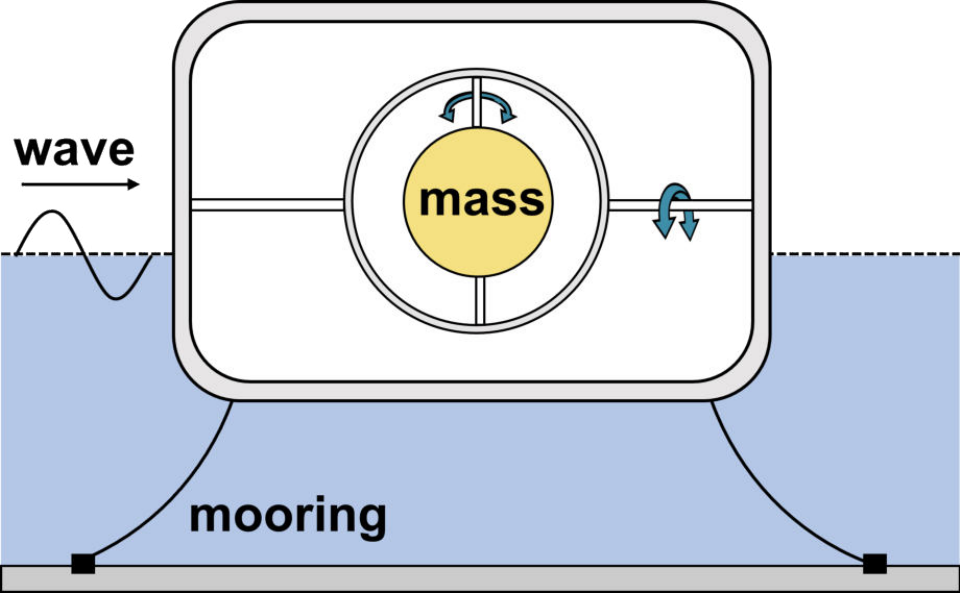}
  \subcaption{}
  \end{minipage}
}
\caption{Mechanical principle of dry-tank En-WEC with (a) horizontal mass-spring system; (b) vertical mass-spring system; (c) horizontal-axis pendulum system; (d) vertical-axis pendulum system; and (e) gyroscopic system}
\label{fig:PTO_dry_tank}
\end{figure}

For dry-tank En-WECs, the mathematical models of their PTO systems can be identified as either translational or rotational oscillators.
The most typical example of a translational oscillator is the mass-spring (MS) system in either horizontal or vertical direction.
Using a MS system, the linear oscillations of the mass can be converted to electricity via the piezoelectricity device \cite{Viet2016, Viet2018}, the rack-pinion-gear or rope-pulley system \cite{Clemente2020}, or the linear electric generator \cite{Jin2020}.
Examples of the rotational oscillator include the pendulum system and the gyroscopic system.
For the pendulum system, the rotational axis between the oscillator and the hull can be either vertical \cite{Jiang2023} or horizontal \cite{Cordonnier2015,Pozzi2018}. 
For the gyroscopic system, a gyroscopic device is installed inside the floater, whose relative inertial motion is used to feed an electric generator through a series of transformation stages \cite{Townsend2013,Vissio2017}.
The aforementioned dry-tank En-WEC concepts with the horizontal/vertical SM system, horizontal/vertical-axis pendulum system, and gyroscopic system are illustrated in Fig. \ref{fig:PTO_dry_tank}.

Compared with conventional Ex-WECs, dry-tank En-WECs naturally have an improved survivability.
However, some inherent challenges of moving mechanical components have not been fully eliminated inside the hull.
First, the mechanical oscillator inside the hull can only be tuned in one frequency, so that the oscillator can resonate out of phase with the floater at the tuned frequency to guarantee an efficient transfer of vibrating energy from the hull to the oscillator.
As the stiffness, friction, or lubrication of the oscillatory system varies with time, the natural frequency of the oscillator inevitably  drifts away from the pre-tuned value.
Thus, retuning maintenance has to be conducted regularly, and the maintenance cost is of course high \cite{Maheen2023}.
Second, since the wave excitation frequency or oscillation frequency of the floater is normally very low, a large stroke for the SMD system or a large length of pendulum is thus required, which may impede its practical applications due to the limited space inside the hull.
Third, the mechanical PTO system directly connects with the hull without buffer medium.
Constantly under irregular impulses of random waves, frequent collisions occurs at the connection or joint of adjacent moving parts of the mechanical PTO system.
All the aforementioned facts harm the durability of an En-WEC.

\begin{figure}[!htp]
\centering
{
  \begin{minipage}{0.32\linewidth}
  \centering
  \includegraphics[width=1.0\linewidth]{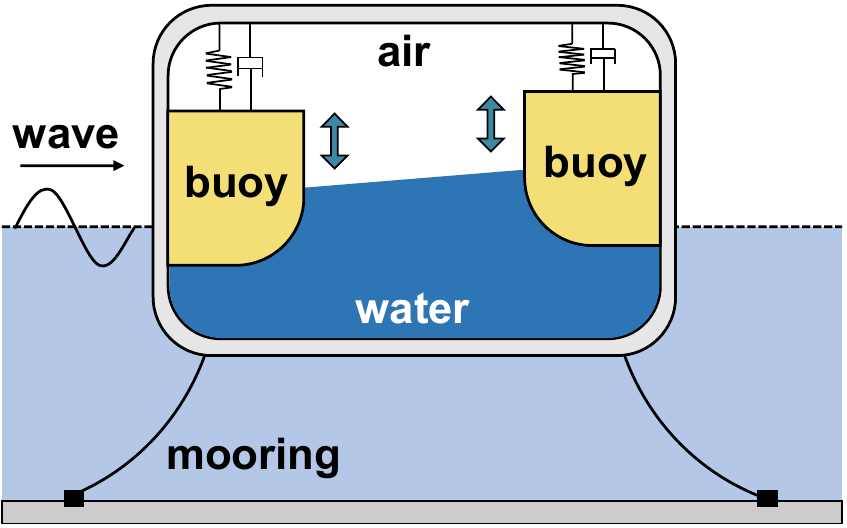}
  \subcaption{}
  \end{minipage}
}
{
  \begin{minipage}{0.32\linewidth}
  \centering
  \includegraphics[width=1.0\linewidth]{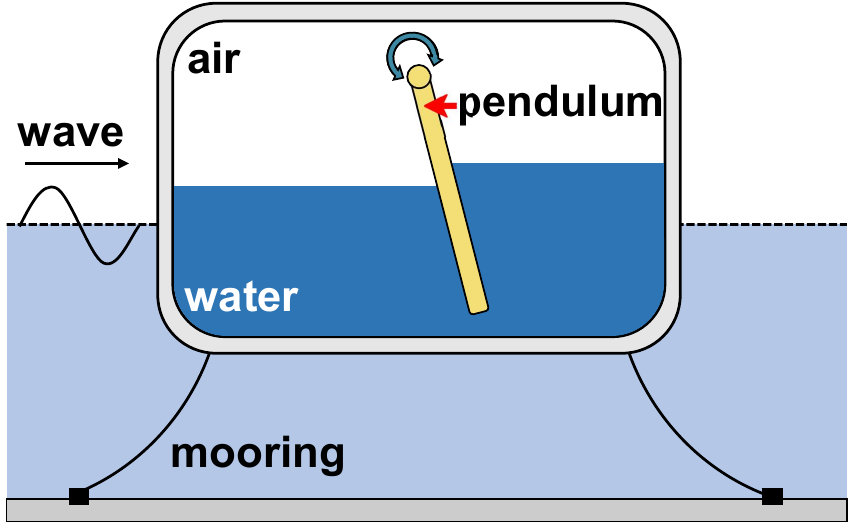}
  \subcaption{}
  \end{minipage}
}
\\
{
  \begin{minipage}{0.32\linewidth}
  \centering
  \includegraphics[width=1.0\linewidth]{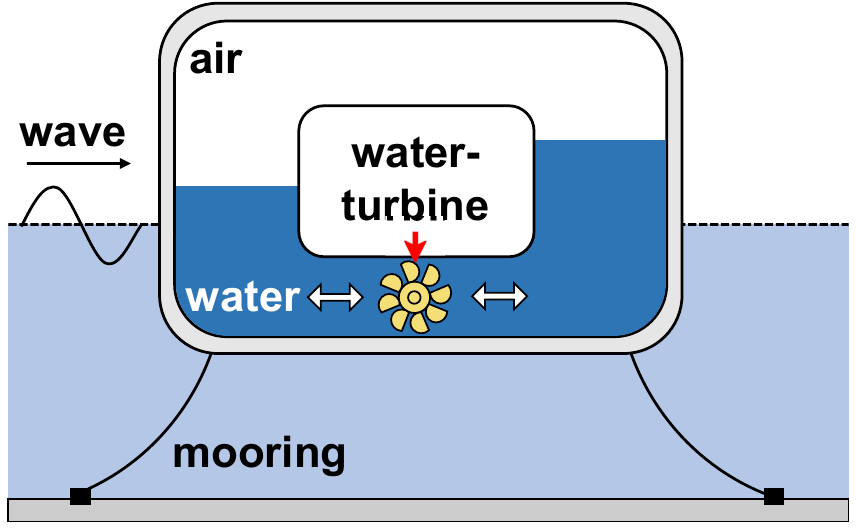}
  \subcaption{}
  \end{minipage}
}
{
  \begin{minipage}{0.32\linewidth}
  \centering
  \includegraphics[width=1.0\linewidth]{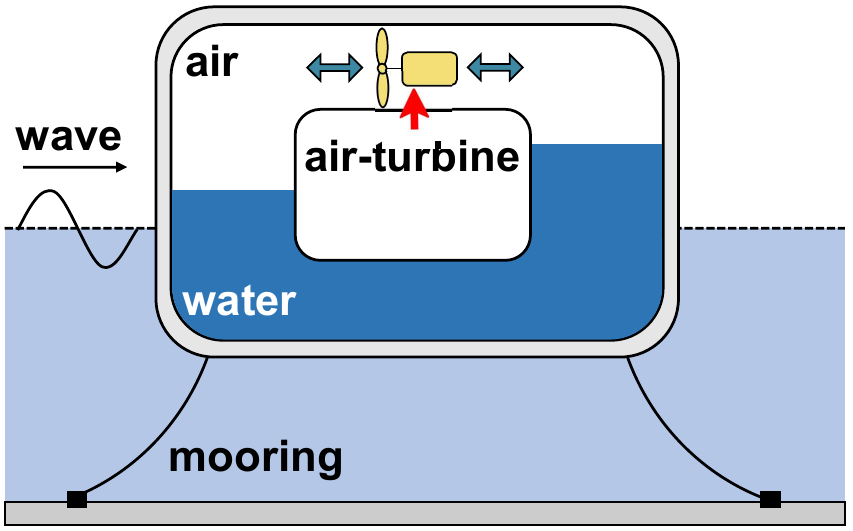}
  \subcaption{}
  \end{minipage}
}
\caption{Mechanical principle of liquid-tank En-WEC with (a) buoy system \cite{Zhang2022}; (b) pendulum system; (c) water-turbine system; and (d) air-turbine system}
\label{fig:PTO_liquid_tank}
\end{figure}

To improve the durability of an En-WEC, a useful strategy is to introduce an extra buffering medium between the floating hull and the internal PTO system.
Unlike the dry-tank En-WEC whose internal PTO system is directly drive by the motion of the hull, the introduced buffering medium can first absorb the kinetic energy of the hull for temporary storage and then drive the PTO system.
Liquids can be filled in the enclosed hull to serve as an buffering medium.
In operation, the floating hull is oscillated by external ocean waves, which subsequently activates the sloshing motion of internal liquid.
The sloshing waves further drive different types of PTO systems for electricity.
Corresponding to each inherent challenge of the dry-tank En-WEC, the liquid-tank En-WEC evidently has the following advantages.
First, natural frequencies of the system can be controlled easily by adjusting the filling level of the liquid tank.
Since the liquid mass is always conserved in an enclosed hull, less tuning maintenance is required compared with the dry-tank En-WEC.
The regular maintenance cost is saved.
Second, the natural oscillation frequency of the liquid body is determined by the geometrical configurations and filling level of the tank. 
A range of low natural frequencies can be easily tuned out within a limited internal space.
Third, the liquid body is a natural frequency modulator of vibrations. The oscillating liquid body can help filter out random frequency components outside the dominant frequency spectra, leading to a smooth hydrodynamic load to the PTO system. 
The liquid tank can be considered as a natural storage of mechanical energy.
Once initiated, sloshing waves in the tank can persist for a long duration, so that less wave energy can be ``wasted'' through wave scattering effects. 
With the aforementioned advantages, the liquid-tank En-WEC theoretically has a good durability.

For liquid-tank En-WECs, only a small number of relevant studies can be found in literature, four categories of which have been recognised, as shown in Fig. \ref{fig:PTO_liquid_tank}.
The first category was first established by Zhang et al. \cite{Zhang2022}, which utilises the buoy-type PTO system to harvest sloshing wave energy in the floating liquid tank, as shown in Fig. \ref{fig:PTO_liquid_tank}(a).
In operation, the floater is oscillated by external ocean waves, which subsequently activates the internal sloshing motion.
After that, the sloshing waves further excite the floating buoys to drive the PTO system.
In the second category (Fig. \ref{fig:PTO_liquid_tank}(b)), a pendulum instead of buoys was used as the PTO system in liquid tank \cite{Gunawardane2021}.
In the third category (Fig. \ref{fig:PTO_liquid_tank}(c)), the liquid tank is U-shaped with a cross-flow water-turbine install in the horizontal tube, converting kinetic energy of bidirectional flows into electricity \cite{Kim20151,Ding2023}.
In the fourth category (Fig. \ref{fig:PTO_liquid_tank}(d)), a pipe is used to connect two air chambers of the U-shaped liquid tank, in which an air-turbine system is used to harvest the energy of bidirectional air flows \cite{Ribeiro2021}.

Among the four categories of liquid-tank En-WECs, the pneumatic type with an air turbine has the simplest machinery system.
Fonseca et al. \cite{Fonseca2010} presented an asymmetric floater with an interior U-shaped liquid tank, named ``UGEN''.
The principle of UGEN is to install an air turbine at a duct connecting two air chambers of UGEN to generate electricity. 
Fonseca and Pessoa \cite{Fonseca2013} further established an numerical model to calculate hydrodynamic responses of the UGEN in waves and the power produced by the PTO system.
Ribeiro e Silva et al. \cite{Ribeiro2016} applied an optimization  procedure to improve the floater geometry, turbine characteristics, and mass distribution of the UGEN.
Ribeiro e Silva et al. \cite{Ribeiro2021} presented an experimental study on the hydrodynamic responses of the UGEN in regular and irregular wave conditions, while perforated plates were used in the model to mimic the pressure loss through the air turbine.
However, the results indicated that the hydrodynamic efficiency of a conventional pneumatic liquid-tank En-WEC is not promising.
Taking a full-scale UGEN of 29.4m long, 24.3m wide and 25.7m tall for example, even at the resonance condition (with dimensionless wave period of 0.4), the Capture Width Ratio (CWR) in regular waves with dimensionless wave amplitudes of $A_{\rm{w}}/L$ = 0.031, 0.062 and 0.103 (with $L$ for wavelength) is only about 0.15, 0.10 and 0.06, respectively \cite{Ribeiro2021}.
This hydrodynamic efficiency is far more lower than an ordinary pneumatic-type Ex-WEC of the similar geometrical dimensions.

The efficiency is a major weakness of the existing pneumatic type liquid-tank En-WECs.
If the system is not designed properly, a large proportion of energy is lost when the liquid's energy transfers to the air and then to the turbine rotor.
Actually, the maximum kinetic energy, which depends on the liquid mass in the tank and the oscillation velocity of the liquid column, has already determined the upper limit of the energy to be harvested.
A deep understanding of the mechanisms behind pneumatic liquid-tank En-WECs is essential for improving the device's efficiency.
However, such a liquid-tank system is a complex multiphase and multiphysics object.
The hydrodynamic behaviour of the liquid column, the aerodynamic status of the air chamber, and the mechanical property of the turbine are closely coupled.
This means the dynamic behaviours of the liquid, air, and turbine mutually determine each other simultaneously.
Previous practices using a perforated disk to represent the impulsive air turbine (e.g. Ribeiro e Silva et al. \cite{Ribeiro2021}) cannot reflect the complex coupling of hydro-aero-turbo dynamics.
At present, there still lacks an integrated study on coupled hydro-aero-turbo dynamics of this specialized liquid tank system.

To bridge this gap, this study develops an integrated numerical model for pneumatic liquid-tank En-WECs. 
The coupled hydro-aero-turbo dynamics and power harvesting properties of the liquid-tank system are investigated.
A scaled prototype of the wave-energy-harvesting (WEH) liquid tank with an impulse air turbine system is created to experimentally validate the numerical model.
A series of multi-layered impulse air turbine systems (MLATS) are introduced to replace conventional Wells turbines in the liquid tank for efficiency improvement.
The optimal parameters of the MLATS are investigated. 
The MLATS also shows promising reliability compared to existing single-rotor turbine designs.
Section \ref{sec:design} describes the overall concept of the WEH liquid tank. 
The parameter details of the MLATS are also provided.
Section \ref{sec:num} offers the mathematics and numerical settings of the coupled hydro-aero-turbo-dynamic model.
Section \ref{sec:exp} presents the physical model of the scaled prototype and the measurement system.
Section \ref{sec:Results} compares the experimental and numerical results, explores the effects of MLATS parameters, and identifies the advantages of the present wave-energy-harvesting liquid tank.
Finally, conclusions are given in Section \ref{sec:concl}.

\section{Concept description of wave-energy-harvesting liquid tank} \label{sec:design}

\begin{figure}[!htp]
\centering
{
  \begin{minipage}{0.55\linewidth}
  \centering
  \includegraphics[width=1.0\linewidth]{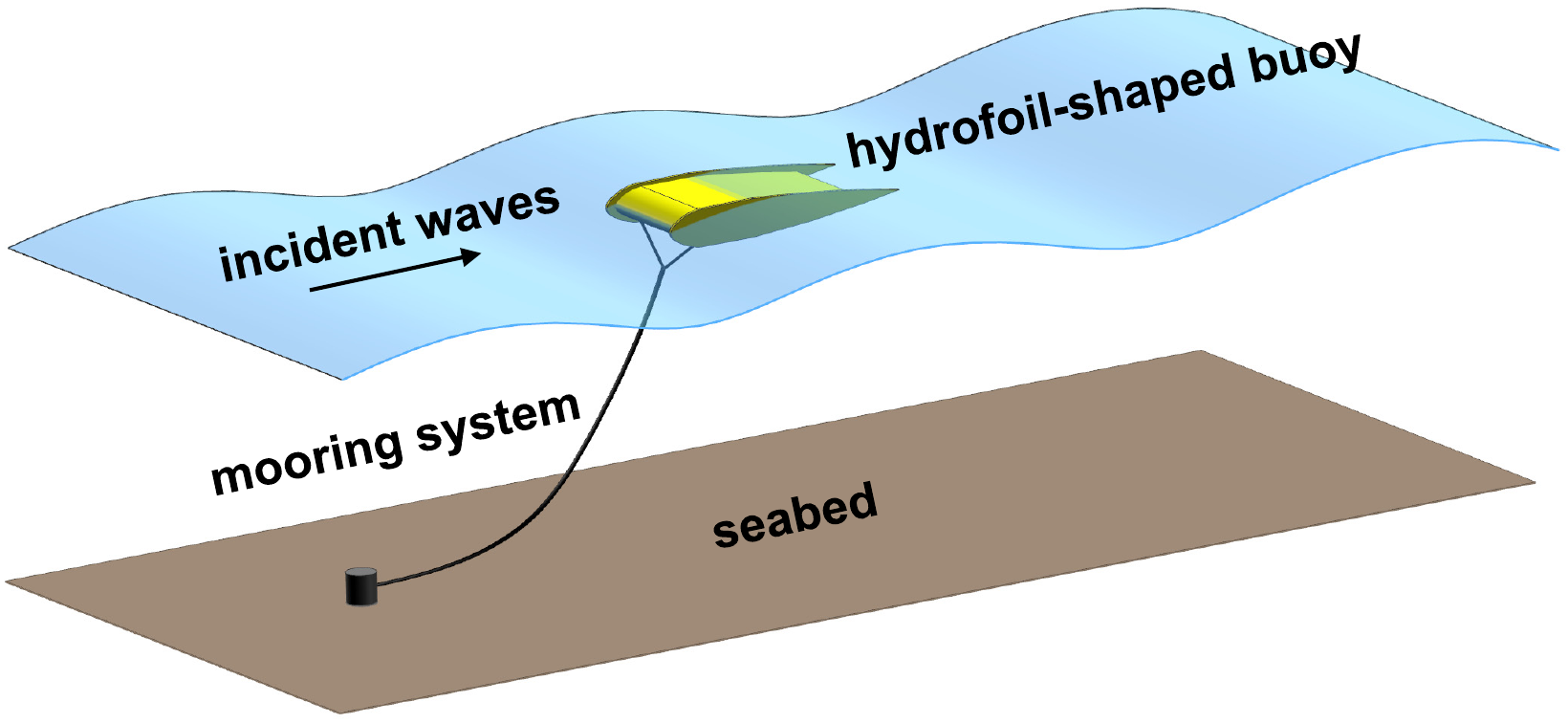}
  \subcaption{}
  \end{minipage}
}
{
  \begin{minipage}{0.4\linewidth}
  \centering
  \includegraphics[width=1.0\linewidth]{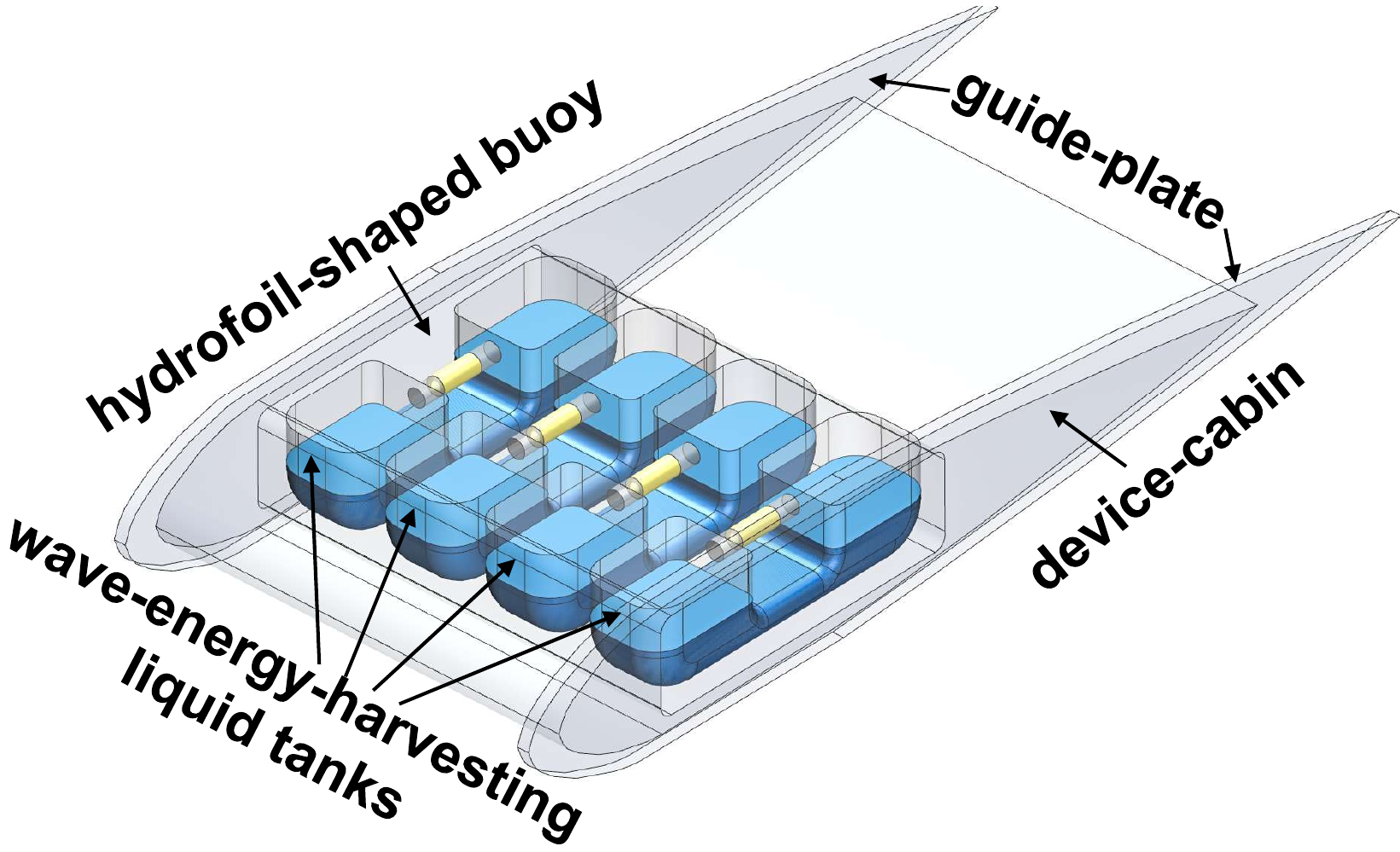}
  \subcaption{}
  \end{minipage}
}
{
  \begin{minipage}{0.35\linewidth}
  \centering
  \includegraphics[width=1.0\linewidth]{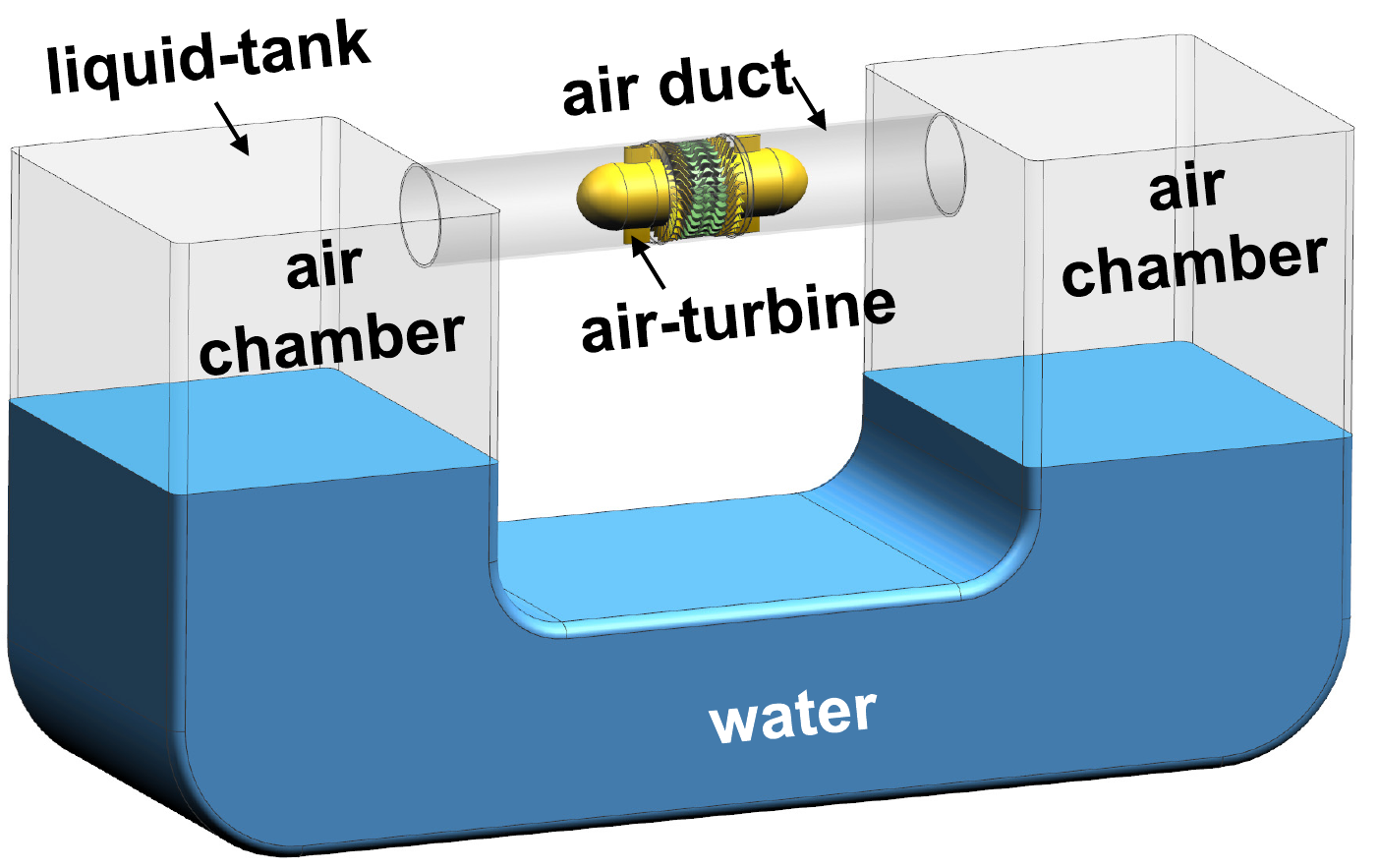}
  \subcaption{}
  \end{minipage}
}
{
  \begin{minipage}{0.35\linewidth}
  \centering
  \includegraphics[width=1.0\linewidth]{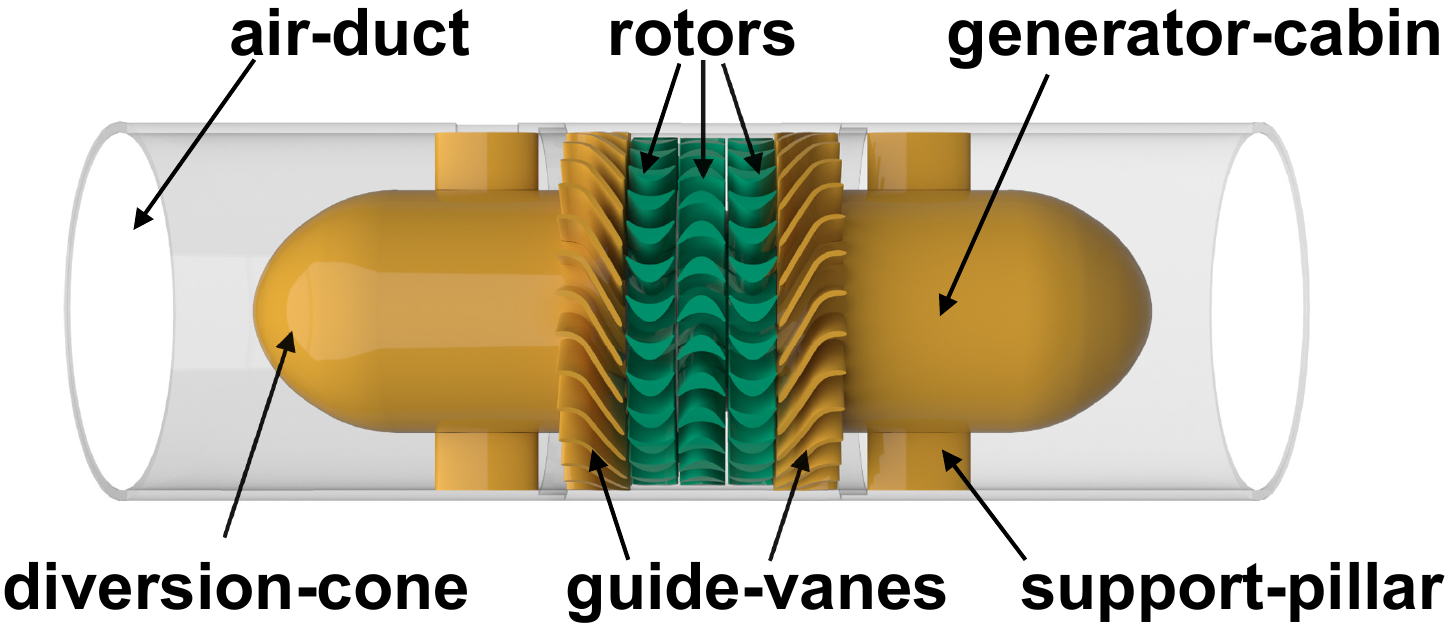}
  \subcaption{}
  \end{minipage}
}
\caption{(a) External appearance of hydrofoil-shaped buoy; (b) concept of wave-energy-harvesting liquid tanks inside the buoy; (c) components of wave-energy-harvesting liquid tank; (d) components of multi-layered impulse air-turbine system}
\label{fig:Enclosed WEC}
\end{figure}

This study considers a novel design of WEC device proposed by the authors in 2020.
The external appearance of the device is a hydrofoil-shaped buoy, as shown in Fig. \ref{fig:Enclosed WEC} (a). 
The buoy is a concealed floater on the free surface, with most of its volume submerged in water. 
Guide-plates are installed on both sides of the buoy to regulate the water-flow direction. 
The buoy is connected to the seabed via a single-point mooring system, ensuring it always faces incident waves and is adaptive to wave directions. 
Multiple U-shaped WEH liquid tanks are arranged in parallel inside the buoy (see Fig. \ref{fig:Enclosed WEC} (b)). 
Each WEH liquid tank has two sealed air chambers, and these two chambers are connected by an air duct. 
The air duct is equipped with a novel MLATS that converts the kinetic energy of air flow into electricity, as shown in Fig. \ref{fig:Enclosed WEC} (c). 
During operation, the hydrofoil-shaped buoy rides on the wave and experiences longitudinal motions along the wave direction. 
The buoy excites internal liquid-body oscillations, which further causes bi-directional air flows through the air duct. 
The airflows can continuously drive the MLATS to produce electricity.
This study mainly focuses on the inherent properties of the coupled hydro-aero-turbo dynamics of the WEH liquid tank.

The details of the proposed MLATS are illustrated in Fig. \ref{fig:Enclosed WEC} (d).
The main body of the air-turbine system is a hollowed cylinder, which serves as the generator cabin.
The generators and control systems are located inside the generator cabin. 
Diversion cones are installed at both ends of the cylinder to compress incoming airflows.
The cylinder is fixed to internal surface of the air duct via two groups of guide-vanes.
The guide vanes are composed of linear and circular arcs to deflect the direction of inlet airflows.
Multiple layers of rotors are installed between two sets of guide-vanes.
Each rotor is connected to an isolated generator inside the generator cabin.
The blade profile of each rotor consists of circular and elliptical arcs.
Adjacent rotors are installed in the opposite way.
As a result, the deflected airflow behind one rotor hits the blades of the downstream rotor, giving a tangential drive to the rotor.
Because of the symmetrical nature of the guide vanes and rotor blade, each rotor always rotates in the same direction, regardless of the airflow direction. 
Ingeniously designed mechanical transmission systems are hidden in the generator cabin. 
Due to technical confidentiality requirements, these mechanical transmission systems not related to hydro-aero-turbo dynamic simulations are not disclosed here.

\begin{figure}[!htp]
\centering
{
  \begin{minipage}{0.65\linewidth}
  \centering
  \includegraphics[width=1.0\linewidth]{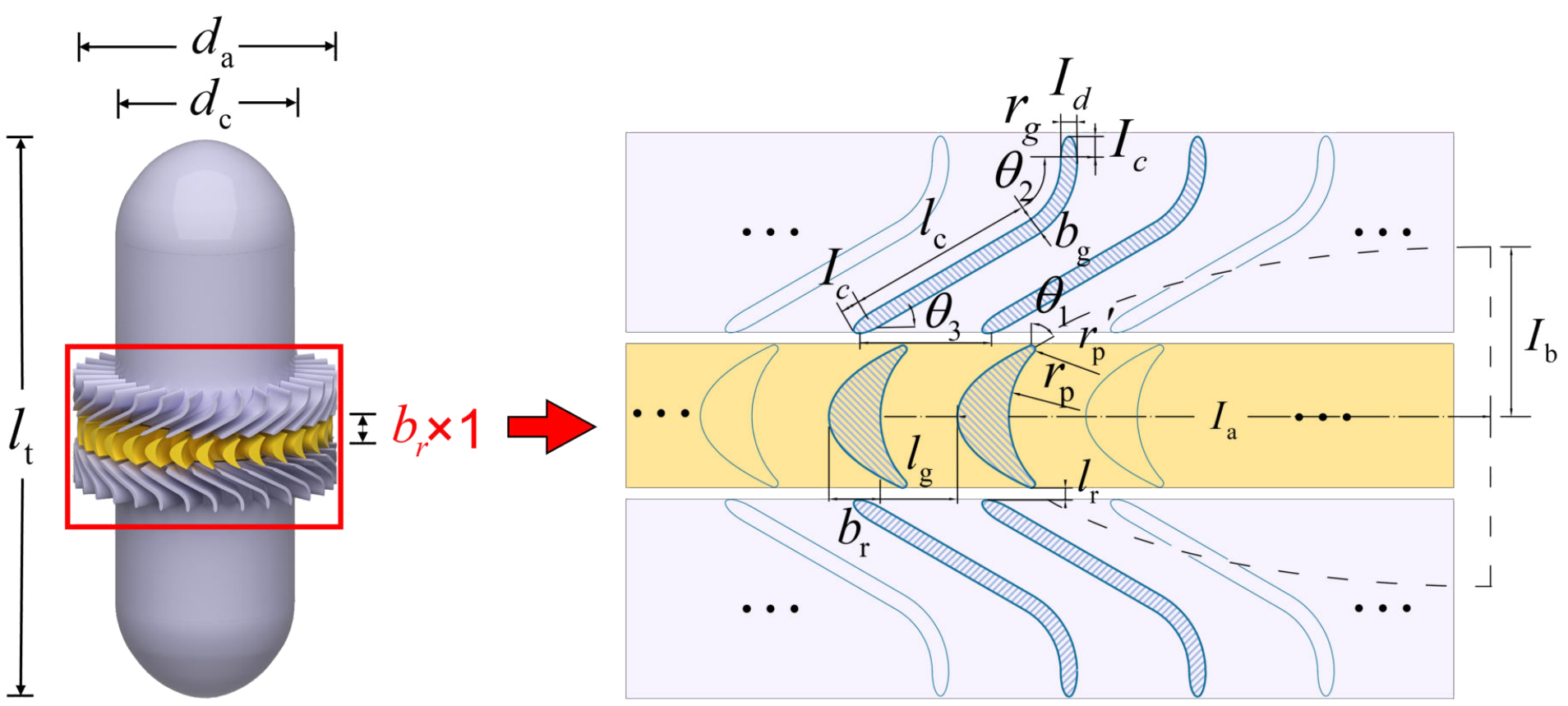}
  \subcaption{}
  \end{minipage}
}
{
  \begin{minipage}{0.65\linewidth}
  \centering
  \includegraphics[width=1.0\linewidth]{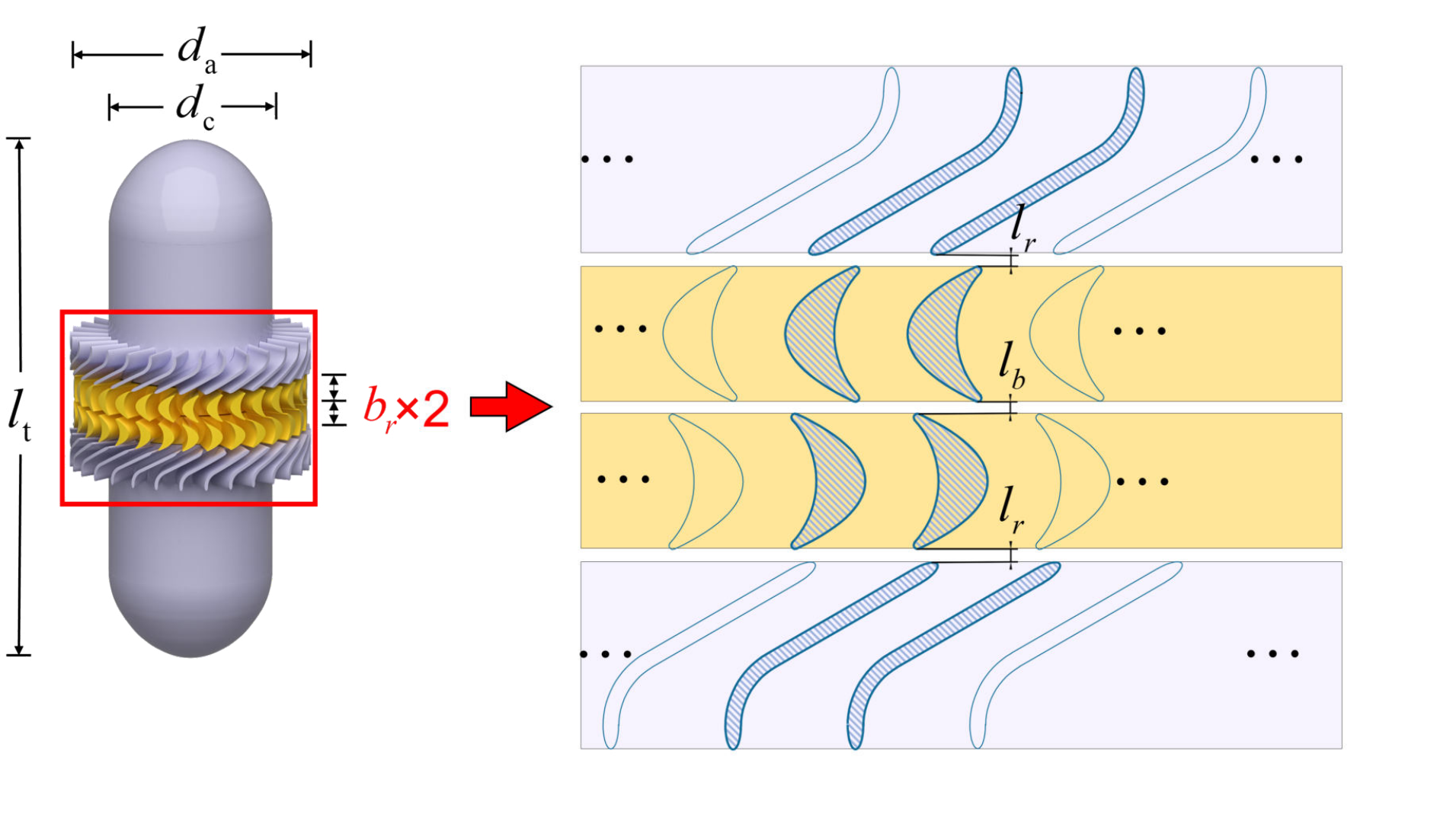}
  \subcaption{}
  \end{minipage}
}
{
  \begin{minipage}{0.65\linewidth}
  \centering
  \includegraphics[width=1.0\linewidth]{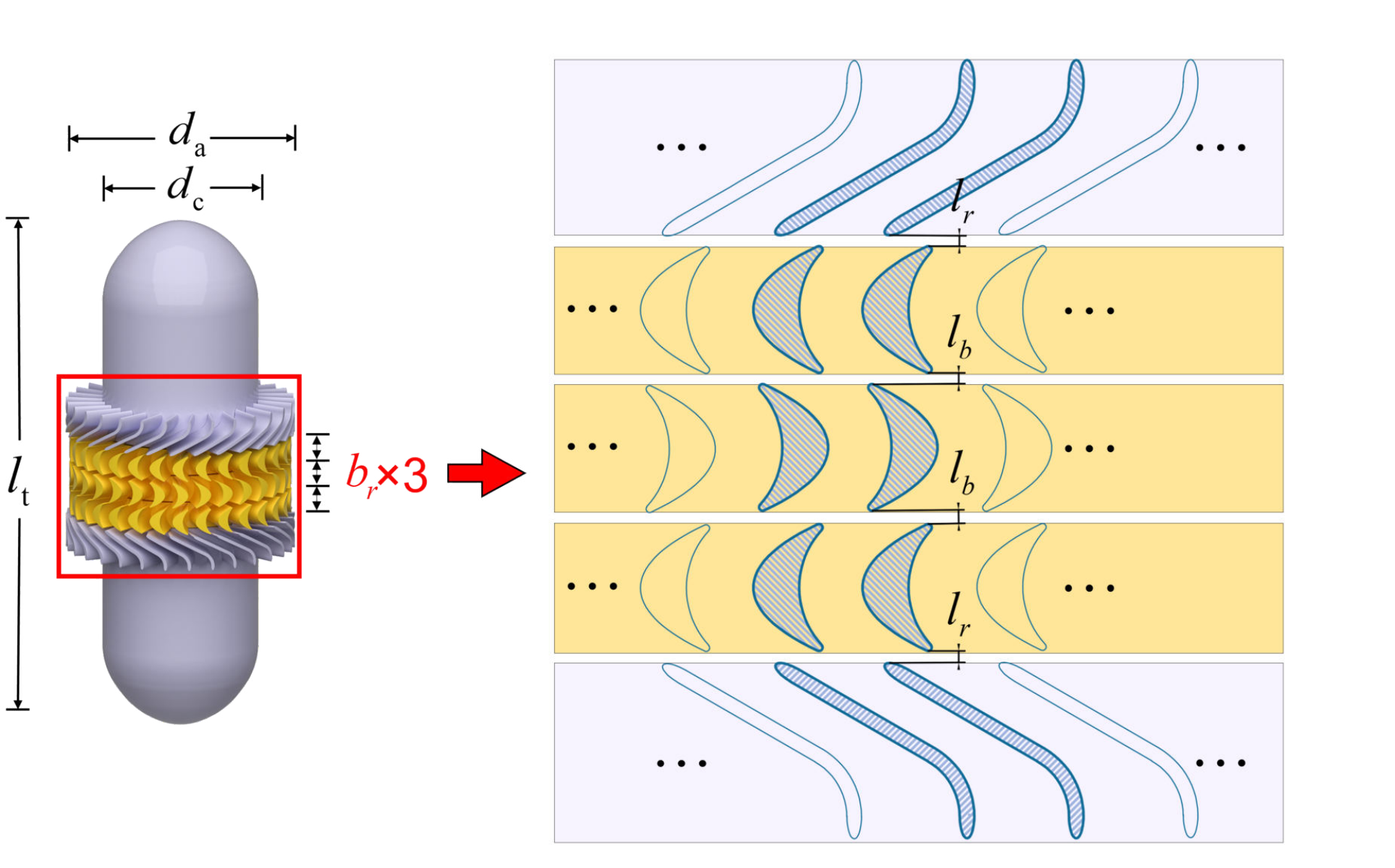}
  \subcaption{}
  \end{minipage}
}
\caption{Design parameters of multi-layered impulse air turbine systems for (a) Turbine-L1, (b) Turbine-L2, and (c) Turbine-L3}
\label{fig:Impulse turbine}
\end{figure}

This study specifically designed three MLATSs, namely Turbine-L1, Turbine-L2, and Turbine-L3, as shown in Fig. \ref{fig:Impulse turbine}.
Turbine-L1, Turbine-L2, and Turbine-L3 are featured with one, two, and three layers of rotors, respectively.
The following parameters are introduced to define the geometry of the turbine system.
The external diameter of guide vanes (also the internal diameter of the air duct) is denoted by $d_\text{a}$. 
The diameter of the cylindrical main body is $d_\text{c}$. 
The total length of the turbine system (the distance between two vertices of diversion cones) is $l_\text{t}$. 
The breadth of each rotor is $b_\text{r}$. 
The gap size between adjacent rotors is $l_\text{b}$. 
The spacing between guide vanes and the adjacent rotor is $l_\text{r}$.
For each rotor blade, the height measured from the blade root on the cylinder to the blade tip is $h_\text{r}$; the cross-sectional profile is formed by an circular arc of radius $r_\text{p}$ and an elliptical arc with semi-major axis $I_\text{a}$ and semi-minor axis $I_\text{b}$; and the angle of attack is $\theta_\text{1}$.
For each guide vane, its has a thickness $b_\text{g}$ and a height $h_\text{g}=(d_\text{a}-d_\text{c})/2$; the cross-sectional profile is formed by a straight segment of length $l_\text{c}$, and an arc of radius $r_\text{g}$ and curvature angle $\theta_2$; both tips are rounded off by ellipses of a semi-major axis $I_\text{c}$ and a semi-minor axis $I_\text{d}$. 
The angle of incidence of each guide vane is $\theta_3$, and the distance between guide vanes is $l_\text{g}$. 

\section{Integrated numerical model for wave-energy-harvesting liquid tank}\label{sec:num}

\begin{figure}[!htp]
\centering
\includegraphics[scale=0.35]{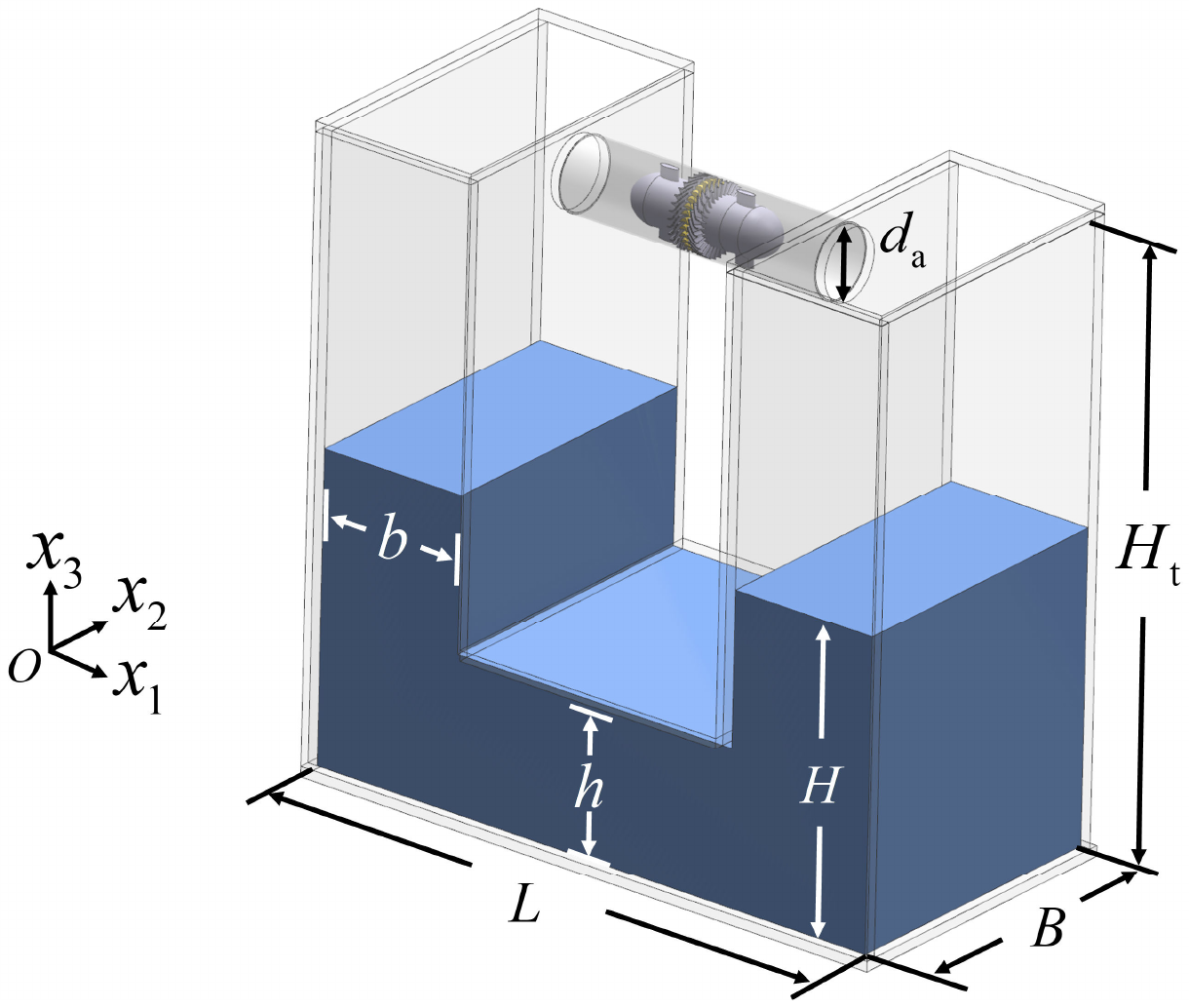}
\caption{Illustration of wave-energy-harvesting liquid tank model}
\label{fig:O-shaped tank}
\end{figure}

The simplified model of WEH liquid tank is considered, as shown in Fig. \ref{fig:O-shaped tank}. 
The liquid tank has a rectangular cross section.
The inner space of the liquid tank has a height of $H_t$ and a breadth of $B$.
The length and height of the horizontal water column in the tank are $L$ and $h$, respectively. 
For each vertical water column, the height is $H$ and the breadth is $b$. 
A circular air duct with a diameter of $d_{\rm{a}}$ connects the tops of two sealed chambers of the liquid tank.
The MLATS is installed in the air duct.
A Cartesian coordinate system $O-x_1 x_2 x_3$ is established, with its origin $O$ coinciding with the center of the bottom of the liquid tank.
The $Ox_1$-axis is along the lengthwise edge of the tank, the $Ox_3$-axis points upward, and the $Ox_2$-axis is determined by the right-hand rule.

The fluid domain consists of both the liquid and air phases. 
The air phase is considered compressible to assess the turbine's energy conversion performance.
The viscous flow theory is used to describe the motion of fluid flows.
Hereafter, all equations are expressed in tensor notation, following the Einstein summation convention.
The Reynolds-Averaged Navier-Stokes (RANS) equations are adopted as the governing equations for the fluid domain:
\begin{equation}
\begin{aligned}
&\frac{\partial \rho}{\partial t}+\frac{\partial (\rho \overline{u}_i)}{\partial x_i} = 0\\
&\frac{\partial (\rho \overline{u}_j)}{\partial t}+\frac{\partial (\rho \overline{u}_i\overline{u}_j)}{\partial x_i}=f_i-\frac{\partial \overline{p}}{\partial x_j}+\frac{\partial \tau_{ij}}{\partial x_i}
\end{aligned}
\end{equation}
where $\rho$ is the density of the fluid, $\overline{u}_i$ ($i = 1,2,3$ corresponding to the spatial directions) is the mean velocity component in the $i$-th direction, and $f_i$ is the body force, $\overline{p}$ is the mean pressure and $\tau_{ij}$ is the stress tensor.
For a Newtonian fluid, the stress tensor can be written as:
\begin{equation}
\tau_{ij}=\mu\left(\frac{\partial \overline{u}_i}{\partial x_j}+\frac{\partial \overline{u}_j}{\partial x_i}-\frac{2}{3}\delta_{ij}\frac{\partial \overline{u}_k}{\partial x_k}\right)
\end{equation}
with $\mu$ being the dynamic viscosity and $\delta_{ij}$ the Kronecker delta.
To close the governing equations, the $k-\epsilon$ turbulence model is used and the ideal gas equation of state is fulfilled in the air phase
\begin{equation}
	\begin{aligned}
	p=\rho R T_\text{0}
	\end{aligned}
	\label{eq:ideal gas equation}
\end{equation}
where $R$ is the universal gas constant, and $T_\text{0}$ is the absolute temperature.
No-slip wall boundaries are applied as the wall boundary condition for all solid surfaces.

The liquid tank experiences horizontal sinusoidal oscillations along the $Ox_1$-axis. 
The body force $f_i$ consists of two components: the gravitational force in the vertical direction and the inertia force in the horizontal direction.
The turbine rotors are rotated passively by the reciprocating airflow. 
According to Newton's second law, the rotational motion of each rotor is governed by
\begin{equation}
	\begin{aligned}
	J {\dot{\omega}}= T_\text{a}-T_\text{r}
	\end{aligned}	
	\label{eq:The torque equation}
\end{equation}
where $J$ represents the moment of inertia of a rotor, $\dot{\omega}$ represents angular acceleration, $T_\text{a}$ represents the total driving torque due to the pressure difference across the rotor, and $T_\text{r}$ refers to the reaction torque from the generator.
The generator's reaction torque consists of inherent resistances and power take-off effects.
For simplification, $T_\text{r}$ is considered proportional to $\omega$ as\cite{Song2016}:
\begin{equation}
	\begin{aligned}
	T_\text{r}=\left(\mu+\mu_\text{pto}\right) \omega
	\end{aligned}	
	\label{eq:The load equation}
\end{equation}
where $\mu$ and $\mu_\text{pto}$ are the inherent resistance and PTO damping coefficients, respectively. 
The PTO damping coefficient $\mu_\text{pto}$ is generated by the reaction force of the stator magnets. 
This occurs when the turbine rotor drives the generator rotor coil to cut magnetic flux lines for power generation. 
The value of $\mu_\text{pto}$ is related to the number of turns of the generator's coil and the magnetic field strength.
In this study, $\mu_\text{pto}$ is considered a constant, with a unit $\text{kg} \cdot \text{m} \cdot \text{s}^{-1}$.
The power generated by each rotor can be calculated through
\begin{equation}
	\begin{aligned}
	 P(t)=\mu_\text{pto} \omega^{2}
	\end{aligned}	
	\label{eq:Power}
\end{equation}

\begin{figure}[!htp]
\centering
\includegraphics[scale=0.4]{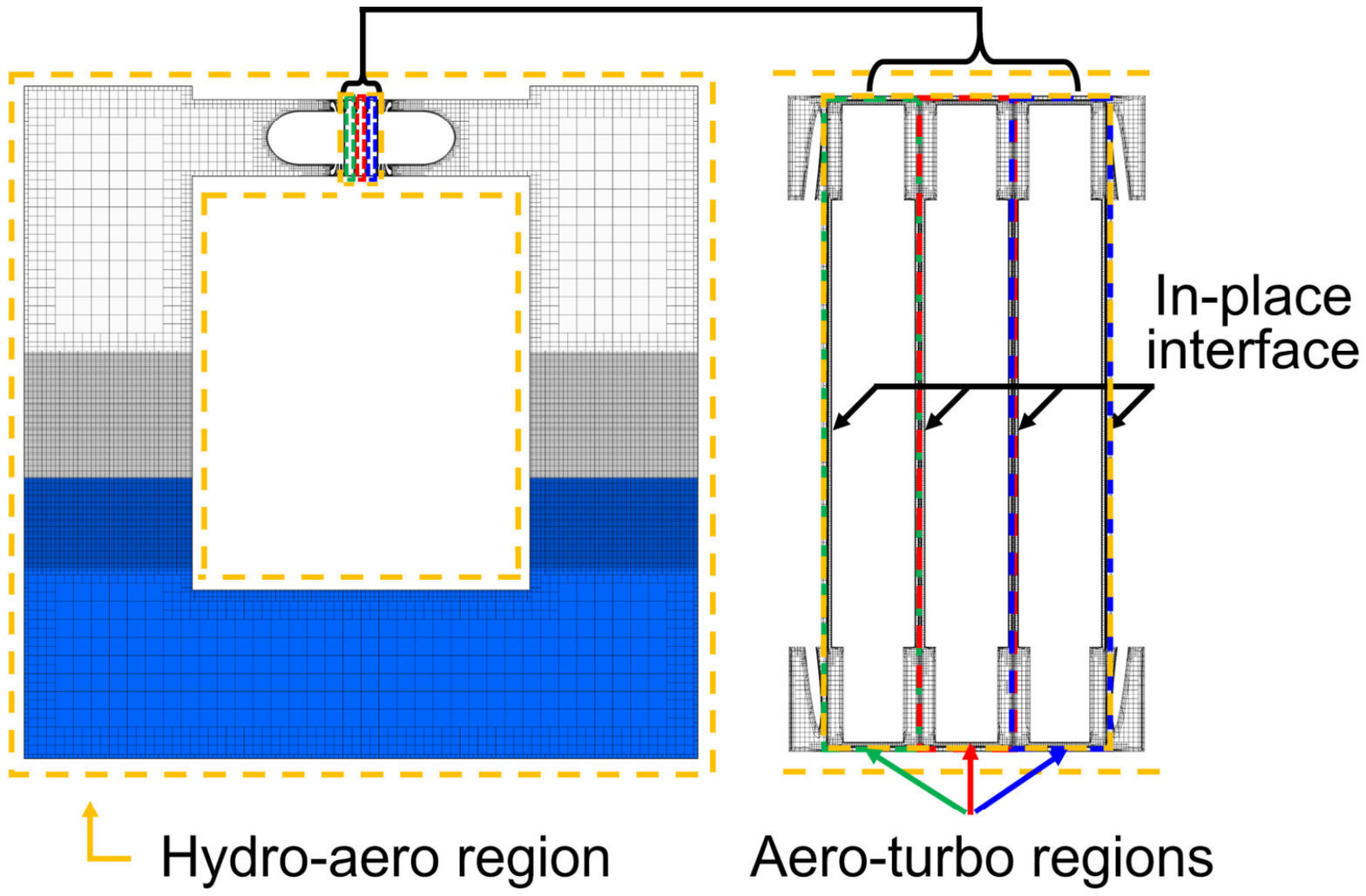}
\caption{Illustration of hydro-aero and aero-turbo regions and zoomed-in view of aero-turbo regions}
\label{fig:Mesh}
\end{figure}

\begin{figure}[!htp]
\centering
\includegraphics[scale=0.4]{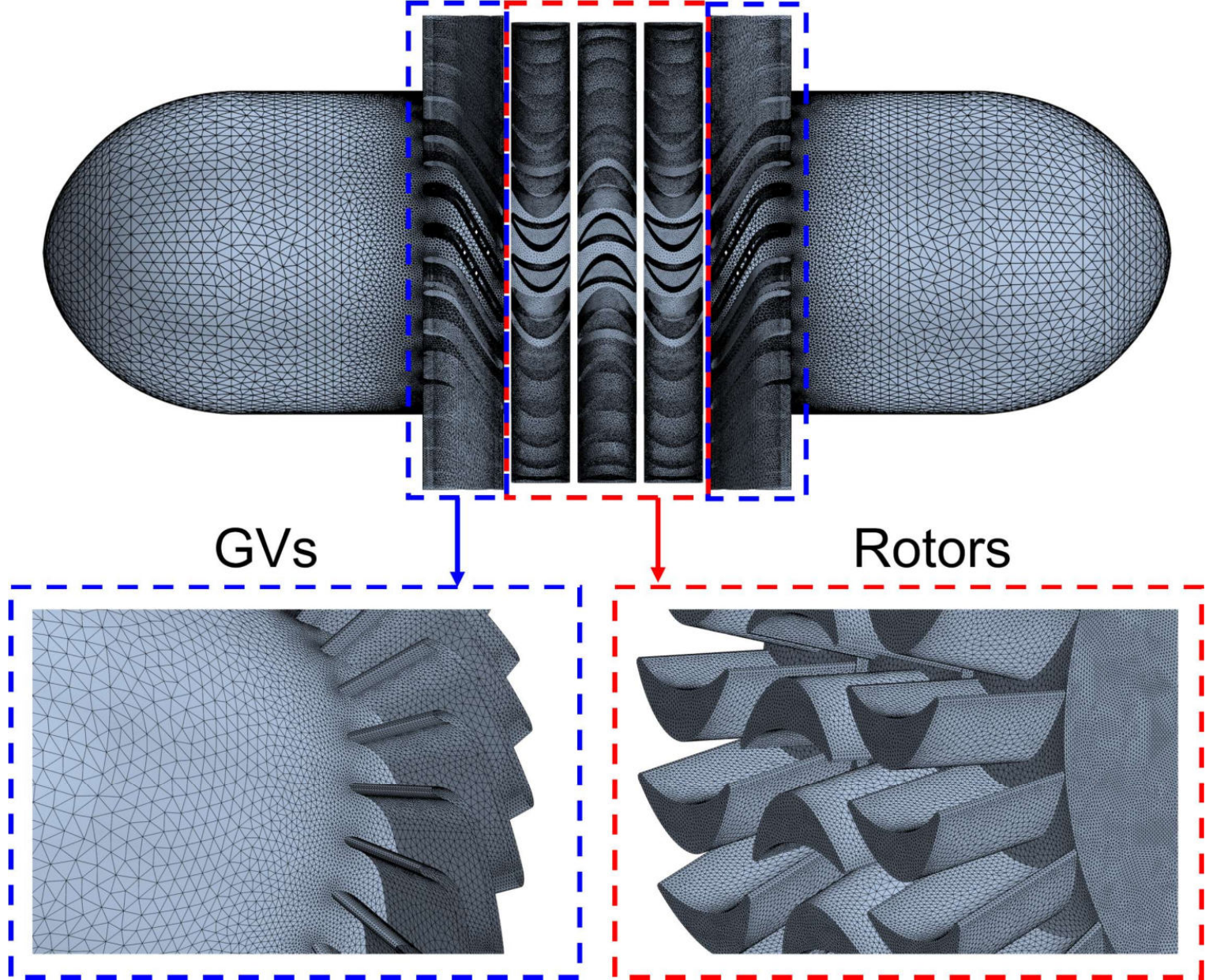}
\caption{Surface mesh of multi-layered air turbine system}
\label{fig:Mesh2}
\end{figure}

The commercial computational fluid dynamics (CFD) software, StarCCM+, is used to implement the aforementioned mathematical process. 
The governing fluid equations are discretized by the finite volume method (FVM). 
The semi-implicit method for pressure linked equations (SIMPLE) scheme is employed as the pressure-velocity coupling algorithm in the solver. 
The second-order upwind scheme is employed to discretize the convection term.
The free-surface motions inside the tank are captured using the High Resolution Interface Capturing (HRIC) method.
The fluid domain is divided into a hydro-aero region and $N$ aero-turbo regions, as shown in Fig. \ref{fig:Mesh}. 
Here, $N$ stands for the number of turbine rotors. 
For each rotor, two circular interfaces are set at its upstream and downstream, separating out an aero-turbo region.
The hydro-aero region is the remaining area of the fluid domain after separating out $N$ aero-turbo regions. 
Two adjacent regions share the same interface.
The numerical mesh is generated for each region. 
Each mesh mainly consists of hexahedral cells. 
Some trimmed polyhedral cells are used near fluid boundaries.
In the hydro-aero region, the mesh near the air-water interface, tank walls, air-duct surface, and guide vanes (GVs) is specifically refined for accuracy. 
In each aero-turbo region, similar mesh cells are created in the air medium. 
The smallest mesh cell is found at the tip of each rotor blade.
The aero-turbo mesh moves with the rotor and is time-invariant relative to the rotor.
At any instant, the interface between adjacent regions facilitates the transfer of physical quantities such as fluid mass, momentum, and energy.
The meshes on the surfaces of GVs and turbine rotors are depicted in Fig. \ref{fig:Mesh2}.

\begin{figure}[!htp]
\centering
\includegraphics[scale=0.5]{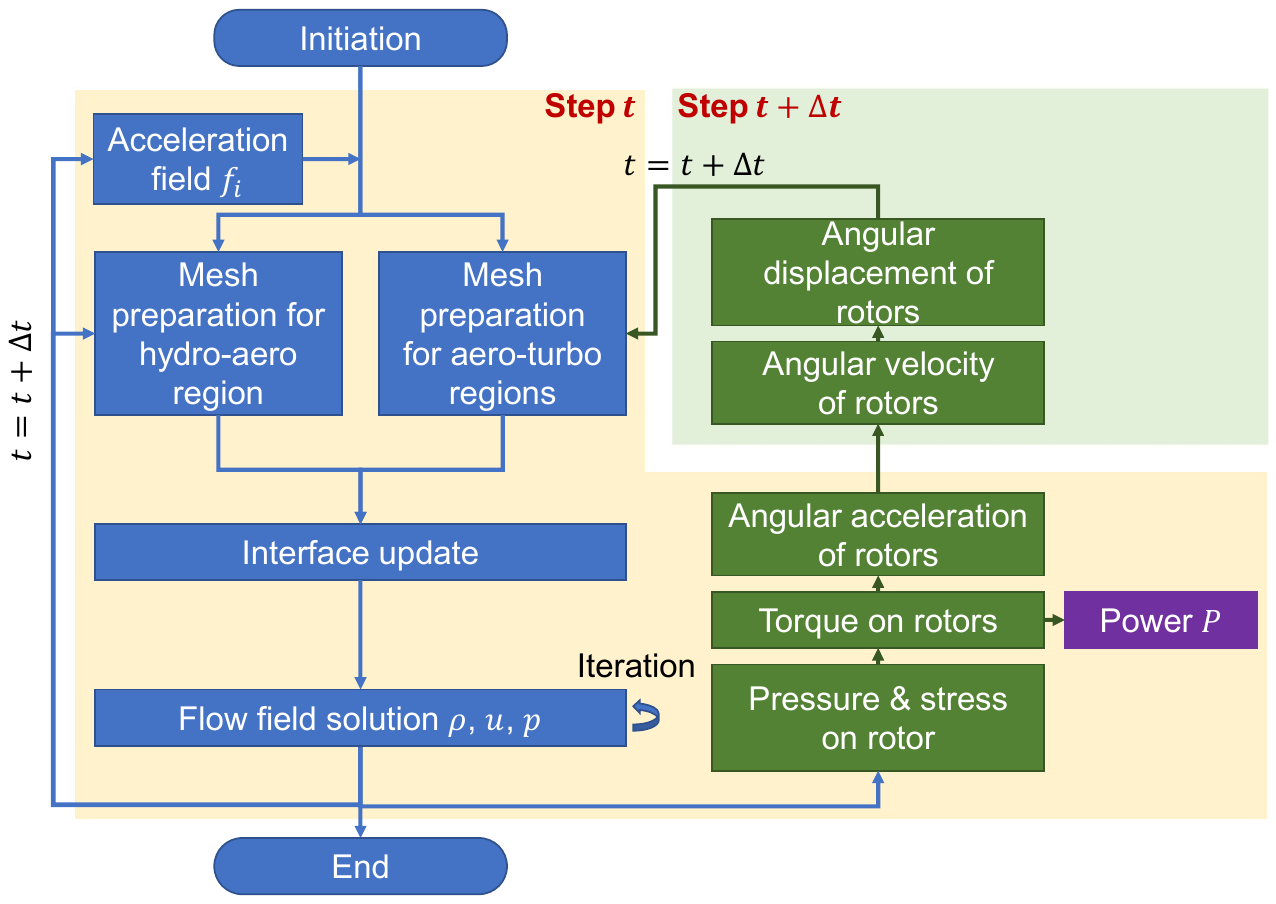}
\caption{Flow chart of integrated numerical model for coupled hydro-aero-turbo dynamics of wave-energy-harvesting liquid tank}
\label{fig:flow}
\end{figure}

A detailed procedure of the integrated numerical model is depicted in Fig. \ref{fig:flow}.
At each time step $t$, the value of $f_i$, which reflects the acceleration field affected by the shaking table, is updated.
The numerical mesh of the hydro-aero region is fixed during the simulation once it is generated in the initial step.
In each aero-turbo region, the mesh rotates according to the angular displacement of the rotor.
The mesh cells in the aero-turbo region are time-invariant relative to the rotor.
A bidirectional transfer of computational quantities is achieved on the interface between each pair of adjacent regions.
Next, the flow field is solved iteratively. 
This results in the distribution of density, velocity, and pressure within the fluid domain.
The pressure and stress on the surface of each rotor can be used to calculate the torque on the rotors through integration.
The angular acceleration of each rotor can be found by using Eq. (\ref{eq:The torque equation}). 
Through time marching, the angular velocity, denoted as $\omega$, and angular displacement, $\phi$, of each rotor are updated via the following implicit scheme:
\begin{equation}
	\begin{aligned}
	\omega(t+\Delta t)=\omega(t)+[\dot{\omega}(t)+\dot{\omega}(t+\Delta)] \Delta t/2 \\
	\phi(t+\Delta t)=\phi(t)+[\omega(t)+\omega(t+\Delta)] \Delta t/2	
	\end{aligned}	
	\label{eq:Newton's Second Law2}
\end{equation}
where $\Delta t$ is the time step.
Then, the calculation enters the loop at the $t+\Delta t$ step. 

\section{Scaled prototype and experimental set-ups}\label{sec:exp}

\begin{figure}[!htp]
\centering
\includegraphics[scale=0.5]{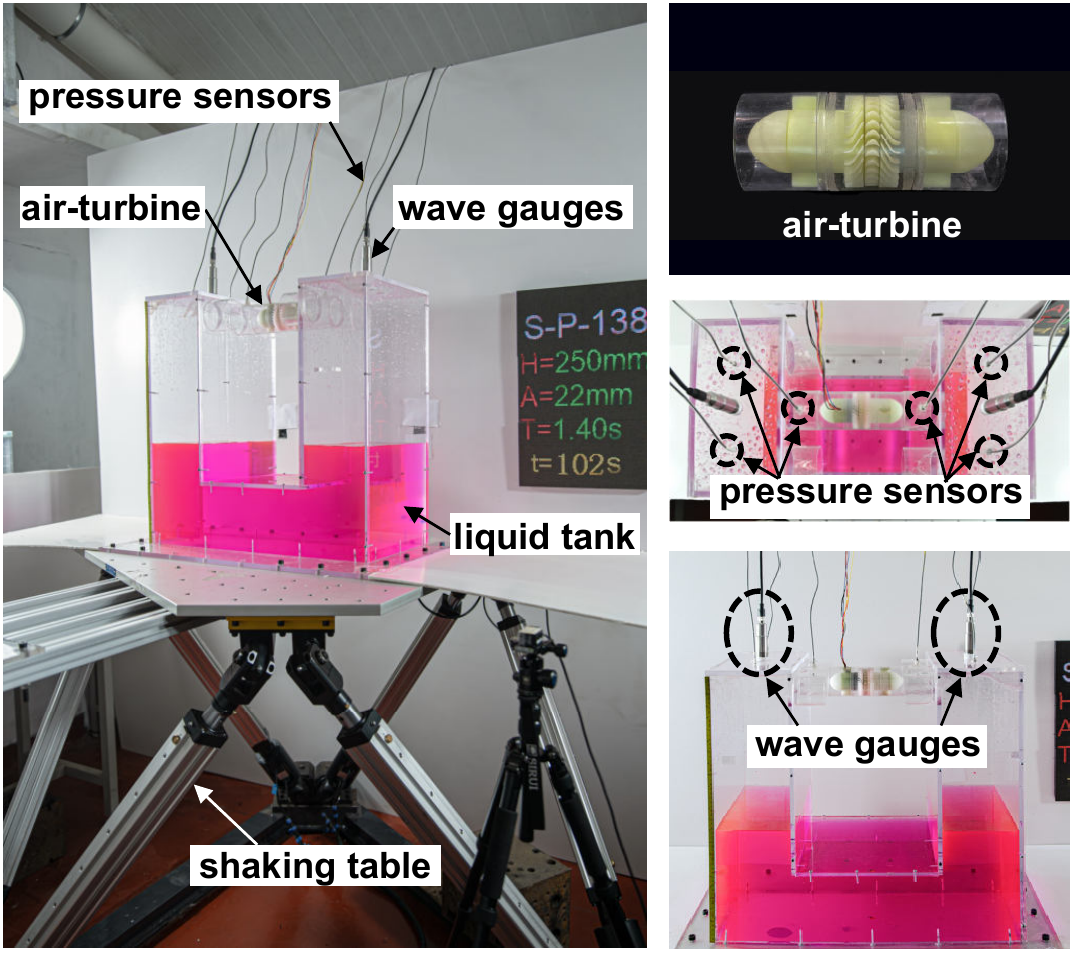}
\caption{Scaled prototype of wave-energy-harvesting liquid tank and measurement system}
\label{fig:Test platform}
\end{figure}

As shown in Fig. \ref{fig:Test platform}, a scaled prototype of the WEH liquid tank with Turbine-L1 is established for experimental evaluation.
The liquid tank consists of polymethyl methacrylate plates and tubes with a thickness of $d_{\rm{w}}=1.0 $ mm.  
The detailed geometry parameters of the liquid tank model are listed in Table \ref{tab:tank}.
For the Turbine-L1 model, all components are made of laser-cured resin materials through 3D printing technology.
The cylindrical main body of the turbine system has two sets of 30 guide vanes.
Each rotor has 30 identical blades. 
The detailed geometry parameters of the air-turbine model are described in Table \ref{tab:turbine}. 
A servo motor (version MF4005 V1 from Lingkong Technology Company) is installed inside the generator cabin of the air-turbine system. 
The servo motor connects to the rotor. 
It acts as the source of the turbine's inherent resistance and measures the rotor's rotational speed. 
\begin{table}
\centering
\caption{Geometry parameters of liquid tank model in physical experiment}
\label{tab:tank}
\begin{tabular}{cccccc}
\hline
Parameter  & Value & Unit & Parameter & Value & Unit \\
\hline
$L$ & 0.600  & m &   $b$        & 0.150  & m      \\
$B$ & 0.300  & m &   $h$        & 0.150  & m     \\
$H_\text{t}$ & 0.600  & m &   $d_\text{a}$ & 0.069  & m   \\
$H$ & 0.250  & m &    $d_\text{w}$  & 1.0    & mm     \\
\hline
\end{tabular}
\end{table}
\begin{table}
\centering
\caption{Geometry parameters of air-turbine model in physical experiment}
\label{tab:turbine}
\begin{tabular}{cccccc}
\hline
Parameter  & Value & Unit & Parameter & Value & Unit \\
\hline
$\theta_1$   & 60   & $\,^{\circ}$   &    $I_\text{d}$  & 1.0   & mm  \\
$\theta_2$   & 60   & $\,^{\circ}$   &    $l_\text{b}$  & 1.1   & mm  \\
$\theta_3$   & 30   & $\,^{\circ}$   &    $l_\text{c}$  & 10.8  & mm  \\
$b_\text{g}$ & 1.0  & mm             &    $l_\text{g}$  & 7.8   & mm  \\
$b_\text{r}$ & 3.1  & mm             &    $l_\text{r}$  & 1.1   & mm  \\
$d_\text{a}$ & 69.0 & mm             &    $l_\text{t}$  & 166.5 & mm  \\
$d_\text{c}$ & 47.0 & mm             &    $r_\text{g}$  & 4.8   & mm  \\
$I_\text{a}$ & 48.0 & mm             &    $r_\text{p}$  & 5.8   & mm  \\
$I_\text{b}$ & 10.3 & mm             &    $r'_\text{p}$ & 0.2   & mm  \\
$I_\text{c}$  & 1.5   & mm           &   \\
\hline
\end{tabular}
\end{table}

Two ultrasonic wave gauges are installed at the centres of the ceilings of two sealed chambers to monitor the elevation histories of the free surface in the liquid tank.
Each ultrasonic wave gauge has an effective measurement range of 1.20 m and a resolution of 0.1 mm. 
Pressure sensors are used to measure the air pressure at different positions of the liquid tanks.
Two pressure sensors are placed on the ceilings of each air chamber.
Another two pressure sensors are installed at the upstream and downstream positions of the air turbine in the air duct.
Three cameras are set to capture the liquid motions from different perspectives.

The WEH liquid tank model is fixed on a shaking table. 
The liquid depth in the tank is set as $H = 0.25$ m.
The shaking table can provide 6-degree-of-freedom excitations for a model that weighs up to 50kg. 
In this study, harmonic excitations in the horizontal direction are generated to simulate the surge motion of the floater. 
The excitation amplitude is $A = 0.022$ m.
To ensure the reliability of the experimental data, each case is tested at least three times.

\section{Results and discussion}\label{sec:Results}

\subsection{Convergence study of numerical model}

To ensure the accuracy of the proposed integrated numerical model for WEH liquid tanks, a convergence study is first conducted to determine spatial and temporal discretization.
The model with Turbine-L1 is taken as an example.
The moment of inertia $\textit{J}$ and the damping coefficient $\mu$ of the rotor are set as $\textit{J}=15\textit{J}_0$ and $\mu=4\mu_0$, respectively.
Hereafter, for the convenience of expression, $\textit{J}_0 = 1 \times 10^{-6} \text{kg·m}^2$ and $\mu_0 = 1 \times 10^{-5} \text{kg·m}^{-1}$ are used as units of measurement. 
The excitation period is $T = 1.10$s.

\begin{figure}[!htp]
\centering
{
  \begin{minipage}{0.23\linewidth}
  \centering
  \includegraphics[width=1.0\linewidth]{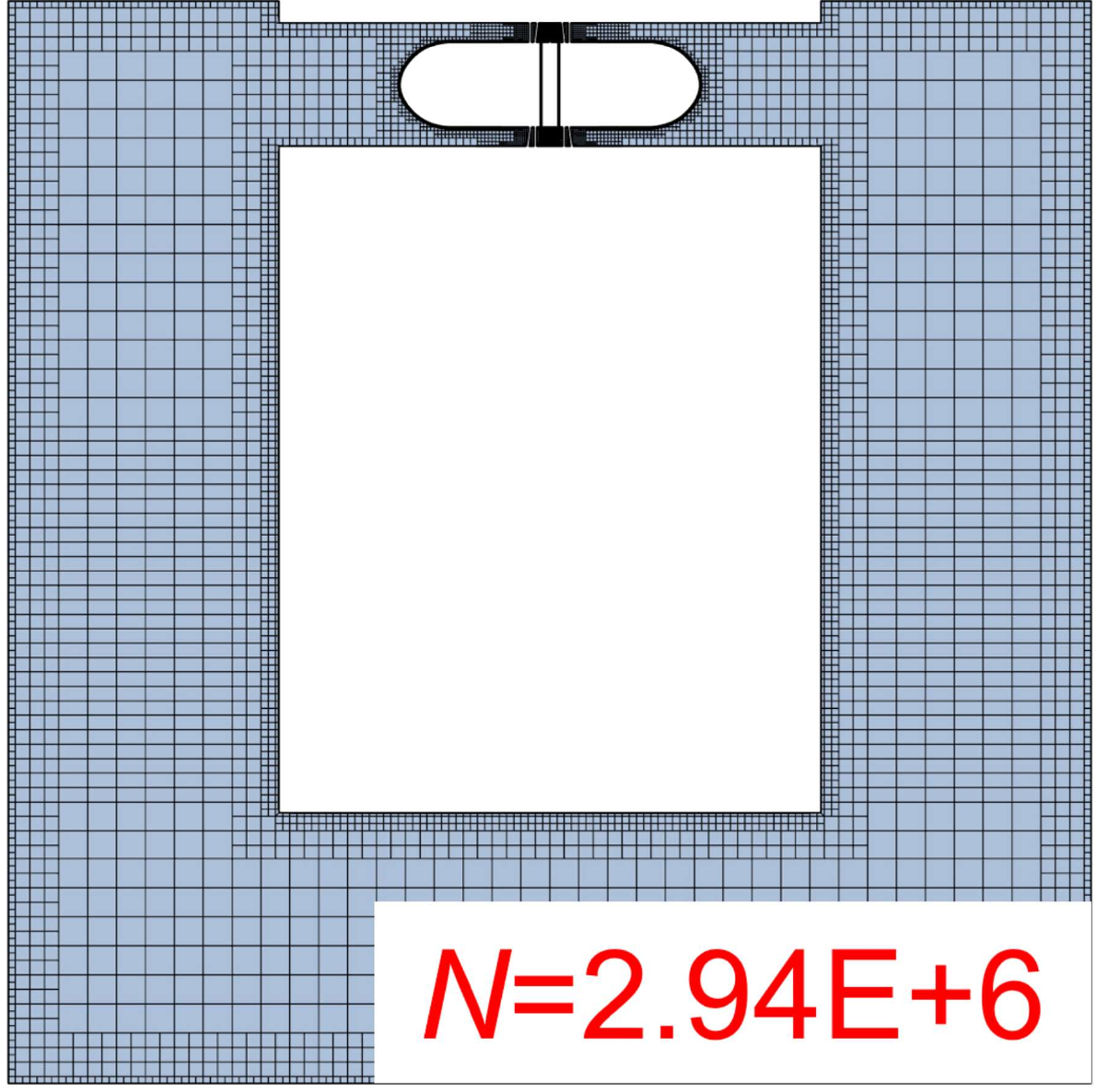}
  \subcaption{}
  \end{minipage}
}
{
  \begin{minipage}{0.23\linewidth}
  \centering
  \includegraphics[width=1.0\linewidth]{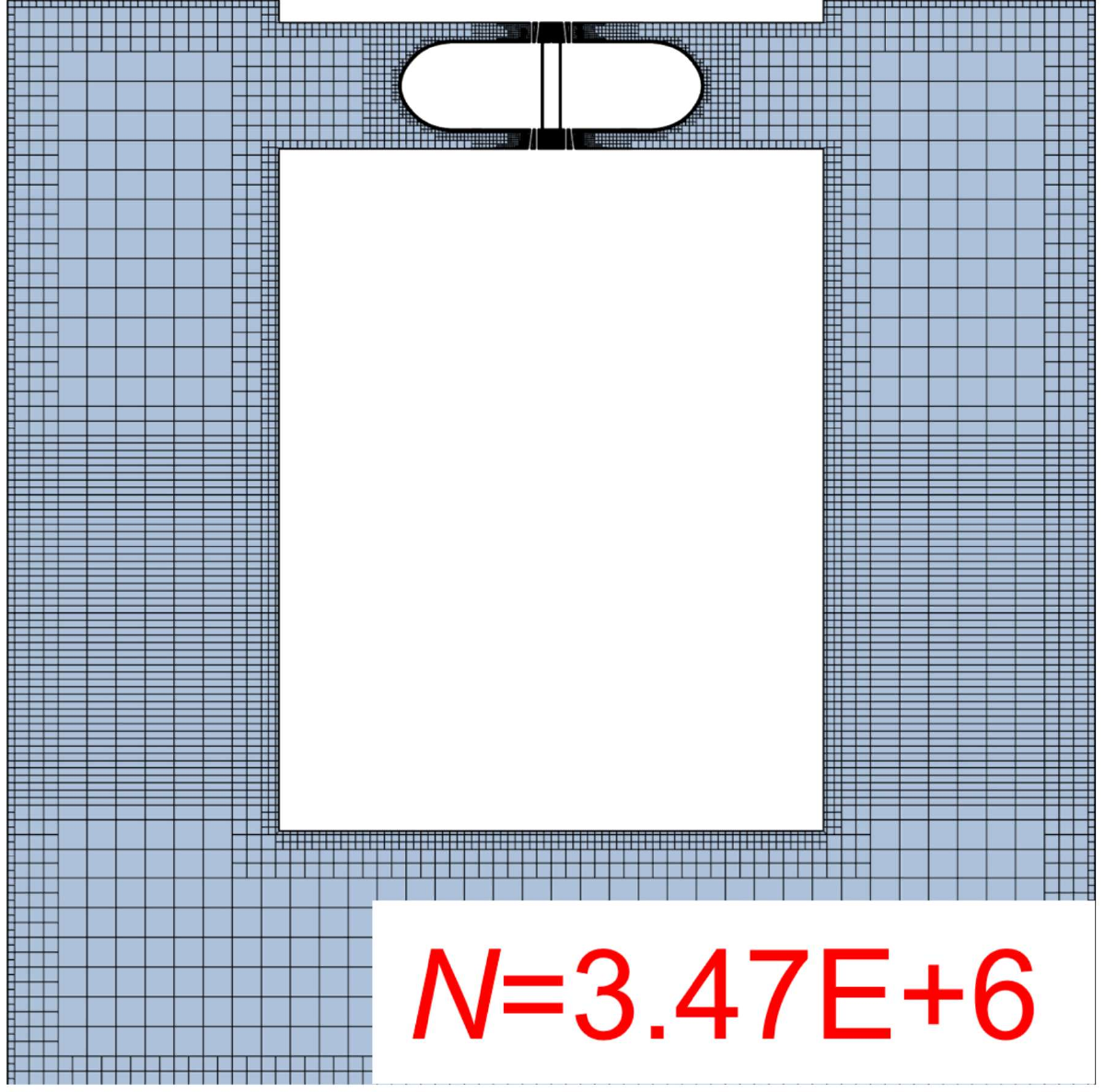}
  \subcaption{}
  \end{minipage}
}
{
  \begin{minipage}{0.23\linewidth}
  \centering
  \includegraphics[width=1.0\linewidth]{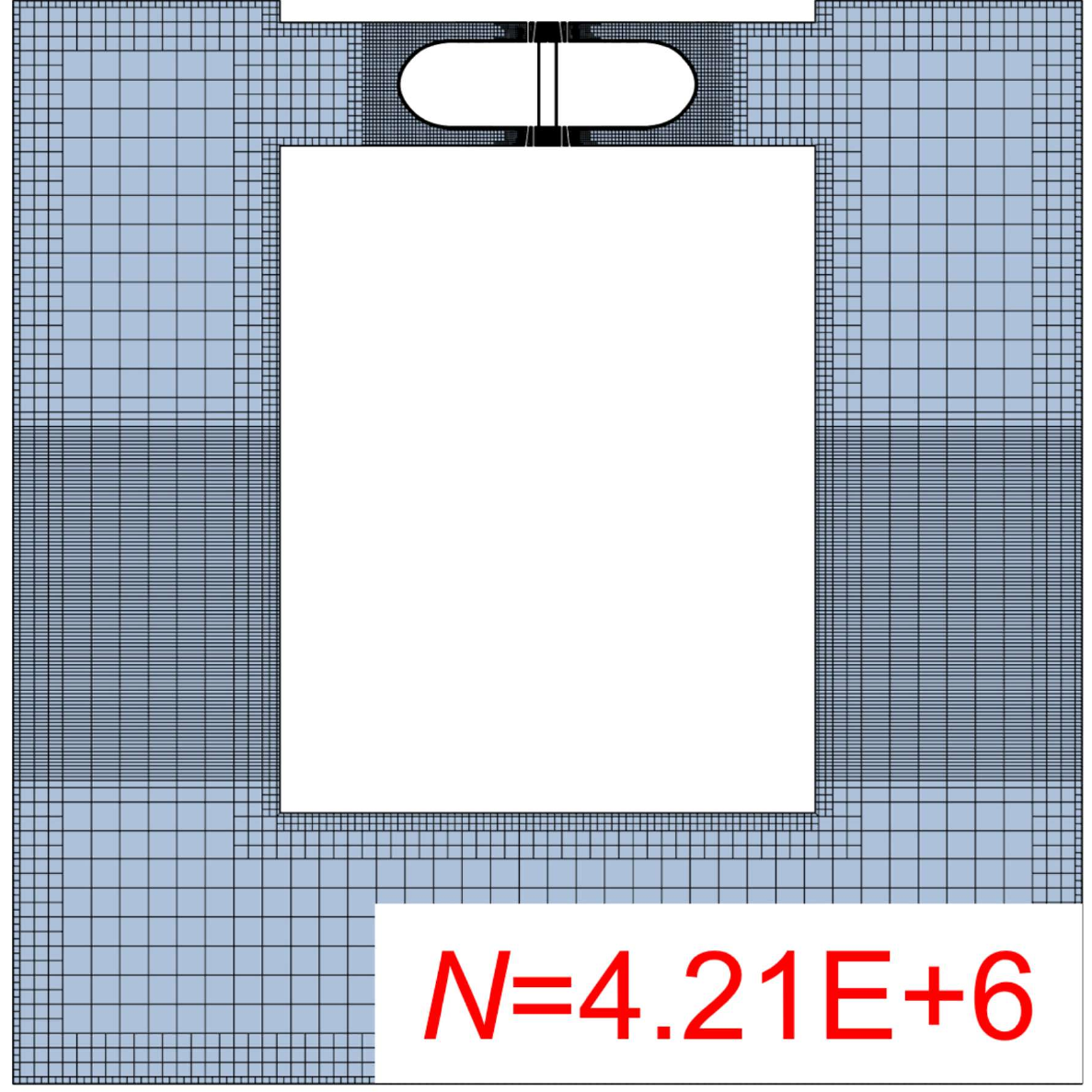}
  \subcaption{}
  \end{minipage}
}
{
  \begin{minipage}{0.23\linewidth}
  \centering
  \includegraphics[width=1.0\linewidth]{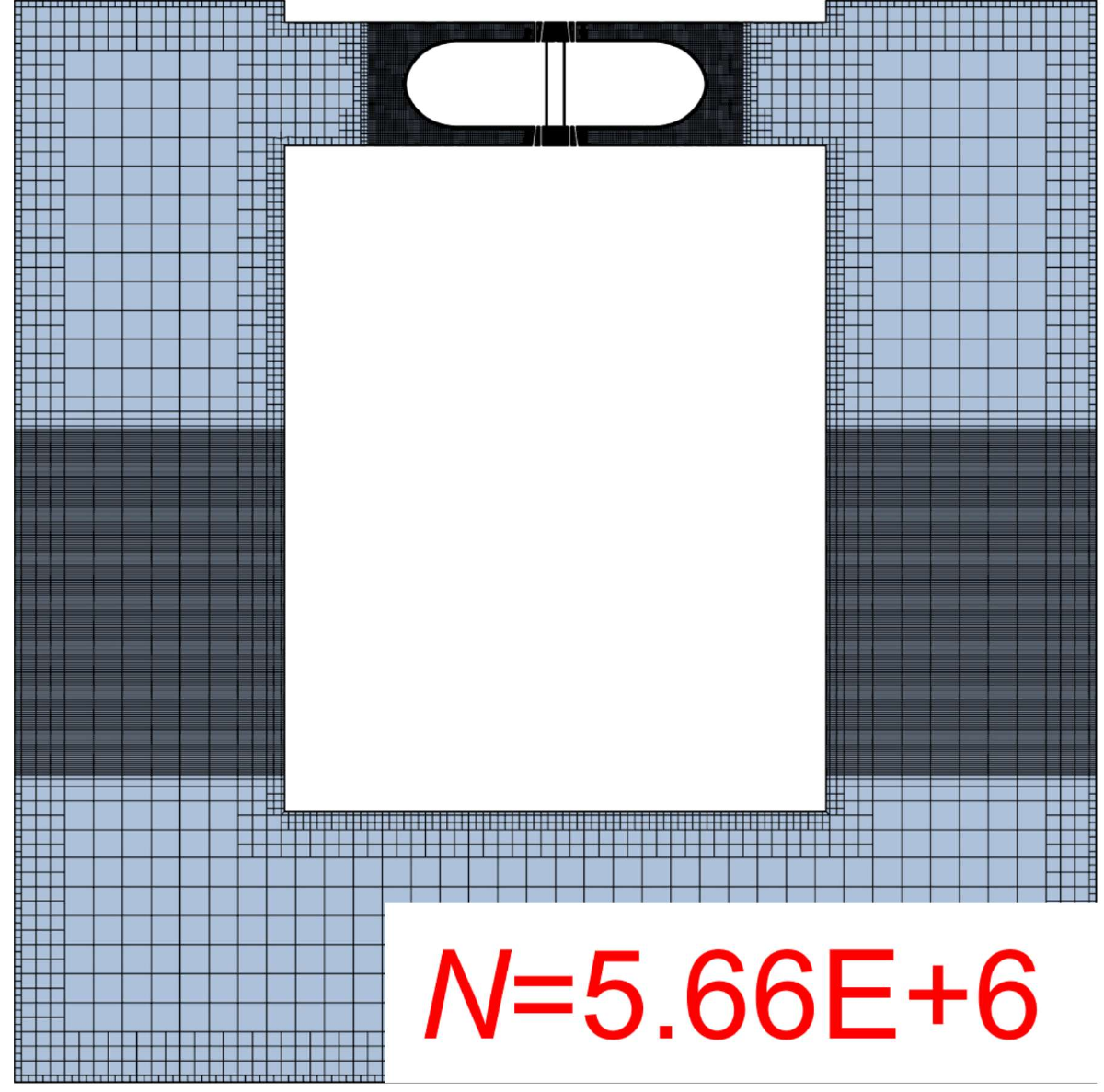}
  \subcaption{}
  \end{minipage}
}
\caption{Mesh models for spatial convergence study: (a) mesh-1; (b) mesh-2; (c) mesh-3; and (d) mesh-4}
\label{fig:mesh_converge}
\end{figure}

Four sets of numerical meshes with different grid densities are established for discretizing the fluid domain, as shown in Fig. \ref{fig:mesh_converge}.
The four meshes have $2.94\times10^6$ (``mesh-1''), $3.47\times10^6$ (``mesh-2''), 4.21$\times10^6$ (``mesh-3''), and 5.66$\times10^6$ (``mesh-4'') cells, respectively.
From mesh-1 to mesh-4, the mesh grids near the free surface and the turbine system are refined.
This refinement is strategically carried out to track the dynamic behaviour of the free surface and capture the flow details during the interaction between the air and the turbine system. 

\begin{figure}[!htp]
\centering
{
  \begin{minipage}{0.45\linewidth}
  \centering
  \includegraphics[width=1.0\linewidth]{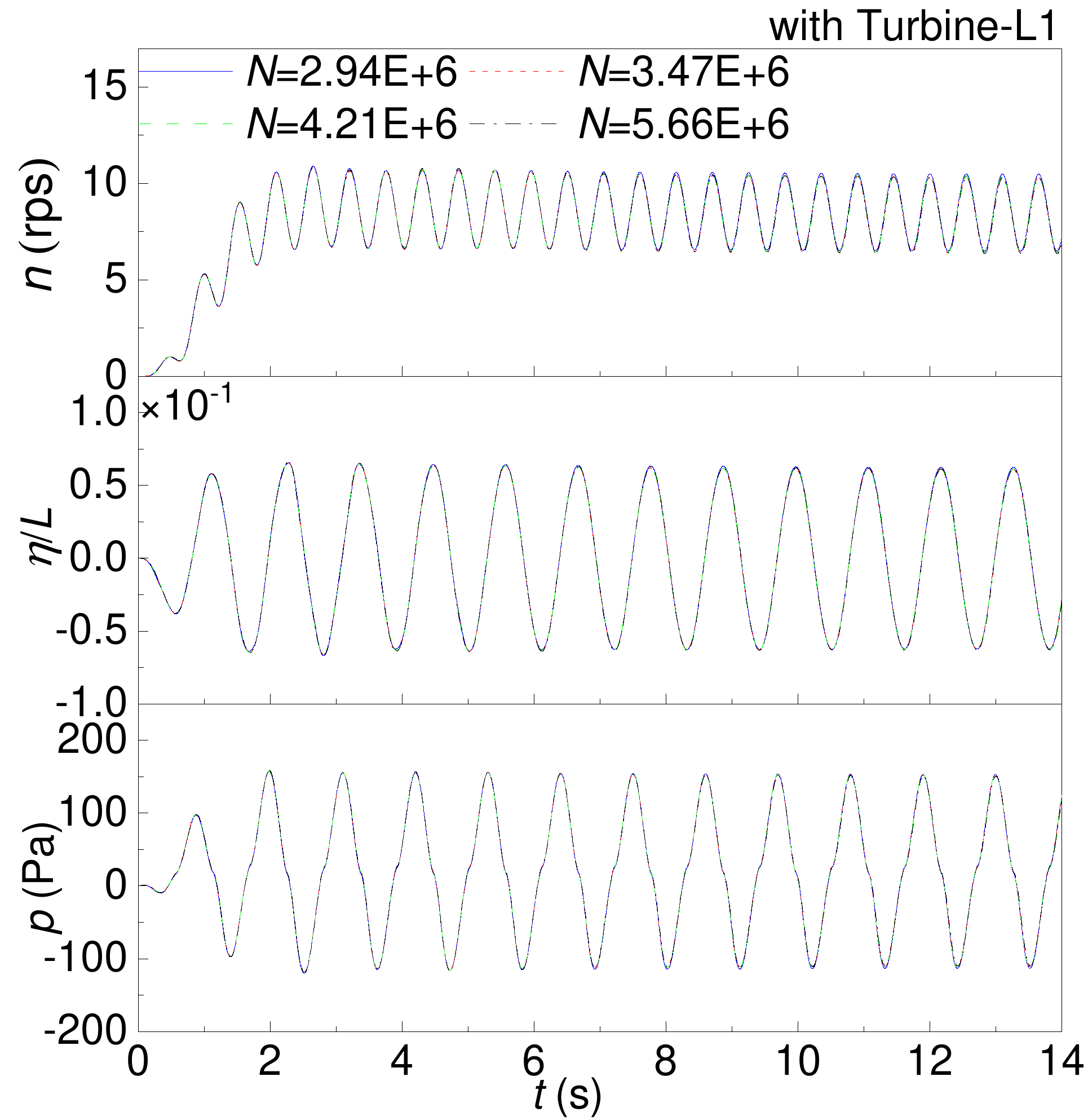}
  \subcaption{}
  \end{minipage}
}
{
  \begin{minipage}{0.45\linewidth}
  \centering
  \includegraphics[width=1.0\linewidth]{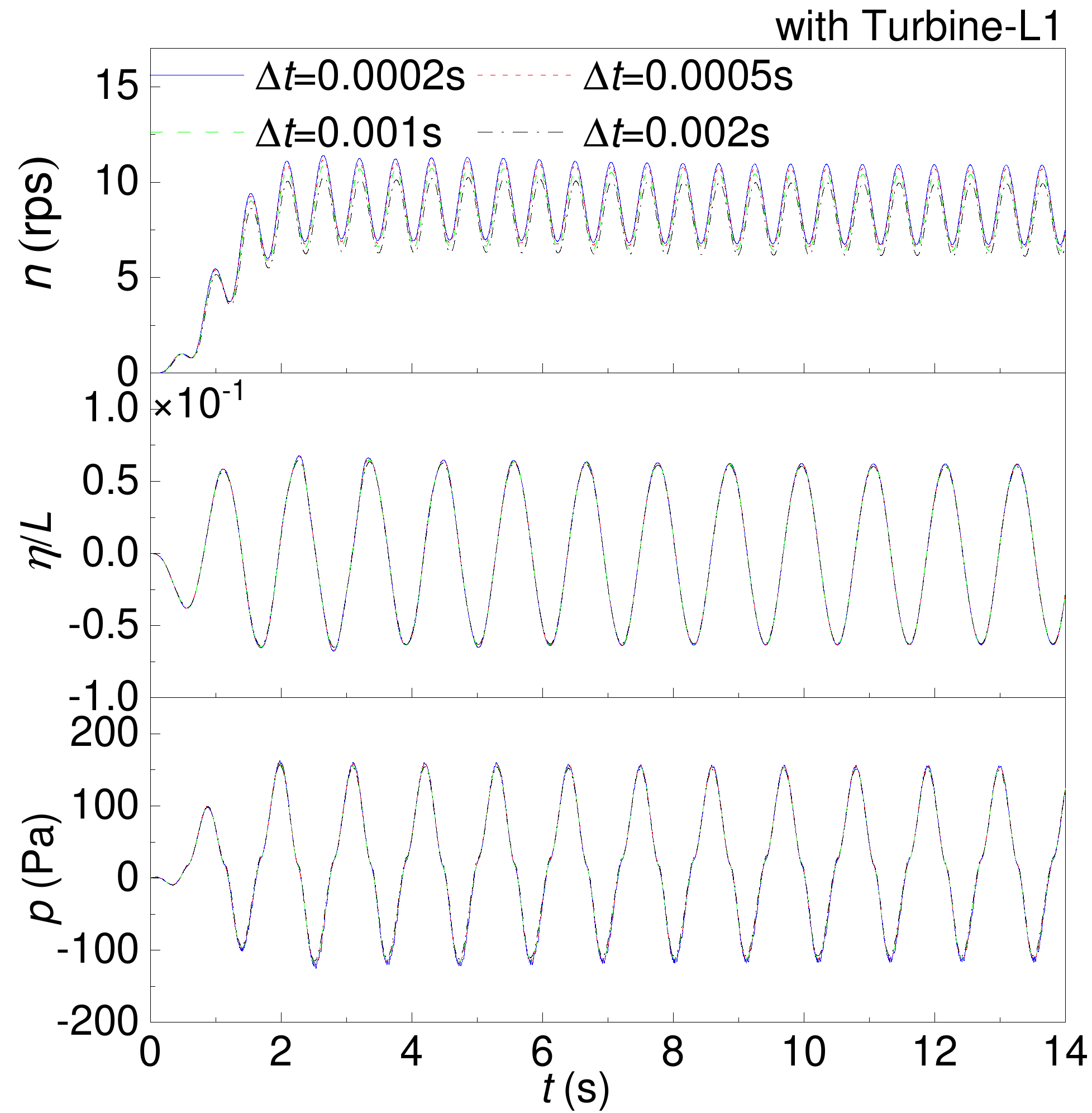}
  \subcaption{}
  \end{minipage}
}
\caption{Convergence study for (a) mesh discretization, and (b) time step}
\label{fig:convergence analysis}
\end{figure}

Fig. \ref{fig:convergence analysis} (a) shows the calculated time histories of the rotor speed, the free-surface elevation, and the pneumatic pressure in the right-hand-side air chamber.
The time step is set as small as $\Delta t=$0.0002s.
It shows that there is an initial stage which lasts for about $2T$.
In the initial stage, the rotor speed increases in a fluctuating manner, while both the free-surface elevation and the pneumatic pressure rapidly increase their amplitudes. 
After that, the system enters a steady stage. 
In this stage, the rotor speed varies sinusoidally around an average value, and both the free-surface elevation and the pneumatic pressure oscillate regularly around zero. 
Note that the period of the rotor speed oscillation is half that of the excitation period, due to the bi-directional symmetry property of the impulse turbine system.
The unit of rotor speed is revolutions per second (rps) hereafter.
All four sets of meshes can lead to stable results. 
Although the differences in results between any two meshes are small enough, it is still clear that the results converge quickly as the number of mesh cells increases.
It is hardly possible to tell the differences in results between mesh-3 and mesh-4 graphically. 
Considering both computational resources and efficiency, the mesh quality of mesh-3 (which has 4.21$\times10^6$ cells) is preferred for numerical calculations hereafter. 

Based on mesh-3, convergence studies are further conducted  for the time step.
Four different time steps are considered, with $\Delta t$ = 0.0002s, 0.0005s, 0.001s, and 0.002s, respectively. 
Accordingly, the calculated time histories of the rotor speed, the free-surface elevation, and the pneumatic pressure are shown in Fig. \ref{fig:convergence analysis} (b).
Compared to the free-surface elevation and pneumatic pressure, the rotor speed is more sensitive to the time step.
When the time step decreases from 0.002s to 0.0002s, the rotor speed history converges noticeably.
A time step of $\Delta t$ = 0.001 s is acceptable for guaranteeing the accuracy of the numerical model.

\subsection{Comparison between numerical and experimental results}

\begin{figure}[!htp]
\centering
{
  \begin{minipage}{1.0\linewidth}
  \centering
  \includegraphics[width=1.0\linewidth]{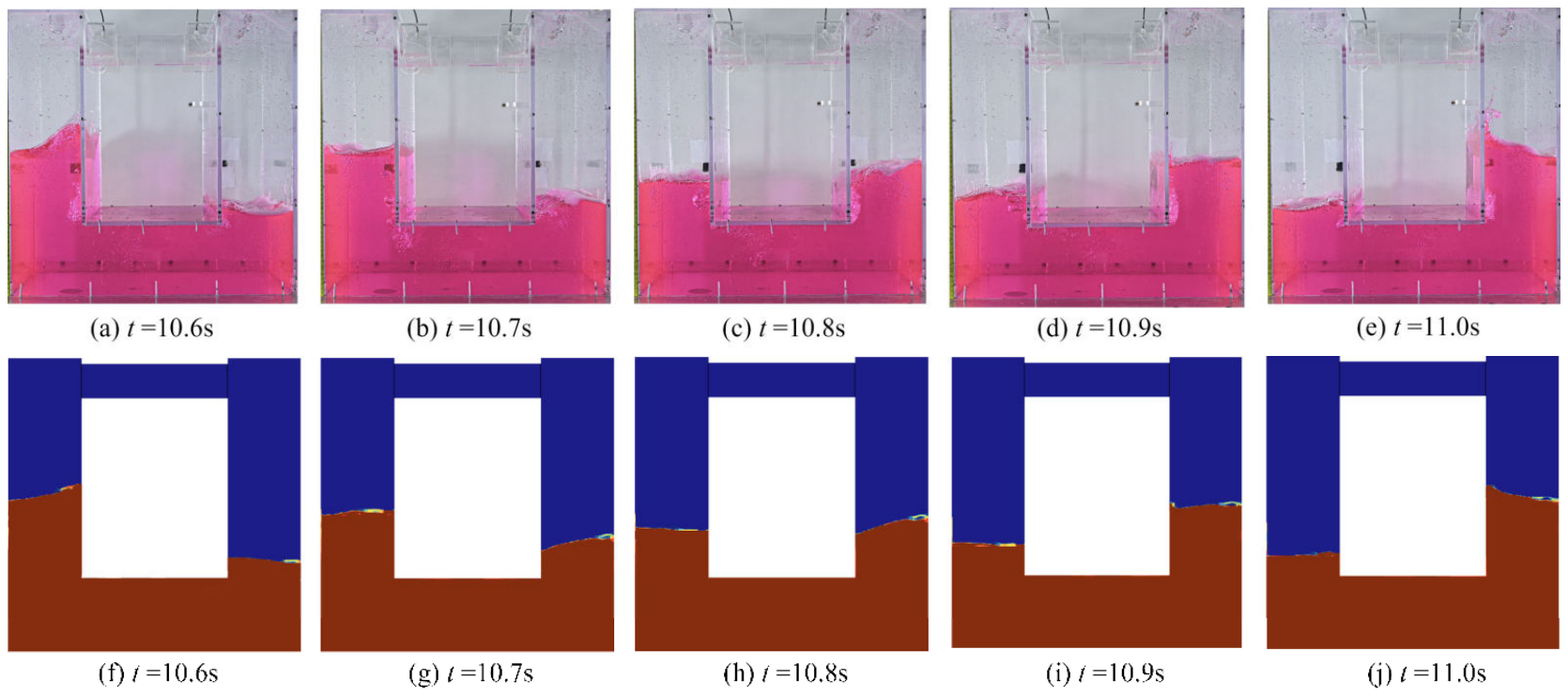}
  \end{minipage}
}
\caption{Comparison of experimental snapshots at $t$= (a) 10.6s, (b) 10.7s, (c) 10.8s, (d) 10.9s, and (e) 11.0s with numerical results at $t$= (f) 10.6s, (g) 10.7s, (h) 10.8s, (i) 10.9s, and (j) 11.0s, for wave-energy-harvesting tank without turbine in excitation condition of $T$=1.10s}
\label{fig:validition of photo_without turbine}
\end{figure}

\begin{figure}[!htp]
\centering
{
  \begin{minipage}{1.0\linewidth}
  \centering
  \includegraphics[width=1.0\linewidth]{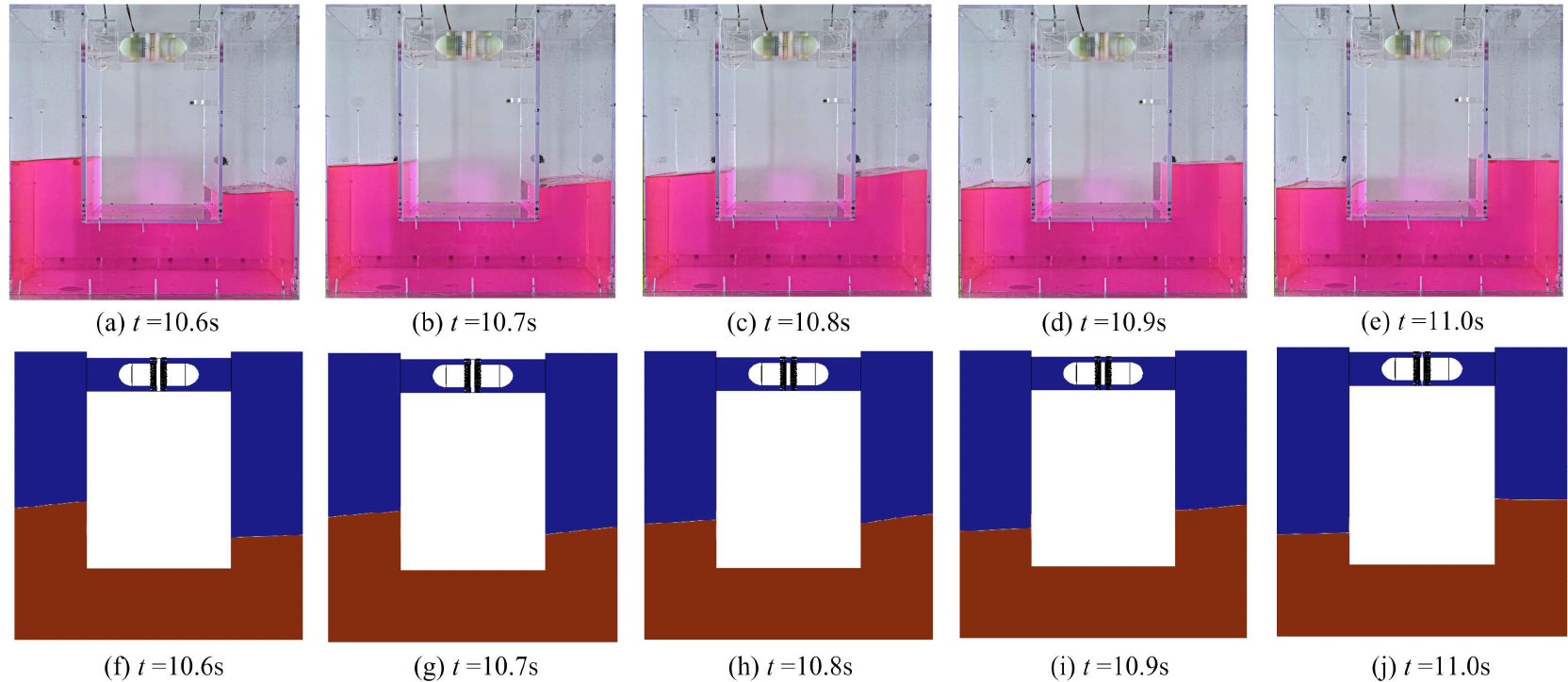}
  \end{minipage}
}
\caption{Comparison of experimental snapshots at $t$= (a) 10.6s, (b) 10.7s, (c) 10.8s, (d) 10.9s, and (e) 11.0s with numerical results at $t$= (f) 10.6s, (g) 10.7s, (h) 10.8s, (i) 10.9s, and (j) 11.0s, for wave-energy-harvesting tank with Turbine-L1 in excitation condition of $T$=1.10s}
\label{fig:validition of photo_with turbine}
\end{figure}

The numerical results are compared with the experimental data to further validate the integrated numerical model.
The excitation condition of $T=1.10\text{s}$ is shown here as an example.
Figs. \ref{fig:validition of photo_without turbine} (a)-(e) illustrate snapshots of the free-surface profile in the WEH liquid tank without the turbine system at \(t = 10.6\)s, \(t = 10.7\)s, \(t = 10.8\)s, \(t = 10.9\)s, and \(t = 11.0\)s respectively. 
Meanwhile, Figs. \ref{fig:validition of photo_without turbine} (f)-(j) show the numerical results at the corresponding instants. 
For the liquid behaviour, the numerical results agree well with the experimental observations at all instants.
The present numerical model can reproduce even the slight free-surface breaking phenomenon under this excitation condition. 
Figs. \ref{fig:validition of photo_with turbine} (a)-(e) further illustrate snapshots of the free-surface profile in the tank with Turbine-L1 at t = 10.6s, 10.7s, 10.8s, 10.9s, and 11.0s respectively.
Figs. \ref{fig:validition of photo_with turbine} (f)-(j) show the numerical results at the corresponding instants.
It can be observed that the present integrated numerical method can also reproduce the liquid behaviour well at all times.

\begin{figure}[!htp]
\centering
{
  \begin{minipage}{0.45\linewidth}
  \centering
  \includegraphics[width=1.0\linewidth]{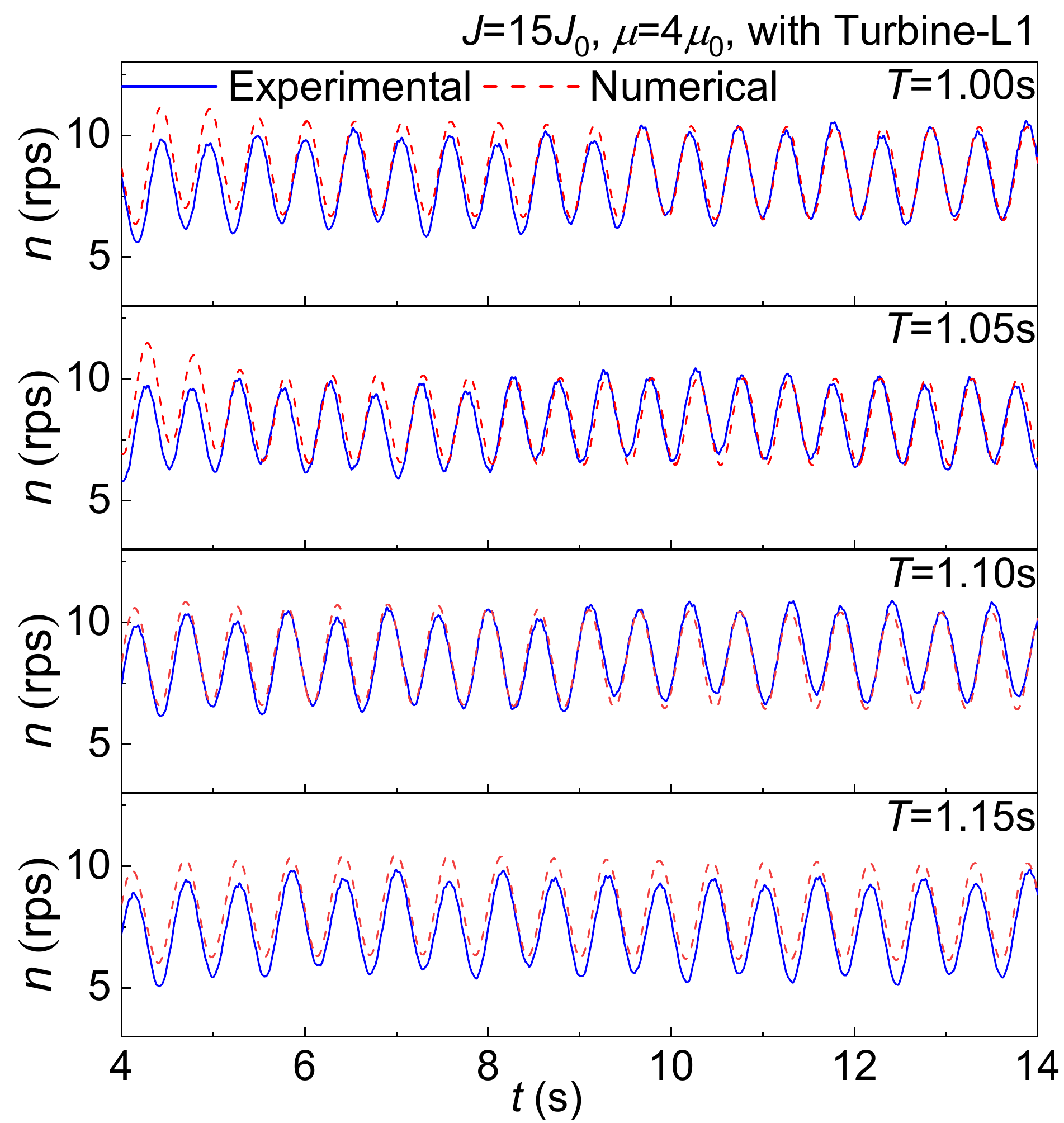}
  \subcaption{}
  \end{minipage}
}
{
  \begin{minipage}{0.45\linewidth}
  \centering
  \includegraphics[width=1.0\linewidth]{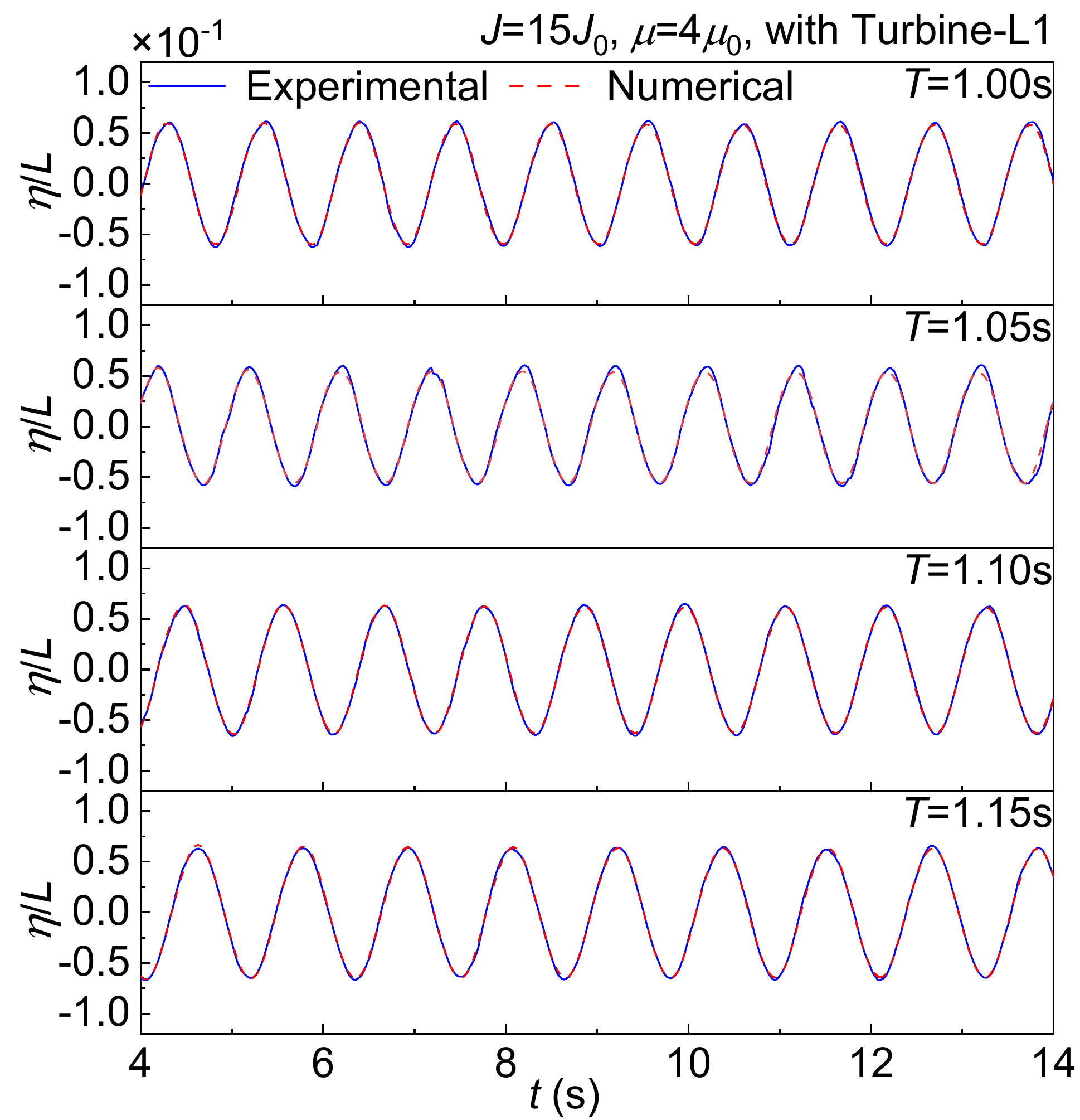}
  \subcaption{}
  \end{minipage}
}
{
  \begin{minipage}{0.45\linewidth}
  \centering
  \includegraphics[width=1.0\linewidth]{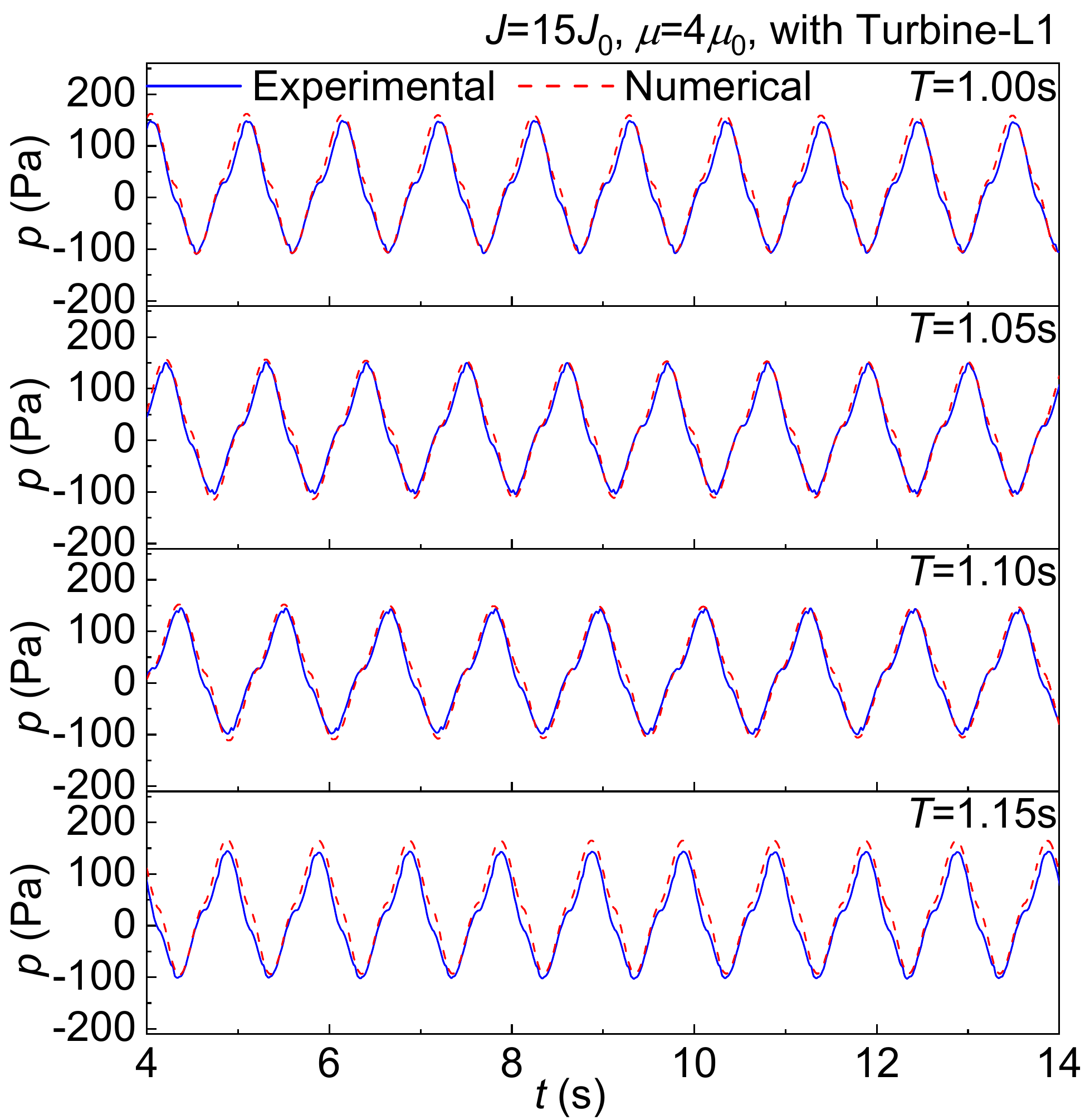}
  \subcaption{}
  \end{minipage}
}
\caption{(a) Validition of rotor speed (b) validiton of liquid surface displacement; (c) validiton of pressure}
\label{fig:comparison}
\end{figure}

Fig. \ref{fig:comparison} further compares the experimental results with the numerical results for more excitation conditions with $T=$1.00s, 1.05s, 1.10s and 1.15s.
Figs. \ref{fig:comparison} (a), (b), and (c) refer to the rotor speed, the free-surface elevation, and the pneumatic pressure in the right-hand-side air chamber, respectively.
In the steady stage of each excitation condition, the numerical results always agree well with the experimental data with sufficient accuracy.
Especially for the free-surface elevation, the experimental results and the numerical results are almost identical.
Based on the aforementioned observations, it can be confirmed that the present integrated numerical model can provide reliable results that accurately reflect the physical reality.

\subsection{Analyses of mechanical parameters of turbine rotor}\label{subsec:rotor_mech}

According to Eq. (\ref{eq:Power}), the rotational behaviour of the turbine rotor is a key factor determining the power output of a WEH liquid tank.
For each rotor, the moment of inertia $J$ and the damping coefficient $\mu$ are inherent dynamic parameters that can affect its rotational behaviour.
Before investigating the power conversion properties of a WEH liquid tank, this subsection first discusses the effects of $J$ and $\mu$ on rotor speed. 

\begin{figure}[!htp]
\centering
{
  \begin{minipage}{0.6\linewidth}
  \centering
  \includegraphics[width=1.0\linewidth]{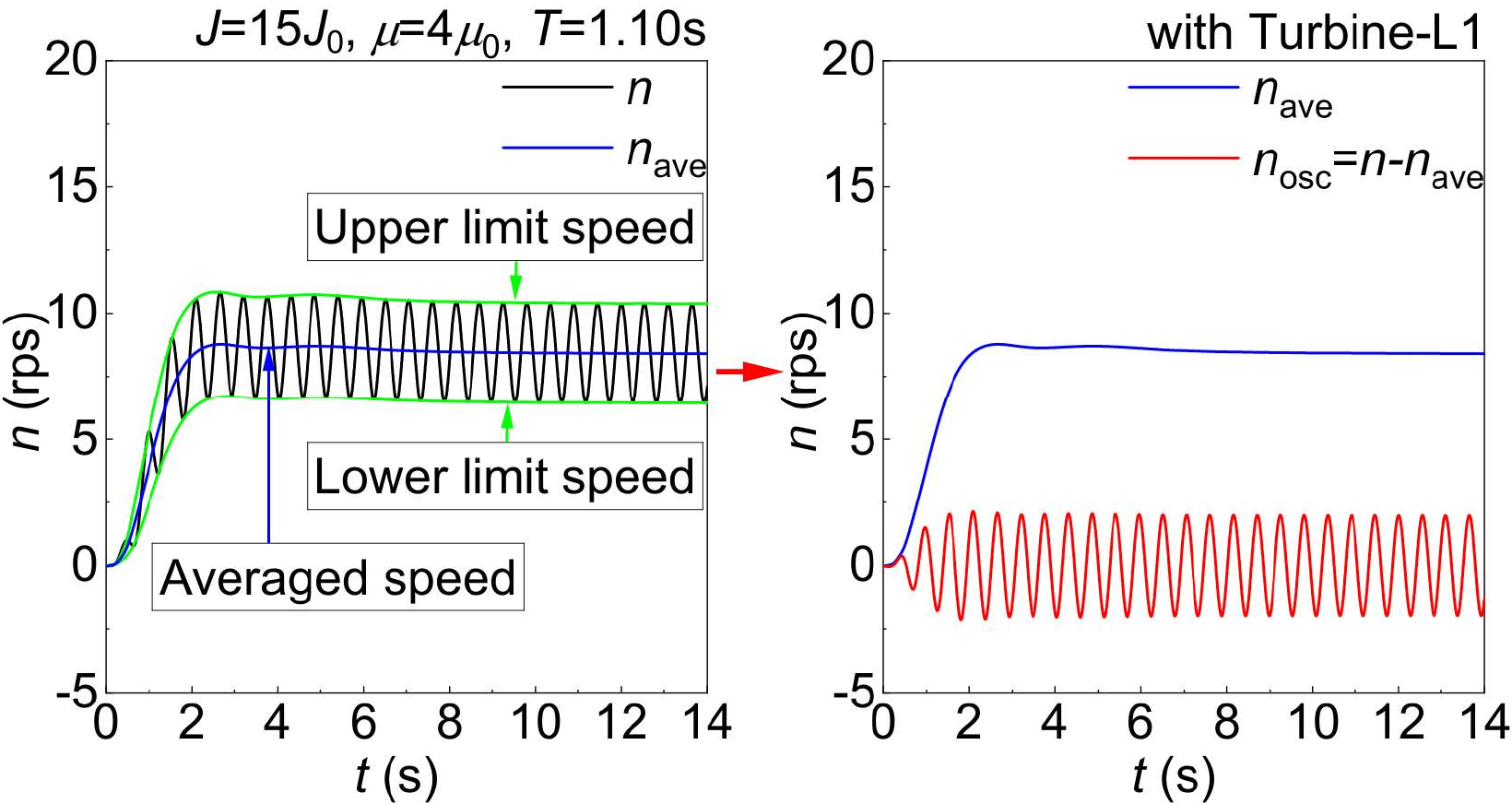}
  \end{minipage}
}
\caption{History of rotor speed and its decomposition}
\label{fig:rotor_history}
\end{figure}

Fig. \ref{fig:rotor_history} gives an example of the rotor speed history, when the tank is excited at $T=1.10$s.
The rotor has $J = 15J_0$ and $\mu = 4\mu_0$.
The rotor speed $n$ first increases from zero over time when the liquid tank starts to move.
After around five periods of excitations, the rotation enters the steady stage.
At this time, the instantaneous rotor speed $n$ fluctuates around a nearly constant average speed $n_{\rm{ave}}$.
The upper and lower limits of the rotor-speed variation are nearly time-invariant.
The value of $n_{\rm{ave}}$ represents the mean of the upper and lower limits.
Subtracting $n_{\rm{ave}}$ from $n$ results in the oscillatory component $n_{\rm{osc}}$.
The oscillatory component in the steady stage is nearly sinusoidal and has an almost constant amplitude.
Half the difference between the upper and lower limits of $n_{\rm{osc}}$ can be defined as the variation range of the rotor (denoted as $|n_{\rm{osc}}|$).
In the steady stage of this case, the averaged rotor speed is about $n_{\rm{ave}} = 8$ rps. 
The maximum speed is 11 rps, and the minimum speed is 5 rps. 
Therefore, the variation range is $|n_{\rm{osc}}| = 3$ rps. 

\begin{figure}[!htp]
\centering
{
  \begin{minipage}{0.45\linewidth}
  \centering
  \includegraphics[width=1.0\linewidth]{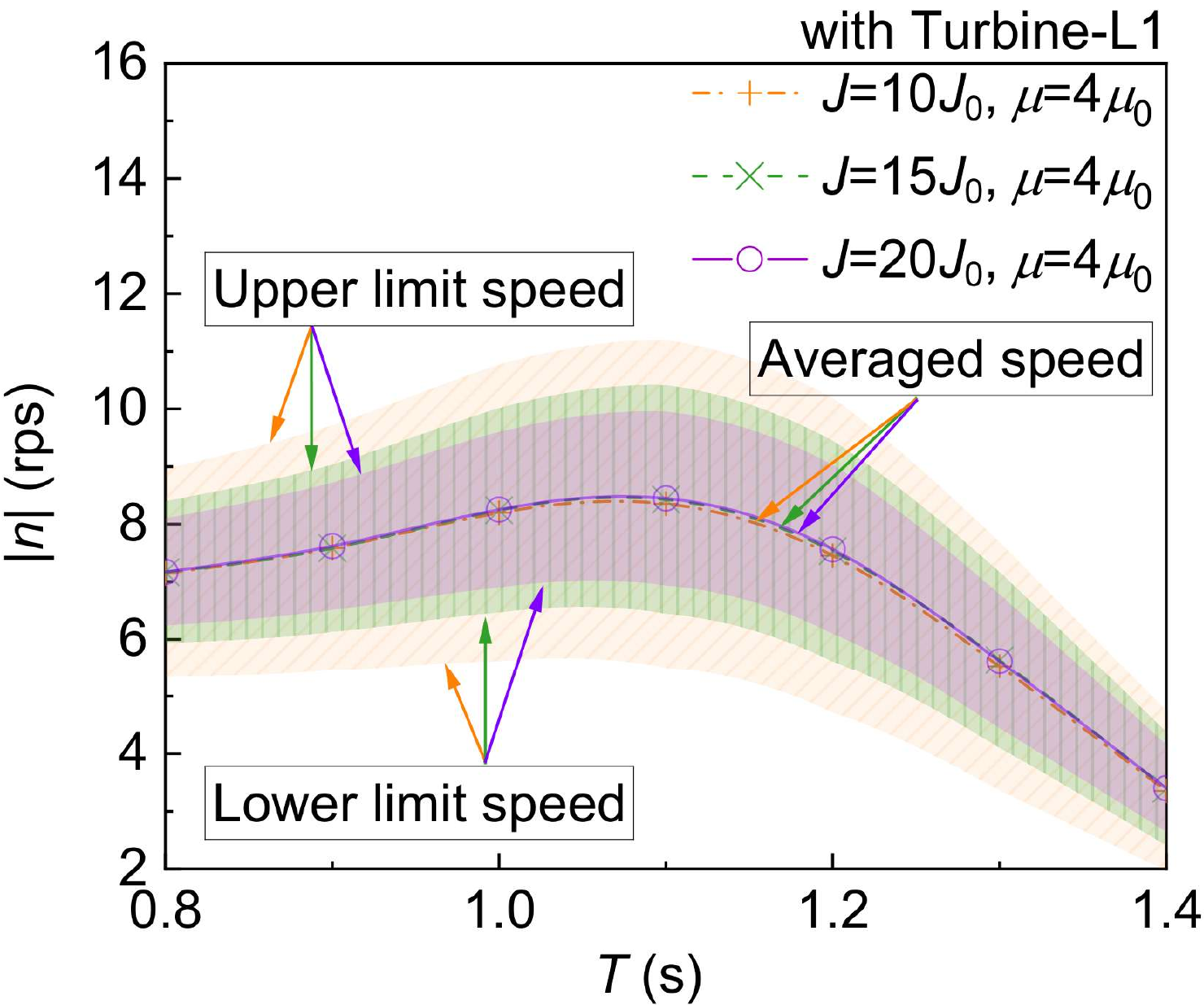}
  \subcaption{}
  \end{minipage}
}
{
  \begin{minipage}{0.45\linewidth}
  \centering
  \includegraphics[width=1.0\linewidth]{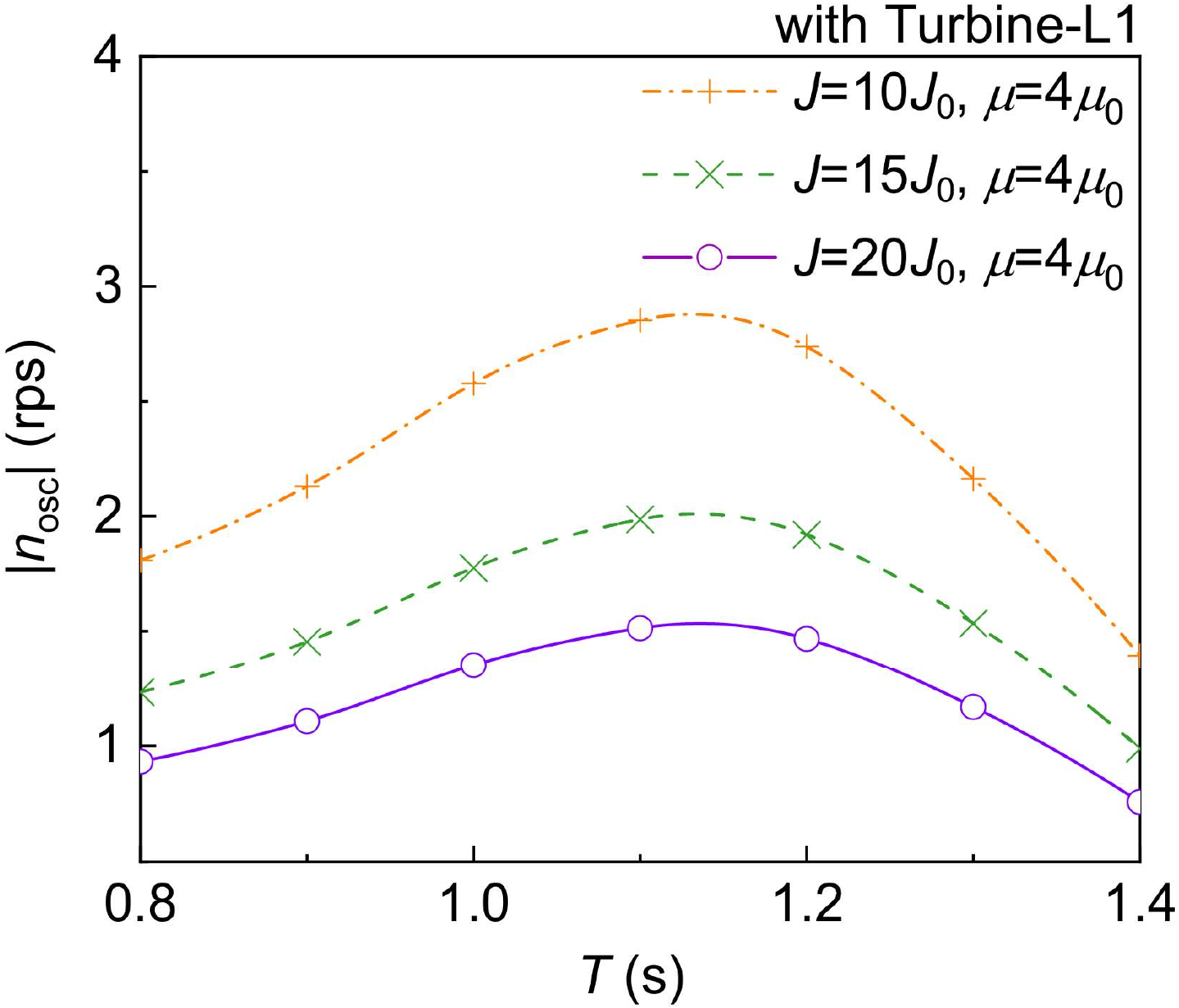}
  \subcaption{}
  \end{minipage}
}
\caption{Effect of moment of inertia on rotor speed: (a) averaged rotor speed and speed limits; (b) variation range of rotor speed}
\label{fig:moment_effect}
\end{figure}

Fig. \ref{fig:moment_effect} (a) illustrates the effects of the moment of inertia on the rotor speed.
The excitation period ranges from $T = 0.80$ s to 1.40 s.
The value of $J$ varies from $10J_0$ to $20J_0$, while the damping coefficient remains constant with $\mu = 4\mu_0$.
For cases with $J = 10J_0$, $15J_0$, and $20J_0$, the averaged speed of the rotor is almost identical in all these excitation conditions.
From $T = 0.80$ s to 1.1 s, the average rotor speed increases slowly from about 7 rps to over 8 rps. 
The maximum averaged rotor speed appears at $T=1.10$s.
In even larger excitation periods, the averaged speed decreases monotonically from $T = 1.10$ s to $T = 1.40$ s, down to 3 rps. 
In Fig. \ref{fig:moment_effect} (b), the variation ranges corresponding to the three moments of inertia are explicitly compared.
The variation ranges of the rotors also first increase with the excitation period, reach the maximum at $T = 1.10$, and then decrease monotonically.
The variation range of the rotor speed is generally smaller when the rotor has a larger moment of inertia.
This indicates that the rotor speed is more stable in the case with a larger moment of inertia.

\begin{figure}[!htp]
\centering
{
  \begin{minipage}{0.45\linewidth}
  \centering
  \includegraphics[width=1.0\linewidth]{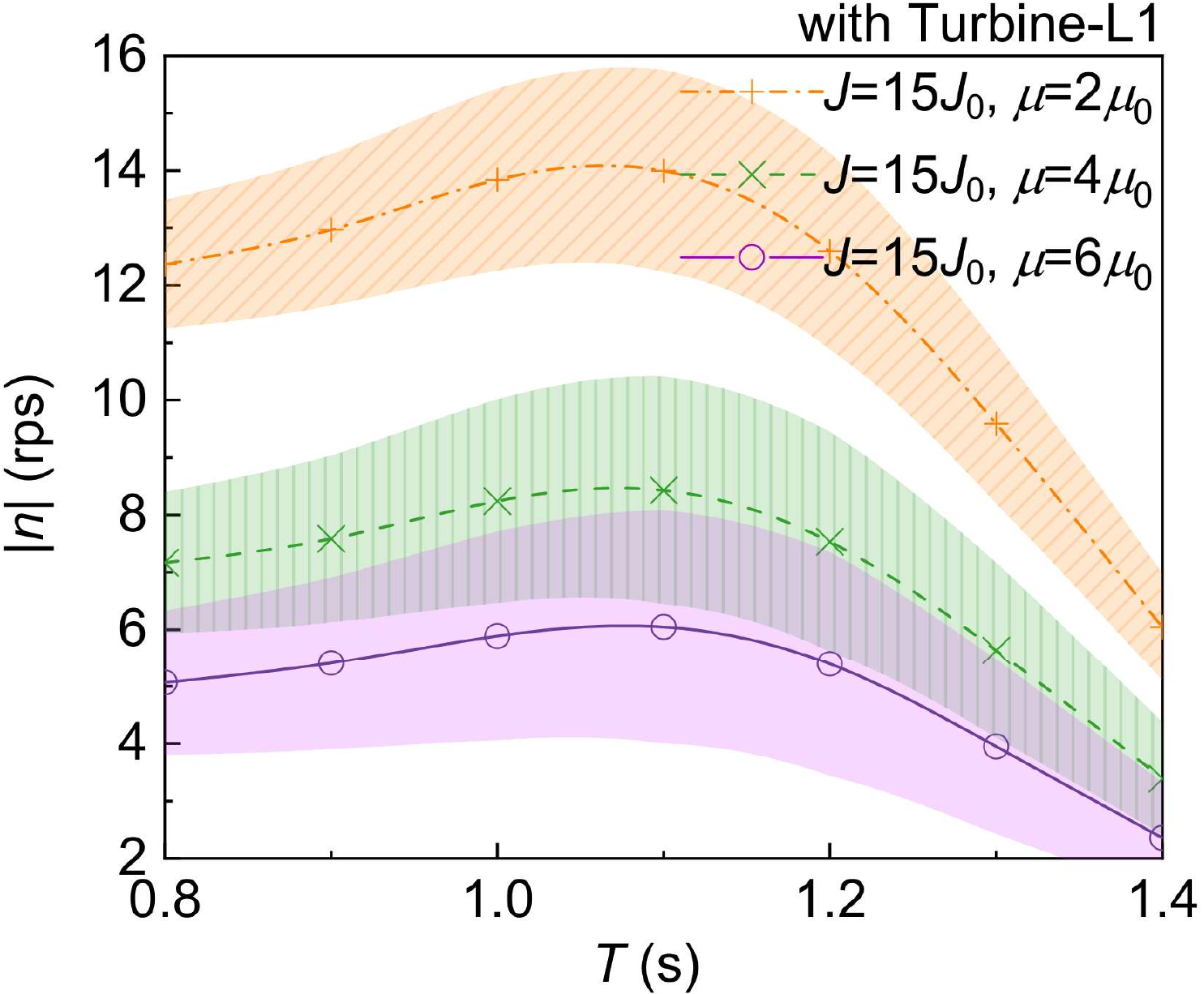}
  \subcaption{}
  \end{minipage}
}
{
  \begin{minipage}{0.45\linewidth}
  \centering
  \includegraphics[width=1.0\linewidth]{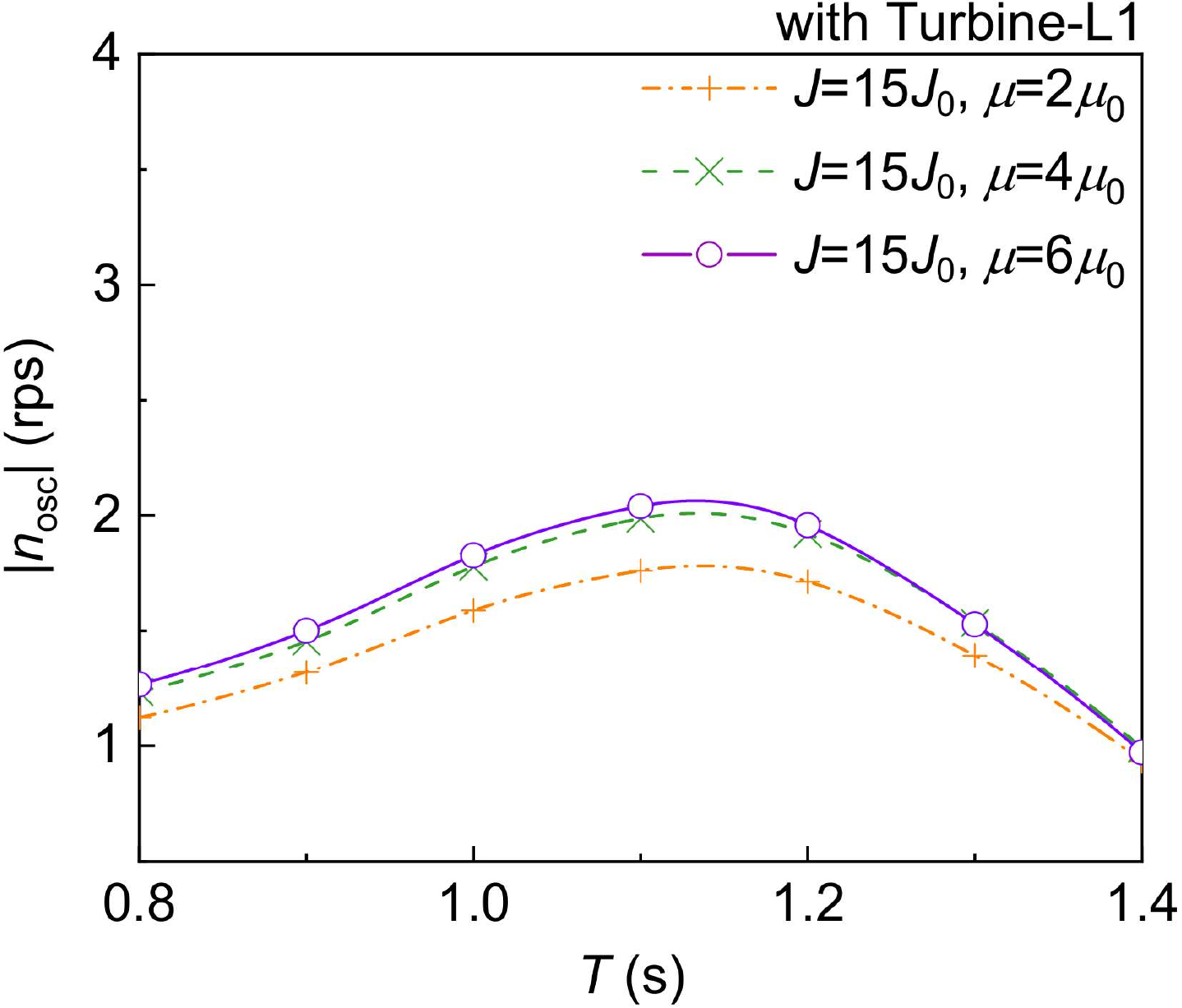}
  \subcaption{}
  \end{minipage}
}
\caption{Effect of damping coefficient on rotor speed: (a) averaged rotor speed and speed limits; (b) variation range of rotor speed}
\label{fig:damping_effect}
\end{figure}

Fig. \ref{fig:damping_effect} shows the effects of the damping coefficient on the rotor speed.
The moment of inertia remains constant, with $J = 15J_0$.
From Fig. \ref{fig:damping_effect} (a), it can be told that as the damping coefficient reduces from $\mu = 6\mu_0$ to $\mu = 2\mu_0$, the maximum averaged rotor speed increases from 6 rps up to over 13 rps.
The averaged rotor speed is generally higher when the damping coefficient of the rotor is reduced.
Fig. \ref{fig:damping_effect} (b) compares the variation ranges of the rotor speed that correspond to three damping coefficients.
Even though a smaller damping coefficient tends to reduce the variation range of a rotor, it does not markedly alter the variation range of the rotor. 
From $\mu = 2\mu_0$ to $\mu = 6\mu_0$, the maximum variation range increases from about 1.6 rps to 2.0 rps.

\subsection{Identification of optimal power take-off damping for wave energy harvesting}\label{subsec:opt_PTO}

The PTO damping of the turbine rotor is a crucial parameter that determines the amount of wave energy harvested.
According to Eqs. (\ref{eq:The load equation}), (\ref{eq:The torque equation}) and (\ref{eq:Power}), the PTO damping and the rotor speed are negatively correlated.
If the PTO damping is too low, only a little torque can be transferred to the rotor to generate electricity, even though the rotor speed is high. 
On the other hand, if the PTO damping is too high, it will impede the free motion of the rotor and reduce the total amount of angular kinetic energy of the rotor.
Therefore, an optimal PTO damping must be figured out to guarantee the efficiency of the present WEH liquid tank.

\begin{figure}[!htp]
\centering
\includegraphics[scale=0.35]{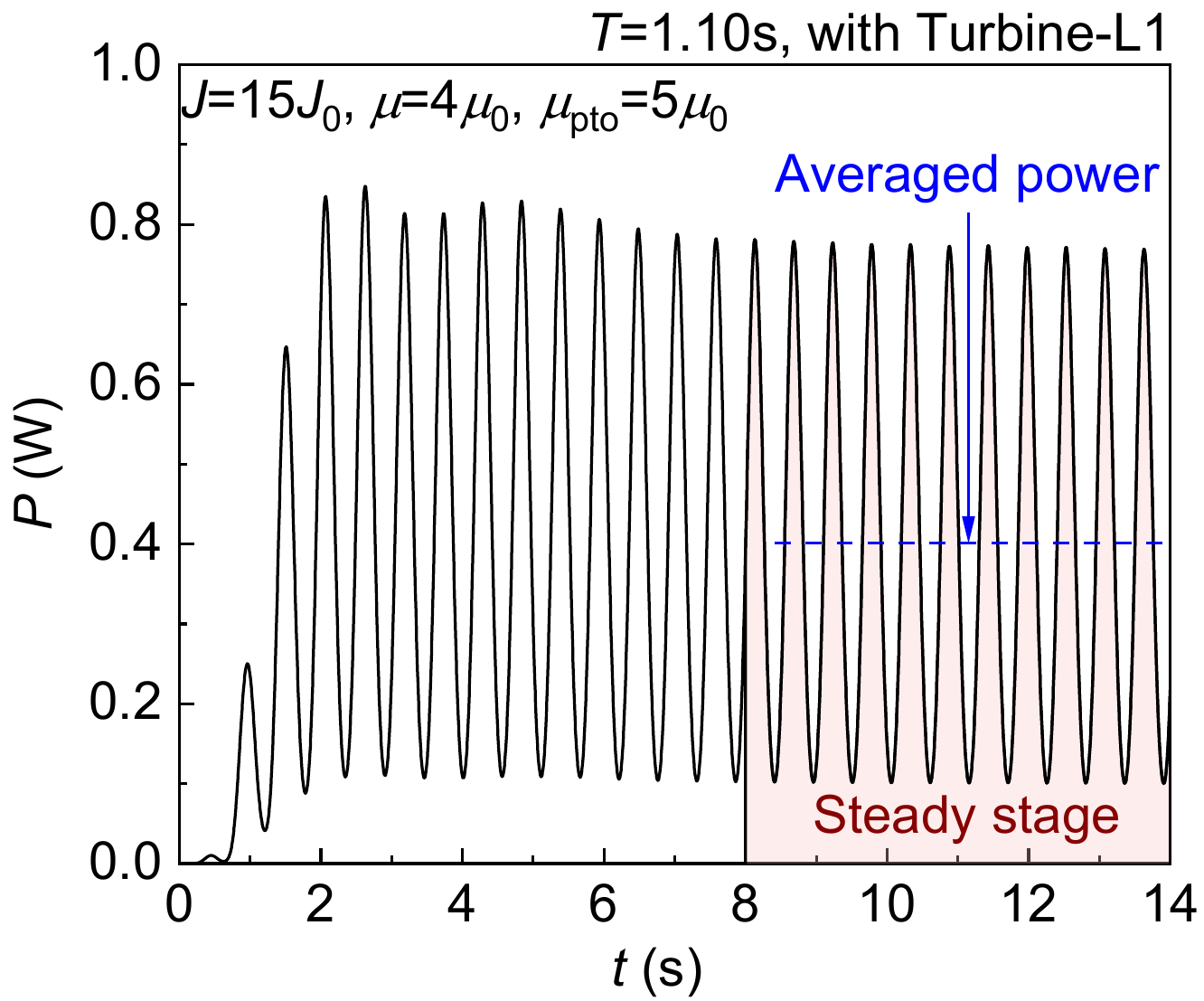}
\caption{History of generated power of wave-energy-harvesting liquid tank}
\label{fig:Power analysis}
\end{figure}

Fig. \ref{fig:Power analysis} shows an example of the real-time power output $P(t)$ of the WEH liquid tank with Turbine-L1.
The excitation period is $T=1.1\text{s}$, the moment of inertia of the rotor is $J=15J_0$, the damping coefficient of the rotor is $\mu=4\mu_0$, and the PTO damping is $\mu_\text{pto}=5\mu_0$.
The power output $P(t)$ increases in a fluctuating manner from zero and eventually stabilizes in a steady stage with a nearly constant maximum amplitude.
The averaged power $P_{\text{ave}}$ in the steady stage can be calculated as follows:
\begin{equation}
\begin{aligned}
P_{\text{ave}}=\frac{\int_{t_{0}}^{t_{0}+mT} P(t) \text{d}t}{mT}
\end{aligned}
\label{eq:power_ave}
\end{equation}
where $t_{0}$ can be any moment in the steady stage, $m$ represents the number of periods to be considered, and $T$ is the excitation period. 

\begin{figure}[!htp]
\centering
{
  \begin{minipage}{0.45\linewidth}
  \centering
  \includegraphics[width=1.0\linewidth]{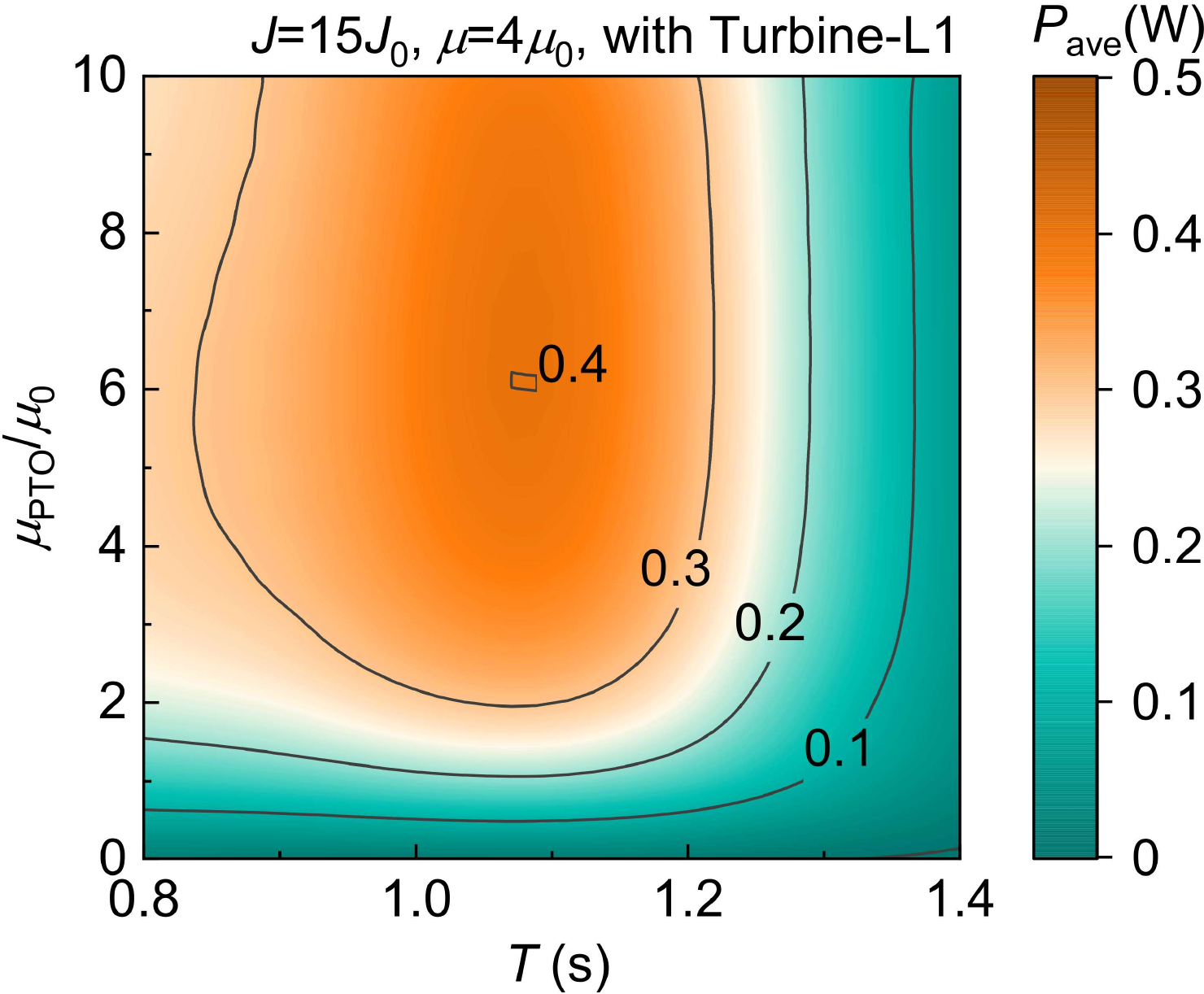}
  \subcaption{}
  \end{minipage}
}
{
  \begin{minipage}{0.45\linewidth}
  \centering
  \includegraphics[width=1.0\linewidth]{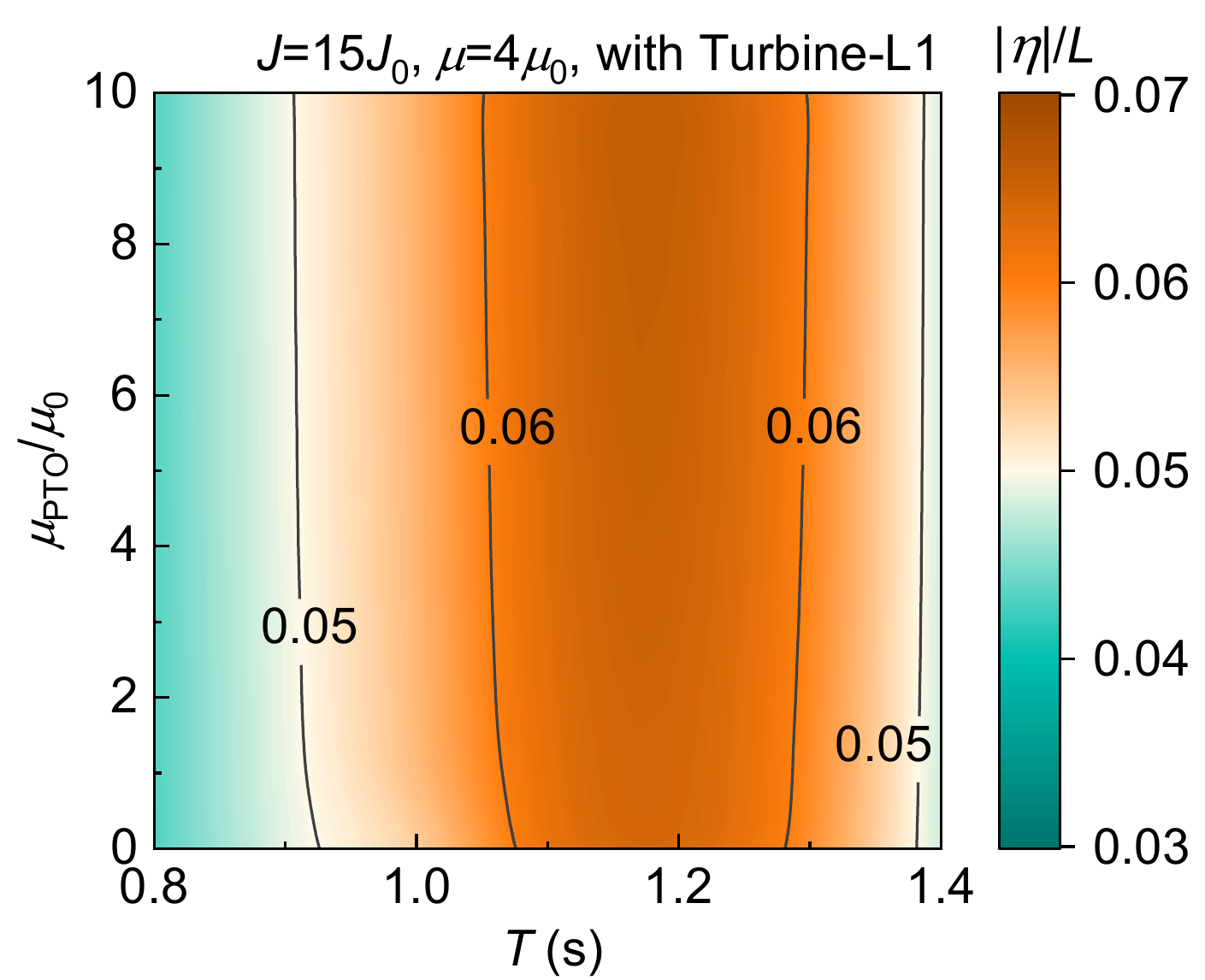}
  \subcaption{}
  \end{minipage}
}
{
  \begin{minipage}{0.45\linewidth}
  \centering
  \includegraphics[width=1.0\linewidth]{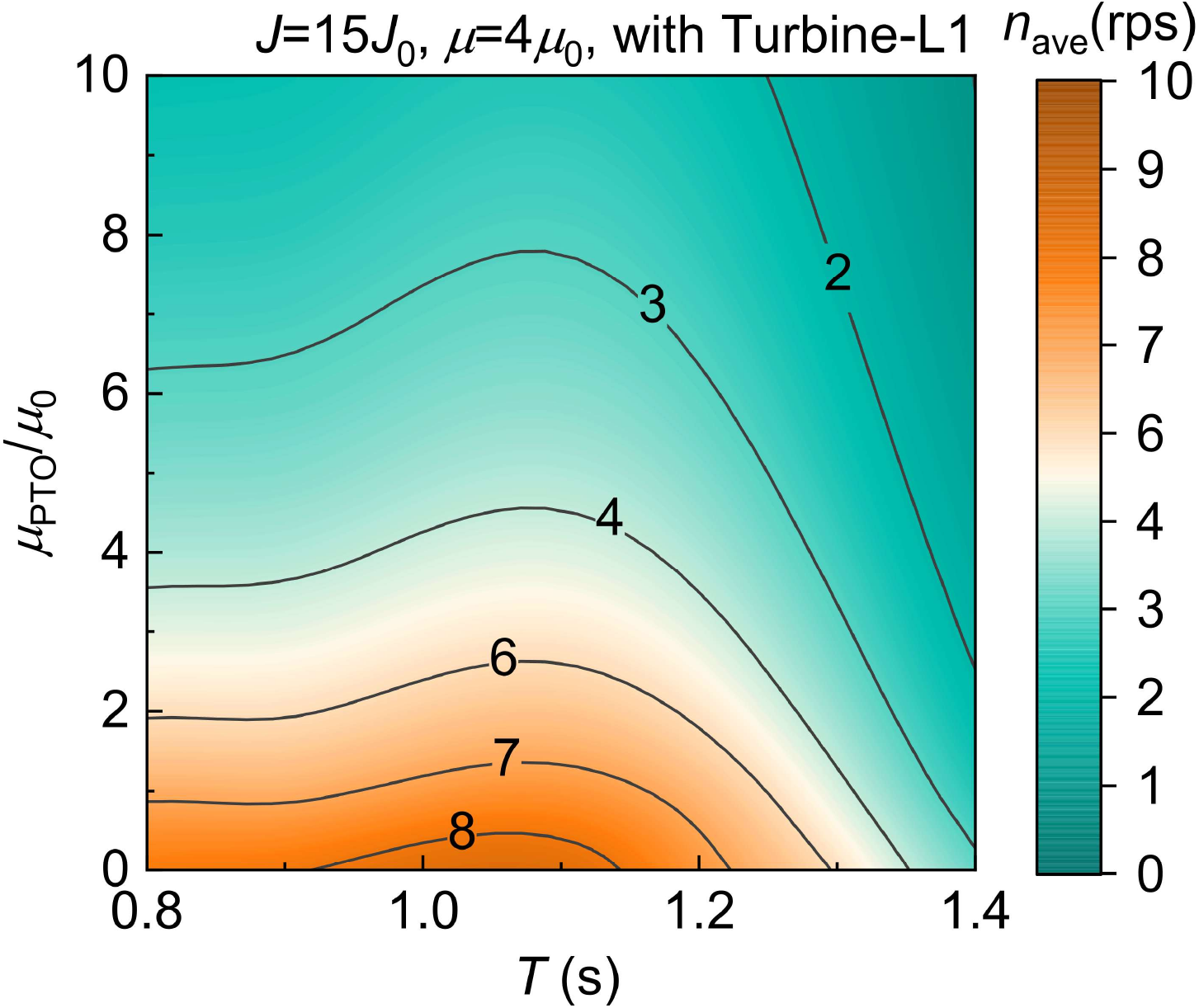}
  \subcaption{}
  \end{minipage}
}
{
  \begin{minipage}{0.45\linewidth}
  \centering
  \includegraphics[width=1.0\linewidth]{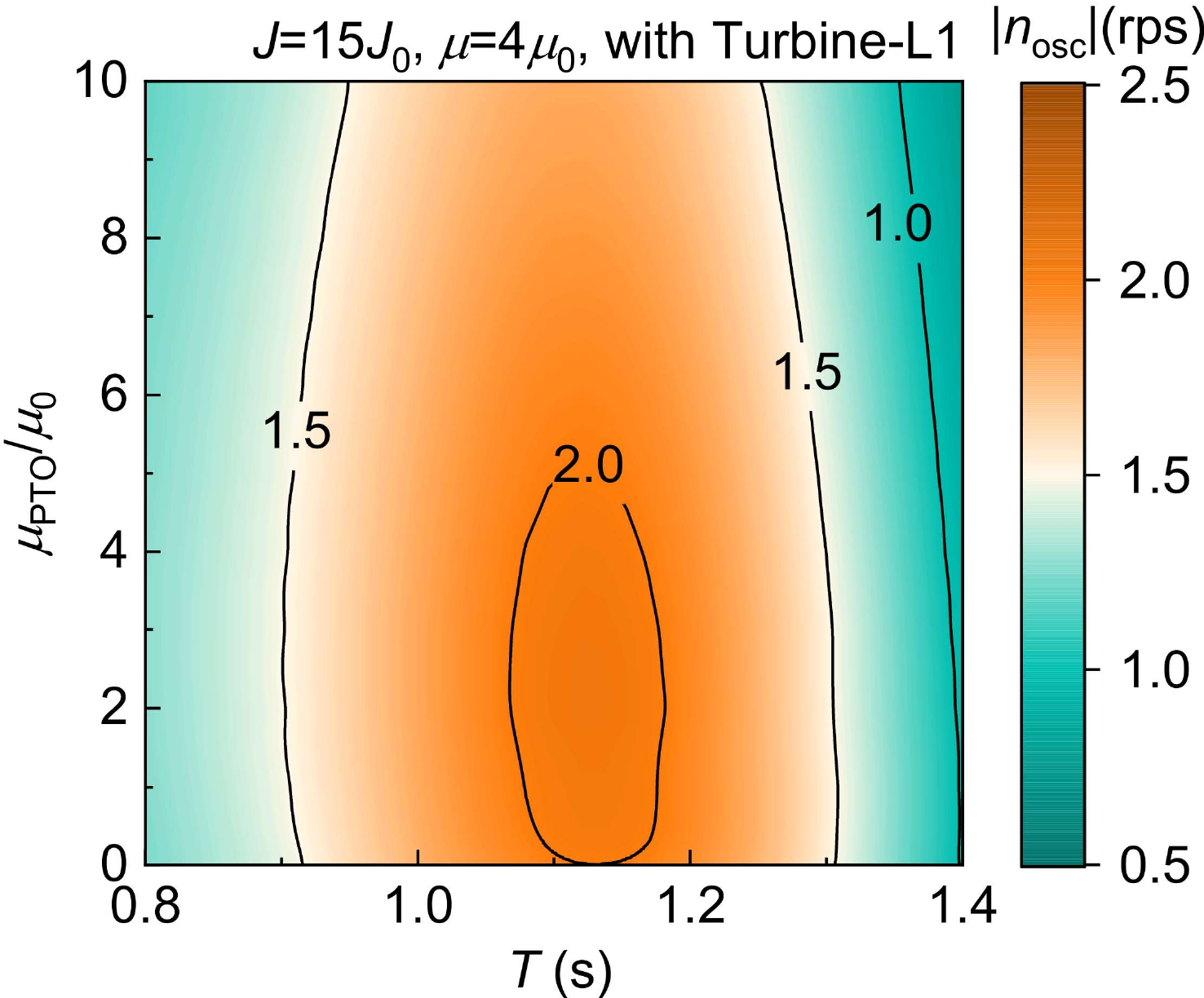}
  \subcaption{}
  \end{minipage}
}
\caption{Effect of power take-off damping on (a) averaged power, (b) free-surface amplitude, (c) averaged rotor speed, and (d) variation range of rotor speed}
\label{fig:optimal PTO damping}
\end{figure}

Fig. \ref{fig:optimal PTO damping} shows the effects of the PTO damping on the averaged power output, the amplitude of free-surface oscillation, the averaged rotor speed, and the variation range of the rotor speed for the WEH liquid tank with Turbine-L1 under various excitation periods. 
The PTO damping $\mu_{\rm{PTO}}$ varies from 0 to $10\mu_0$.
From Fig. \ref{fig:optimal PTO damping} (a), an optimal region of working conditions can be found for power harvesting.
The optimal region is formed from $T = 0.80$ s to 1.25 s and from $\mu_{\rm{PTO}} = 2.0\mu_0$ to $10\mu_0$. 
The maximum averaged power is approximately 0.4W near the condition with $T = 1.10$ s and $\mu_{\rm{PTO}} = 6.0\mu_0$. 
Beyond the optimal region, when $T$ increases from 1.2s to 1.4s, or when $\mu_{\rm{PTO}}$ decreases from $2.0\mu_0$ to $0\mu_0$, the averaged power output continuously decreases from around 0.2W. 
Fig. \ref{fig:optimal PTO damping} (b) indicates that the PTO damping does not affect the free-surface oscillation amplitude evidently under any excitation period.
With a specific PTO damping, the free-surface amplitude first increases with the excitation period from $T = 0.80$ s to around 1.15 s and then decreases with the period.
The maximum free-surface amplitude occurs around $T = 1.15$s, close to the resonant period of the liquid in the U-shaped tank. 
It is known that the oscillation of the liquid body is the source of turbine rotations and power conversion.
For Turbine-L1, changing the PTO damping of the single rotor has little impact on the free-surface amplitude. 
This indicates that the harvested and dissipated energy of Turbine-L1 accounts for a small proportion of the liquid mechanical energy. 
There is still great potential to harvest the liquid-motion energy in the WEH liquid tank by improving the turbine system. 
According to Fig. \ref{fig:optimal PTO damping} (c), the maximum averaged rotor speed exceeds 8 rps, which appears in the condition of around $T = 1.05$ s and $\mu_{\rm{PTO}} = 0$. 
Under each excitation period, the averaged rotor speed decreases as the PTO damping increases.
The averaged rotor speed is between 2 and 6 rps in the optimal region.
In larger period conditions beyond $T = 1.20$ s, the averaged rotor speed decreases rapidly.
Even with $\mu_{\rm{PTO}} = 0$, the averaged rotor speed is only about 3 rps at $T = 1.14$ s. 
Fig. \ref{fig:optimal PTO damping} (d) shows the variation range of the rotor speed.
The maximum variation range is about 2 rps. It appears in the region formed from $T$ = 1.10 s to 1.20 s and from  $\mu_{\rm{PTO}}$ = 0 to $\mu_0$.
Outside this region, the PTO damping does not have much impact on the variation range of the rotor speed. 
\subsection{Efficiency performance of multi-layered impulse air-turbine system}\label{subsec:efficiency}

This subsection compares the performance of the other two MLATSs, namely Turbine-L2 and Turbine-L3, with that of Turbine-L1.
The design details of the three turbines are introduced in Fig. \ref{fig:Impulse turbine}. 
According to the parametric studies in subsections \ref{subsec:rotor_mech} and \ref{subsec:opt_PTO}, the following optimal parameters are set for each rotor: $J = 15J_0$, $\mu = 4\mu_0$, and $\mu_{\rm{PTO}} = 5\mu_0$. 

\begin{figure}[!htp]
\centering
{
  \begin{minipage}{0.45\linewidth}
  \centering
  \includegraphics[width=1.0\linewidth]{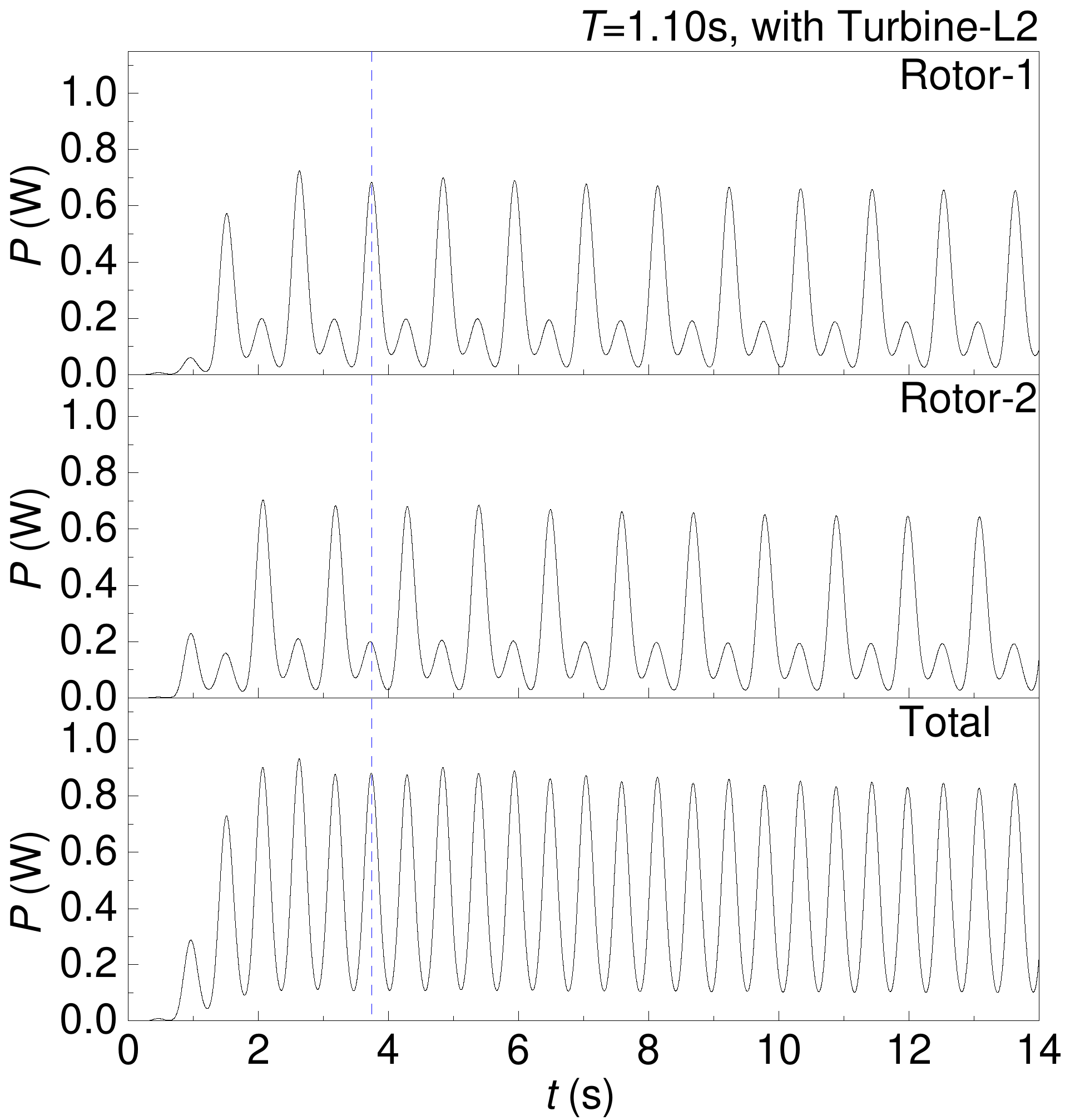}
  \subcaption{}
  \end{minipage}
}
{
  \begin{minipage}{0.45\linewidth}
  \centering
  \includegraphics[width=1.0\linewidth]{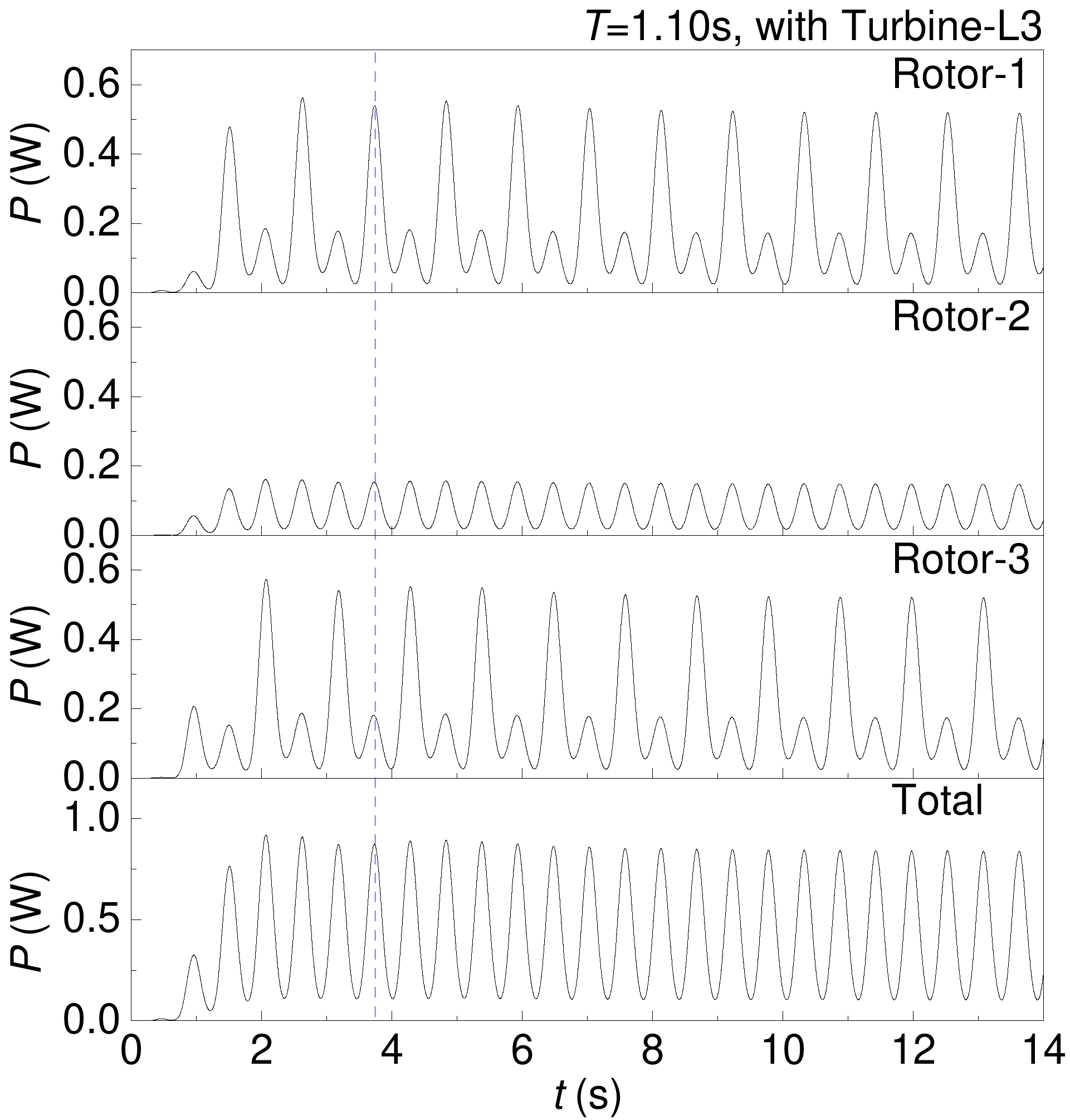}
  \subcaption{}
  \end{minipage}
}
\caption{Power output histories of rotors in wave-energy-harvesting liquid tank with (a) Turbine-L2 and (b) Turbine-L3, for $T=1.10\text{s}$}
\label{fig:power_history}
\end{figure}

The excitation condition with $T = 1.10\text{s}$ is considered for illustration purposes.
Fig. \ref{fig:power_history} (a) shows the power output histories of each rotor and the power summation of two rotors for Turbine-L2.
In the steady stage, during each excitation period, the power output history of each rotor alternately shows a larger peak and a smaller peak. 
Under a certain air-flow direction, the upstream rotor always has a larger power peak. 
When the upstream rotor reaches its larger power peak, the downstream rotor reaches its lower power peak simultaneously.
Since the two rotors work independently, the power histories of the two rotors can be summed up without difficulty.
The summation of the power histories reveals a steady power amplitude, which is similar to the power history of Turbine-L1 in Fig. \ref{fig:Power analysis}. 
Eq. (\ref{eq:power_ave}) can be used to calculate the averaged power of the Turbine-L2.
Fig. \ref{fig:power_history} (b) shows the power output histories of each rotor and the power summation of three rotors for Turbine-L3.
Rotor-2 lies between the other two rotors.
The power output histories of Rotor-1 and Rotor-3 in Turbine-L3 are similar to those of the rotors in Turbine-L2.
In the steady stage, during each excitation period, the power output history of Rotor-1 or Rotor-3 alternately shows a larger peak and a smaller peak. 
Rotor-2 has a steadier yet much smaller amplitude of power output compared with the other two rotors.
The power summation of the three rotors in Turbine-L3 also has a steady amplitude, similar to that of Turbine-L2. 

\begin{figure}[!htp]
\centering
{
  \begin{minipage}{0.45\linewidth}
  \centering
  \includegraphics[width=1.0\linewidth]{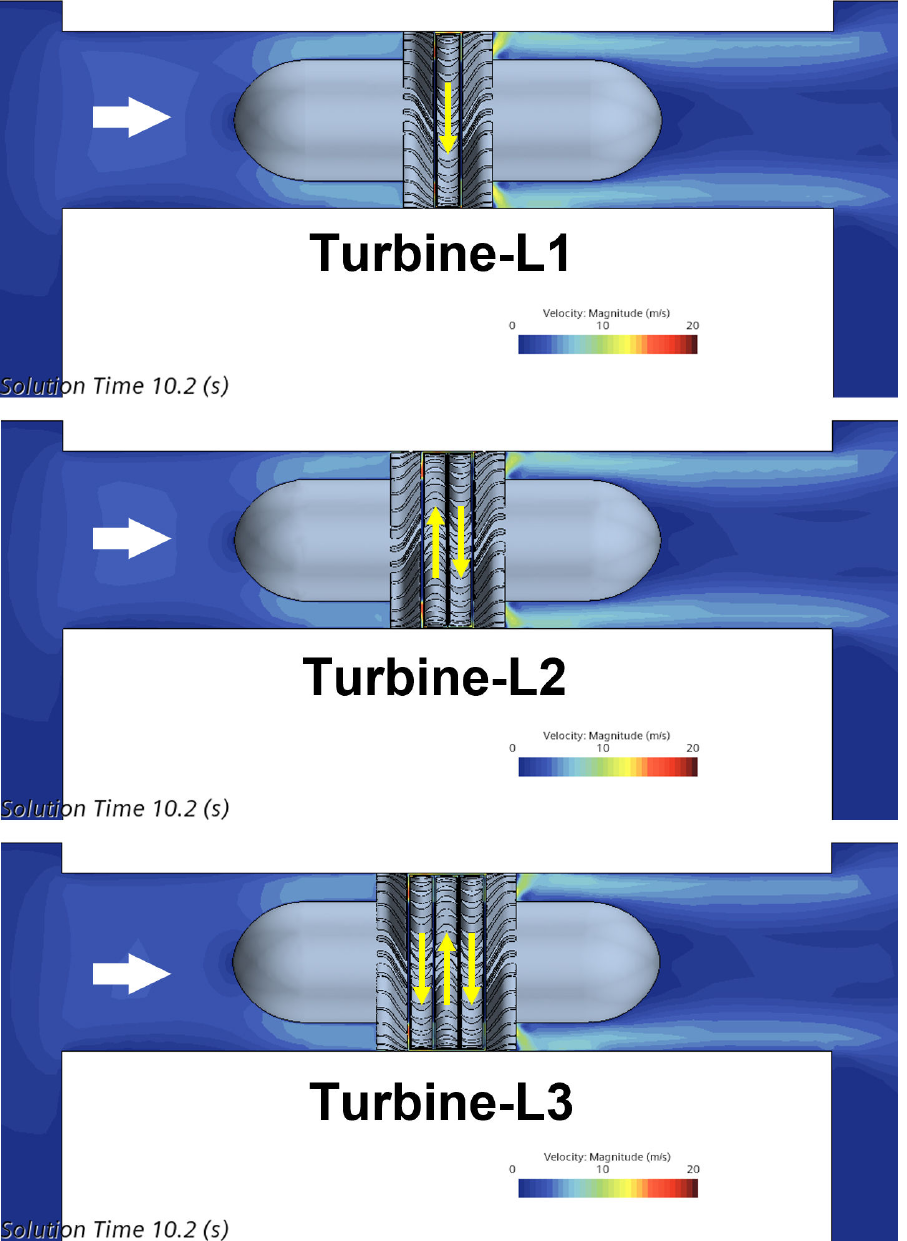}
  \subcaption{}
  \end{minipage}
}
{
  \begin{minipage}{0.45\linewidth}
  \centering
  \includegraphics[width=1.0\linewidth]{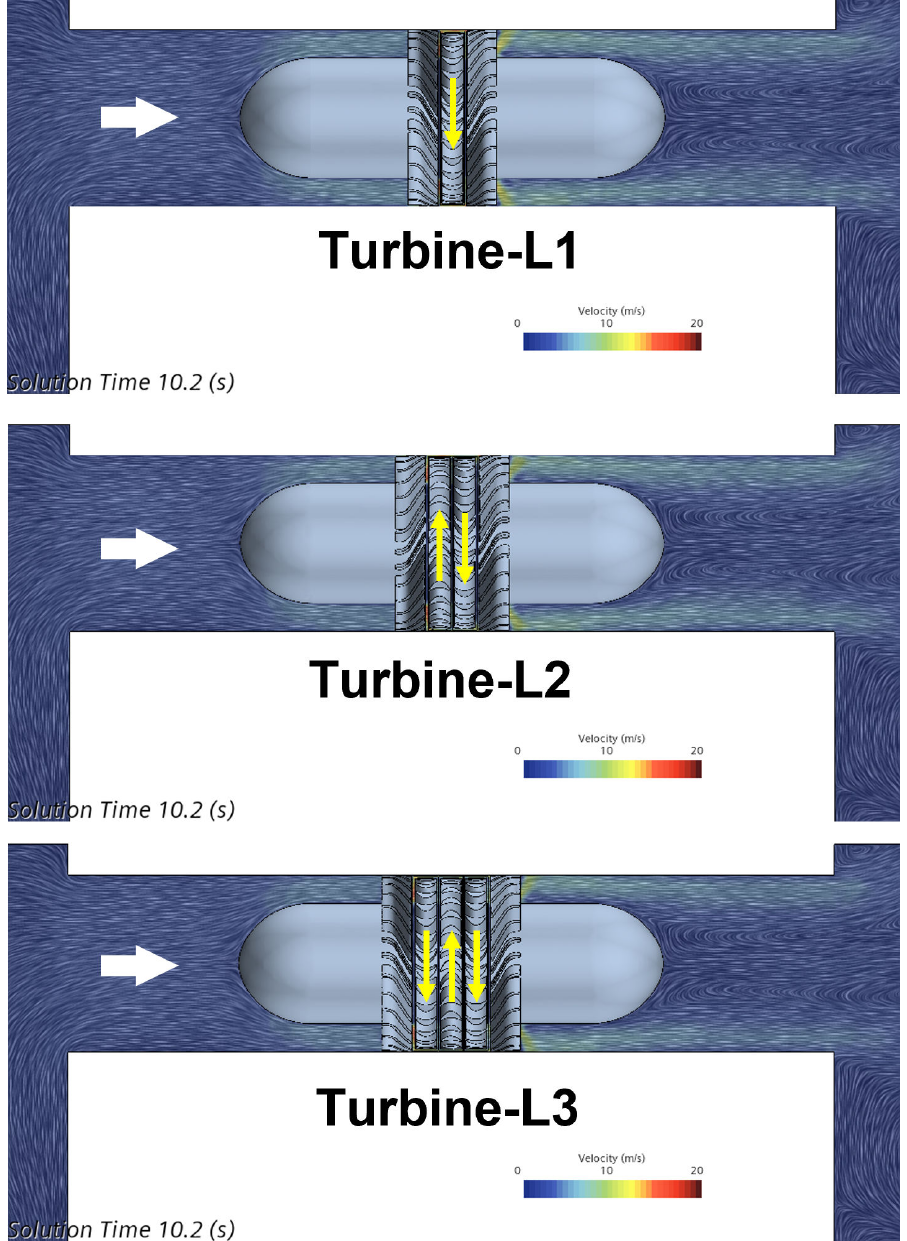}
  \subcaption{}
  \end{minipage}
}
\caption{(a) Amplitude distribution and (b) vector field of air-flow velocity near the turbine system, in excitation condition of $T=1.10\text{s}$}
\label{fig:field}
\end{figure}

Fig. \ref{fig:field} illustrates the amplitude distribution and the vector field of the air-flow velocity near the turbine system.
At the same instant, the velocity amplitude distributions near Turbine-L1, Turbine-L2, and Turbine-L3 display similar patterns.
The upstream vector field in front of each turbine is fairly uniform.
Vortices of similar size can be seen at the downstream side of each turbine. 

\begin{figure}[!htp]
\centering
{
  \begin{minipage}{0.45\linewidth}
  \centering
  \includegraphics[width=1.0\linewidth]{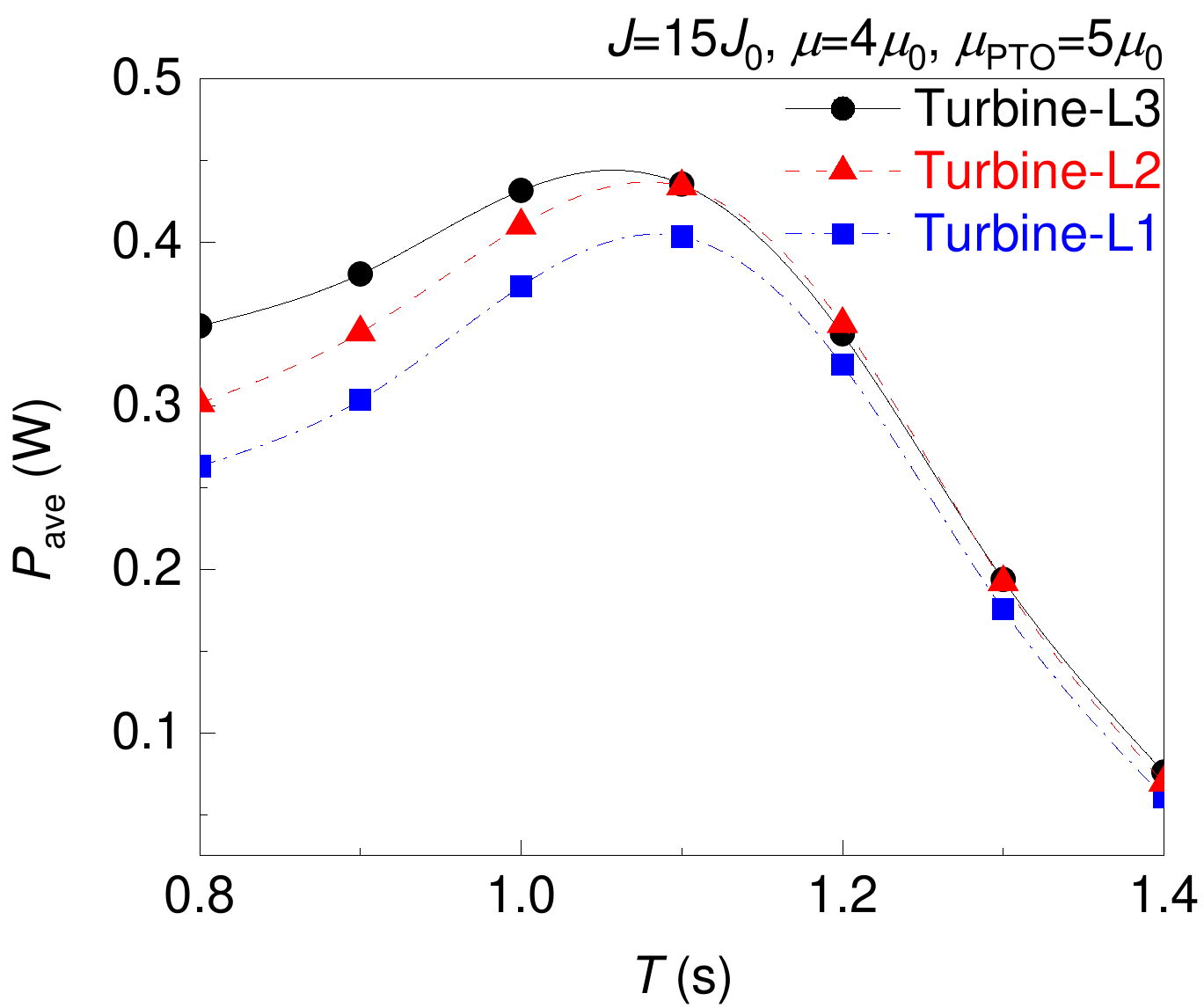}
  \subcaption{}
  \end{minipage}
}
{
  \begin{minipage}{0.45\linewidth}
  \centering
  \includegraphics[width=1.0\linewidth]{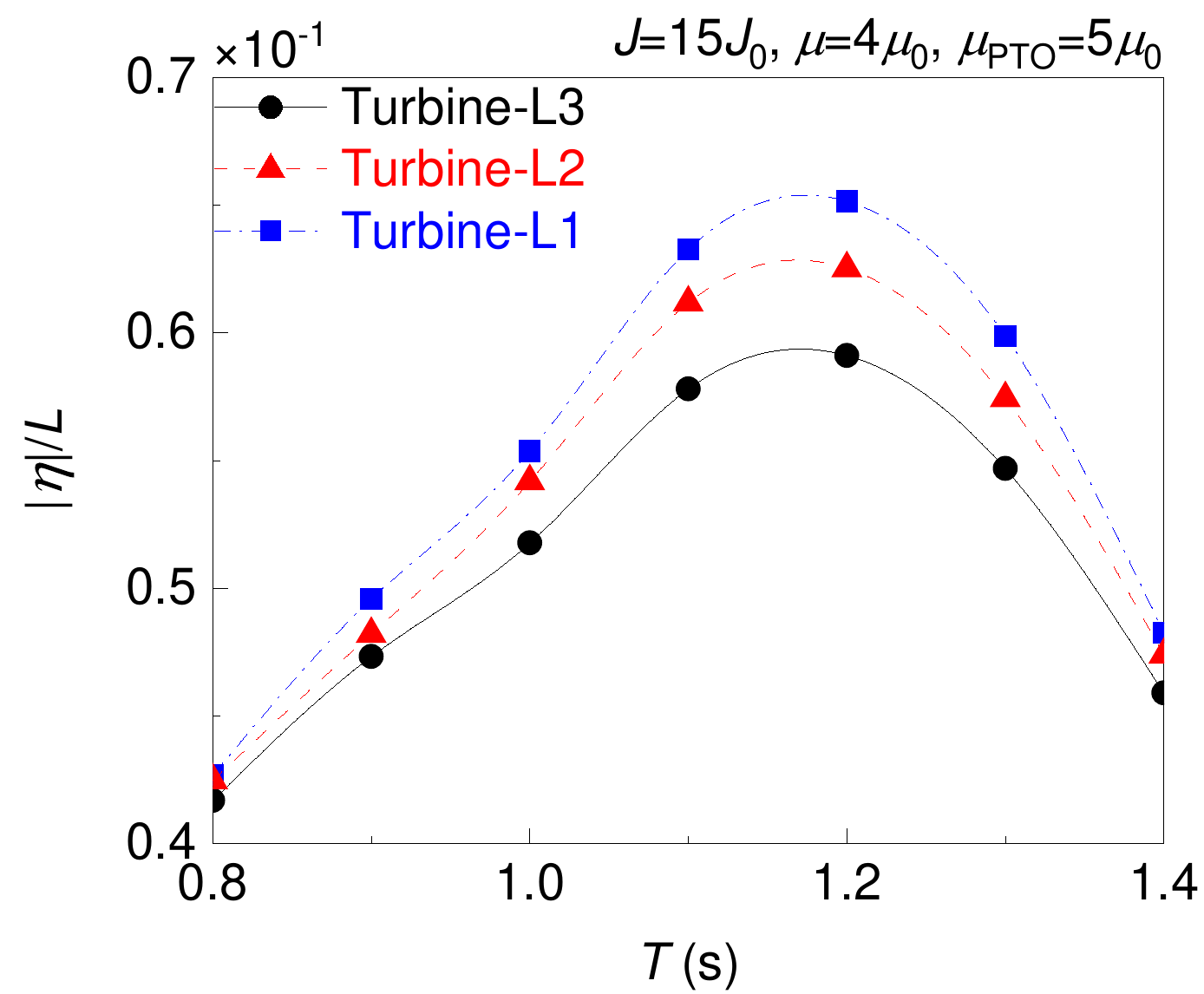}
  \subcaption{}
  \end{minipage}
}
\caption{(a) Averaged power outputs and (b) free-surface amplitudes in wave-energy-harvesting liquid tank with Turbine-L1, Turbine-L2 and Turbine-L3}
\label{fig:L3_power}
\end{figure}

Fig. \ref{fig:L3_power} (a) compares the total averaged power outputs of Turbine-L1, Turbine-L2, and Turbine-L3.
All three turbines have their maximum averaged power appear in the excitation condition of $T = 1.10$s.
The largest power output is $P_{\rm{ave}} = 0.45$ W.
For excitation conditions with $T>1.10$s, improving Turbine-L1 to Turbine-L2 or Turbine-L3 does not increase the averaged output of the WEH liquid tank evidently.
Under conditions of smaller-period excitation, the averaged power output clearly increases from Turbine-L1 to Turbine-L3.
In the excitation condition where $T = 0.80$ s, if Turbine-L2 is used, the averaged power output of Turbine-L1 increases by about $25\%$. 
Meanwhile, if Turbine-L3 is used, the increase ratio is about $40\%$. 
Fig. \ref{fig:L3_power} (b) shows the free-surface oscillation amplitude in the liquid tank when Turbine-L1, Turbine-L2, and Turbine-L3 are used, respectively.
As the number of rotors increases from one to three from Turbine-L1 to Turbine-L3, the free-surface amplitude decreases accordingly. 
From the energy perspective, the energy loss of the liquid motion mainly results from two factors: One is energy dissipation, and the other is energy absorption by the turbine system. 
By comparing the results of $T > 1.10$ s in Figs. \ref{fig:L3_power} (a) and (b), it can be understood that Turbine-L3 and Turbine-L2 dissipate more energy than Turbine-L1 under these large-period conditions. 

\begin{figure}[!htp]
\centering
{
  \begin{minipage}{0.45\linewidth}
  \centering
  \includegraphics[width=1.0\linewidth]{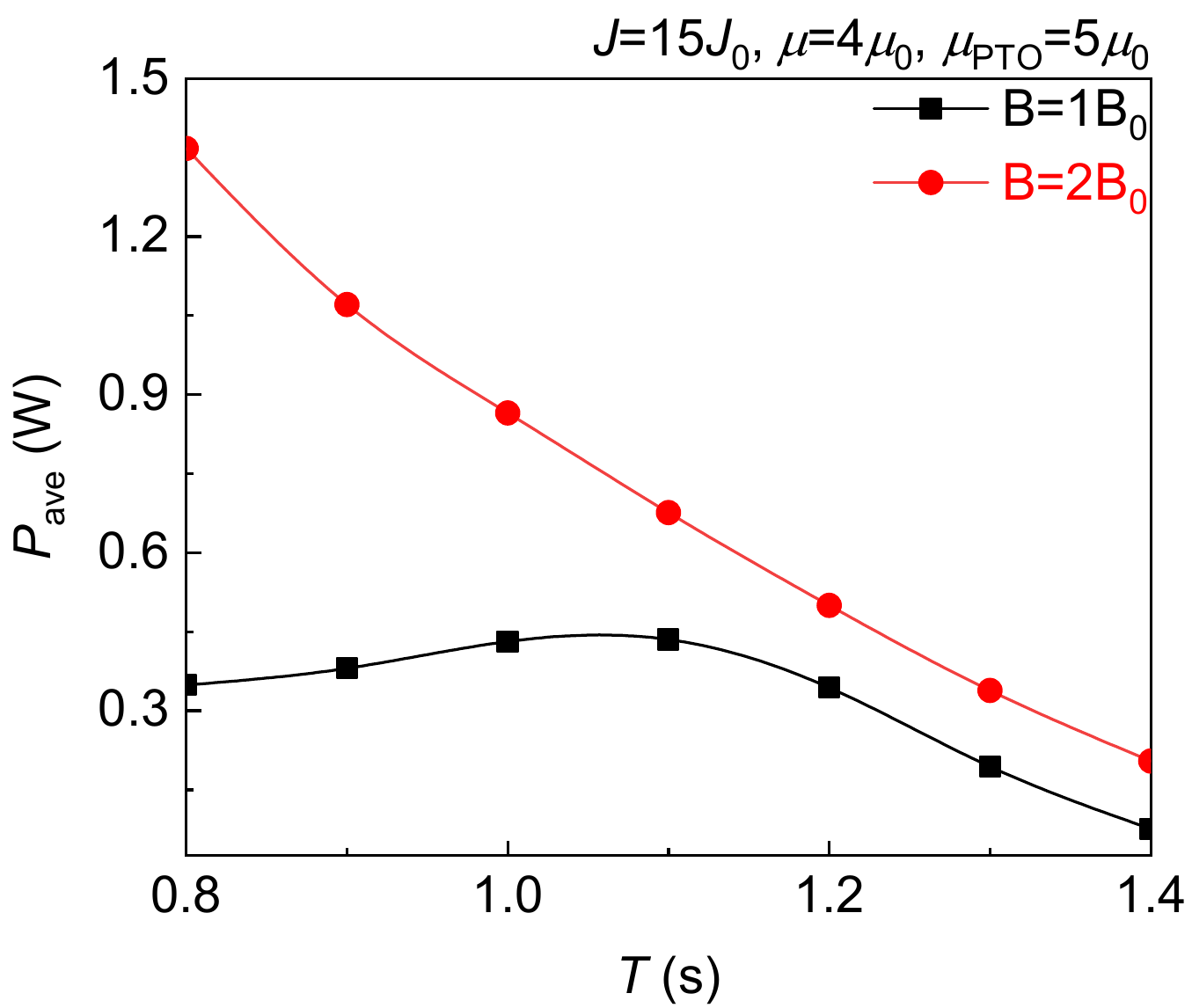}
  \subcaption{}
  \end{minipage}
}
{
  \begin{minipage}{0.45\linewidth}
  \centering
  \includegraphics[width=1.0\linewidth]{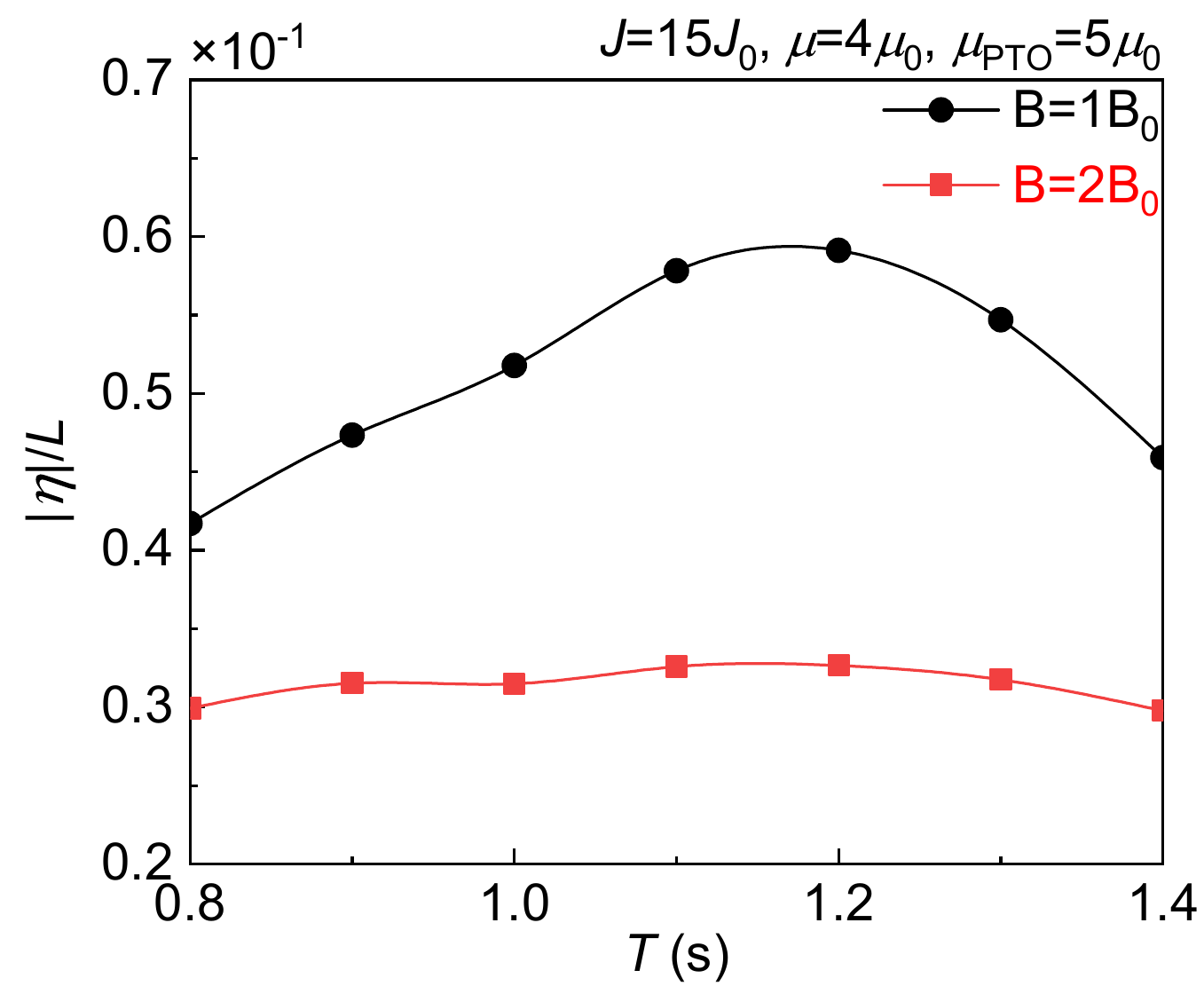}
  \subcaption{}
  \end{minipage}
}
\caption{(a) Averaged power output, and (b) free-surface amplitude in wave-energy-harvesting liquid tank with $B=B_0$, and $B=2B_0$}
\label{fig:2B}
\end{figure}

To further release the power generation potential of the WEH liquid tank with Turbine-L3, the tank breadth $B$ is increased from $B_0=0.300\text{m}$ to $2B_0$.
Fig. \ref{fig:2B} (a) compares the averaged power outputs of the improved WEH liquid tank with that of the original one.
Unlike the original tank whose maximum power output appears at $T = 1.10$ s, for the improved liquid tank, the averaged power output exhibits an increasing trend as the excitation period decreases.  
The averaged power output of the improved liquid tank is always larger than that of the original one.
In the excitation condition where $T = 0.80$ s, the averaged power output is enlarged by approximately four times.
For excitation conditions where $T>$ 1.10 s, the increment of the averaged power generation exceeds 0.15 W.
This observation shows that the effect of the tank breadth on the power output of the WEH liquid tank is nonlinear.
In wave conditions with a small period, increasing the tank's breath can effectively amplify the power generation capability of the liquid tank.
Fig. \ref{fig:2B} (b) compares the corresponding free-surface amplitudes in two liquid tanks. 
The free-surface amplitude is severely suppressed in the tank with $B=2B_0$.
In all excitation conditions, the free-surface amplitude in the improved liquid tank is almost identical.
Compared with the original tank, the improved liquid suppresses the free-surface oscillations in the liquid tank more strongly.

\subsection{Reliability analyses of multi-layered impulse air-turbine system}

\begin{figure}[!htp]
\centering
{
  \begin{minipage}{0.45\linewidth}
  \centering
  \includegraphics[width=1.0\linewidth]{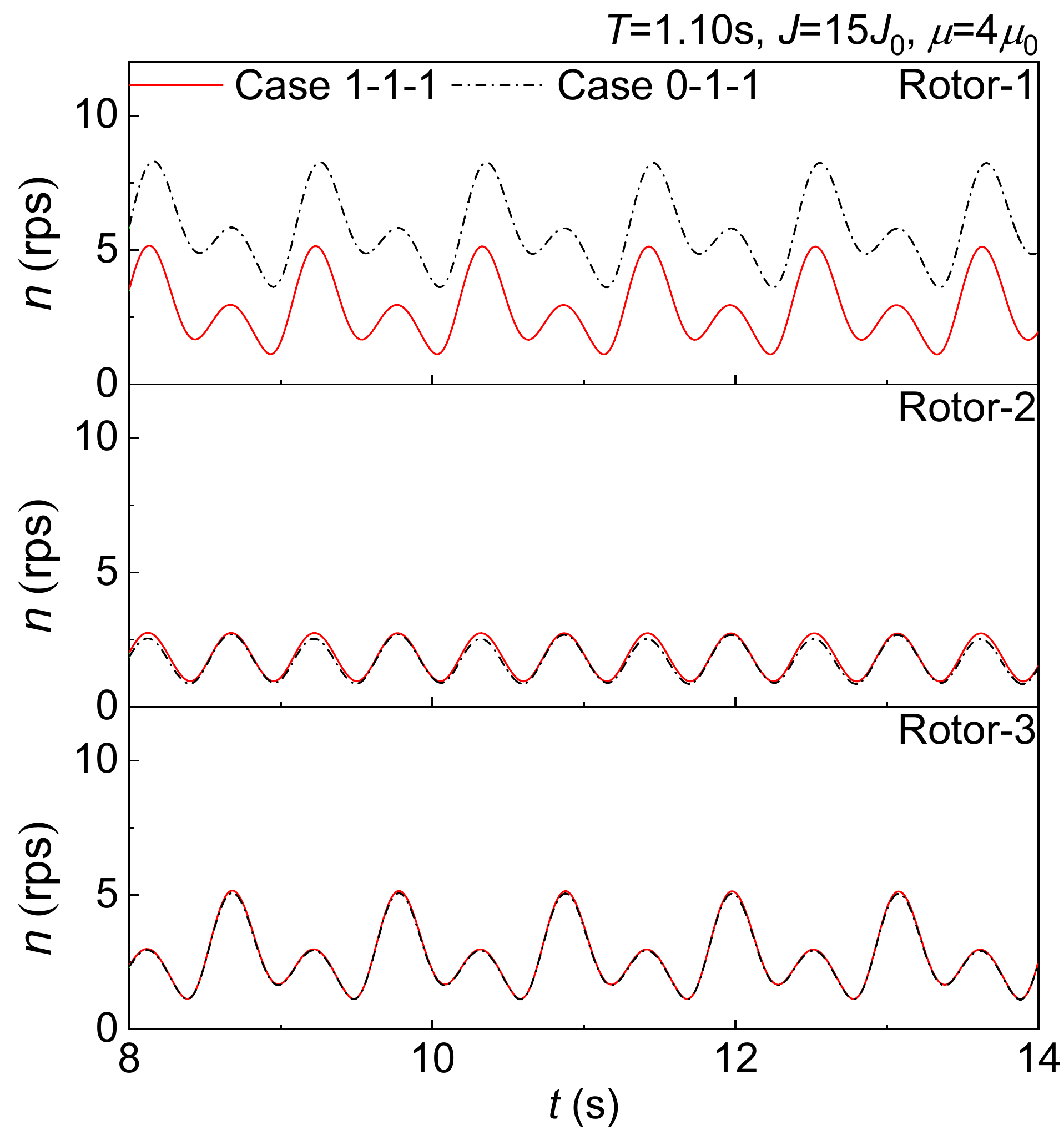}
  \subcaption{}
  \end{minipage}
}
{
  \begin{minipage}{0.45\linewidth}
  \centering
  \includegraphics[width=1.0\linewidth]{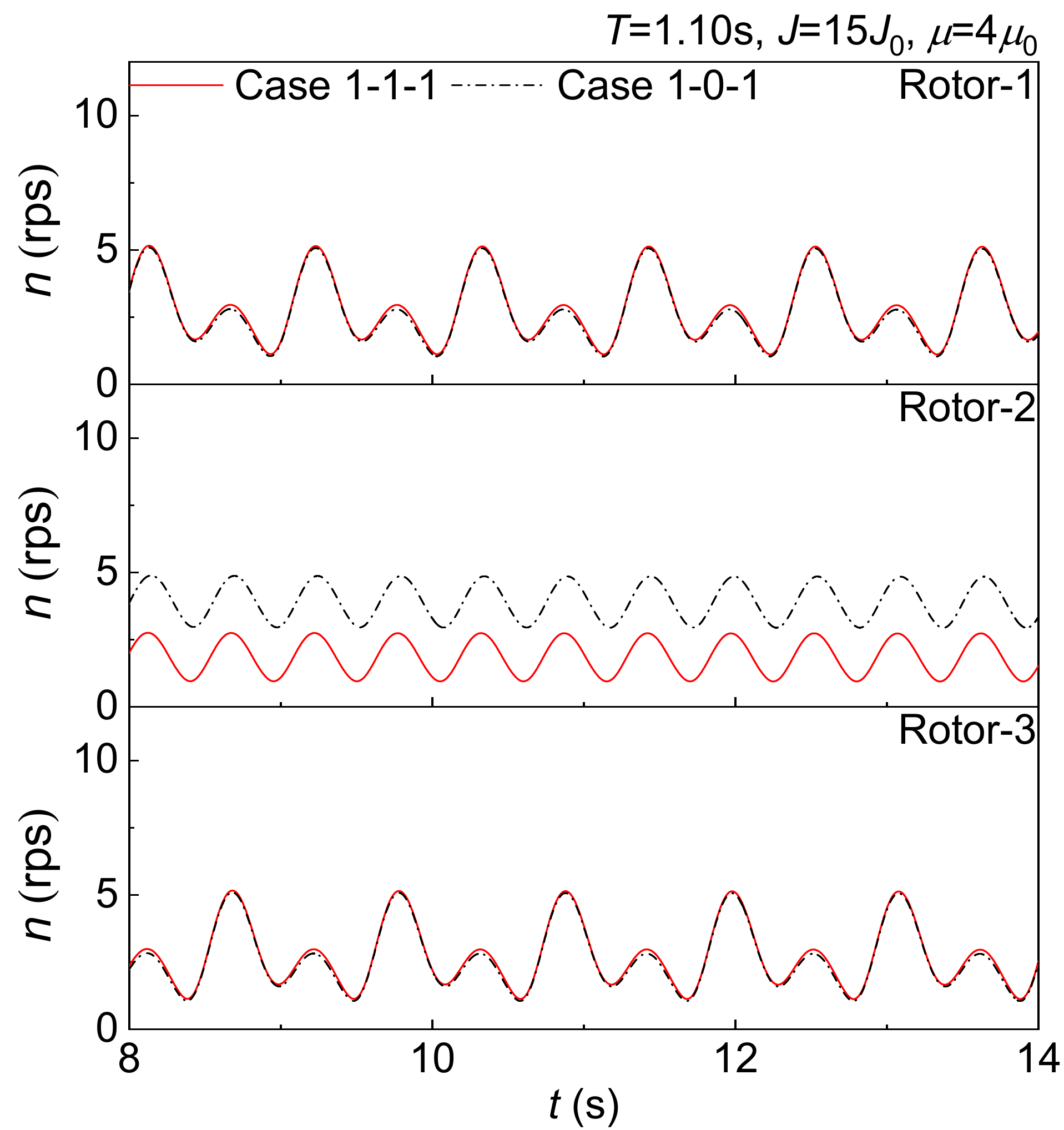}
  \subcaption{}
  \end{minipage}
}
{
  \begin{minipage}{0.45\linewidth}
  \centering
  \includegraphics[width=1.0\linewidth]{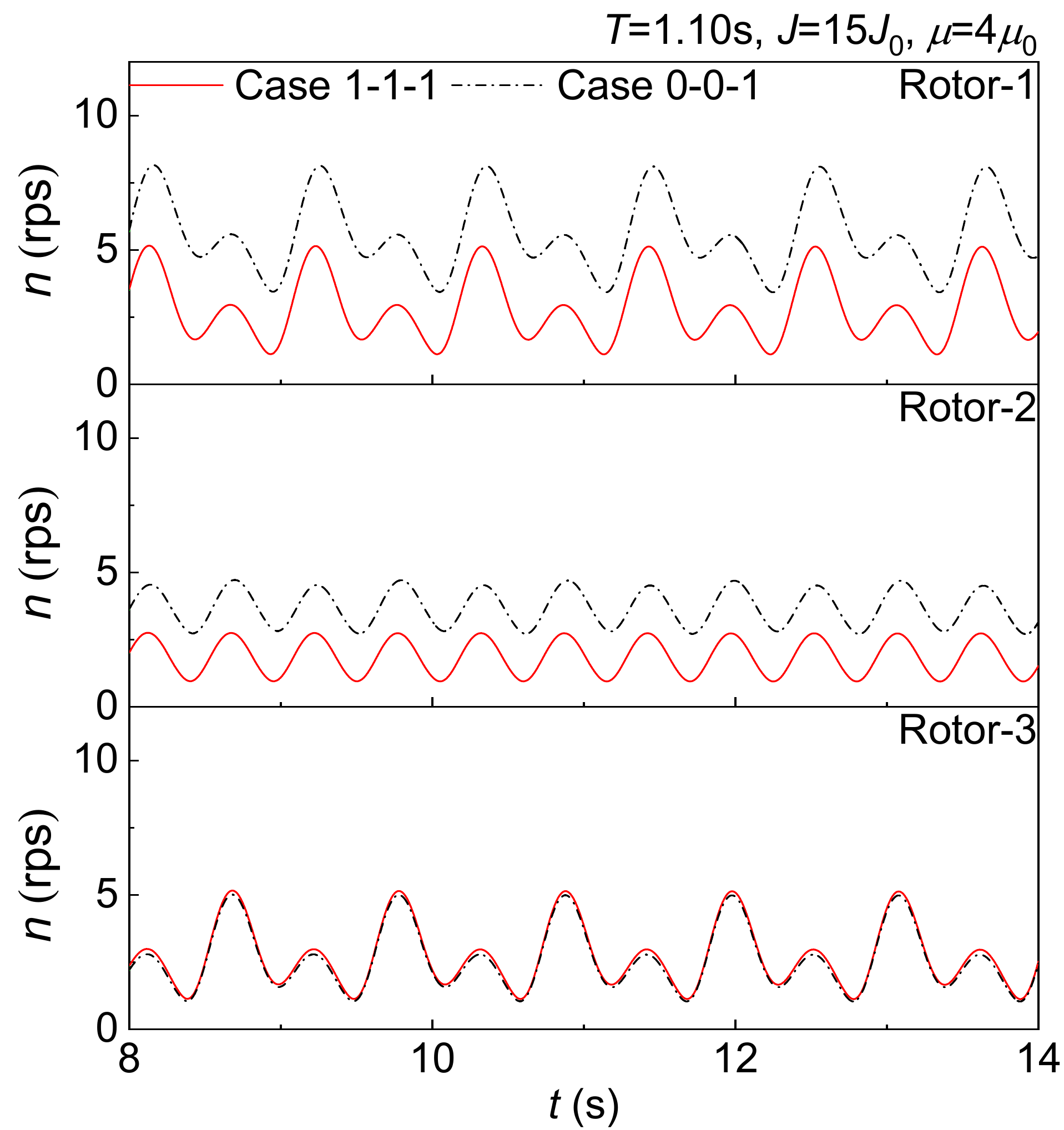}
  \subcaption{}
  \end{minipage}
}
{
  \begin{minipage}{0.45\linewidth}
  \centering
  \includegraphics[width=1.0\linewidth]{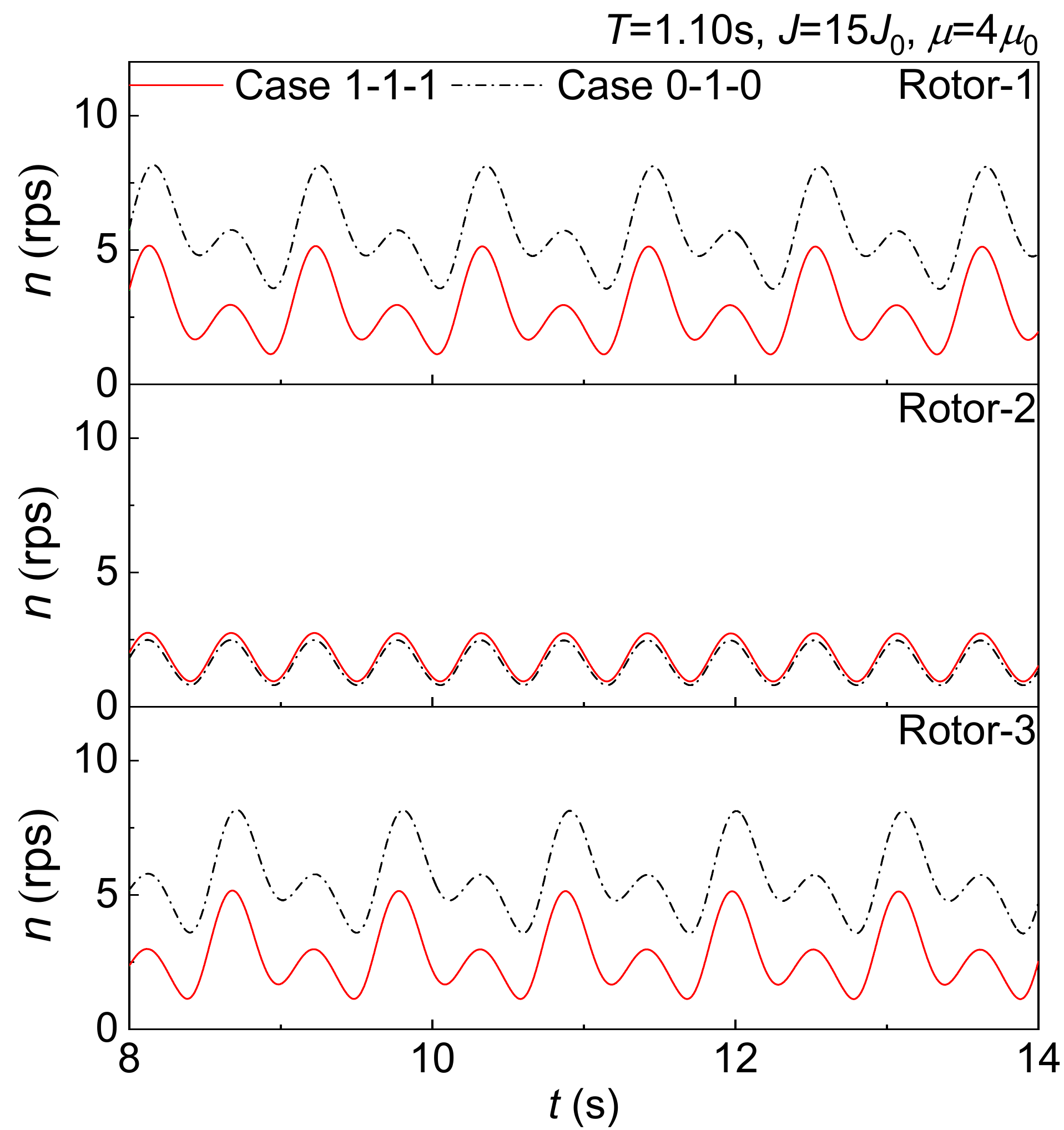}
  \subcaption{}
  \end{minipage}
}
\caption{Rotor speed status of Turbine-L3, in failure cases of (a) 0-1-1, (b) 1-0-1, (c) 0-0-1, and (d) 0-1-0}
\label{fig:rotor speed of each case}
\end{figure}

Reliability analyses are conducted on Turbine-L3 through a series of failure tests.
The cases are referred to by using the index system $i-j-k$.
Here, the indices $i$, $j$, and $k$ correspond to the status of Rotor-1, Rotor-2, and Rotor-3, respectively.
If Rotor-$i$ is damaged, $i = 0$; Otherwise, $i = 1$.
Four failure cases, namely case 0-1-1, case 1-0-1, case 0-0-1, and case 0-1-0, are considered as examples. 
A damaged rotor has zero PTO damping and thus does not generate power.
Case 1-1-1 refers to the status when the turbine works well and all three rotors are healthy.
Fig. \ref{fig:rotor speed of each case}(a) shows the rotor speed status of case 0-1-1 when Rotor-1 is damaged.
Fig. \ref{fig:rotor speed of each case}(b) refers to the status when Rotor-2 fails.
Fig. \ref{fig:rotor speed of each case}(c) corresponds to the situation where both Rotor-1 and Rotor-2 are damaged.
In Fig. \ref{fig:rotor speed of each case}(d), both Rotor-1 and Rotor-3 are damaged at the same time.
Each damaged rotor can rotate freely without the PTO damping.
In all these failure cases, the maximum rotor speed of a damaged rotor increases by approximately 2 to 3 rps compared to its healthy status.
Moreover, the failure of any rotor does not affect the rotor speeds of the undamaged rotors. 
This observation confirms that Turbine-L3 is inherently robust to adapt to damage accidents.

\begin{figure}[!htp]
\centering
\includegraphics[scale=0.3]{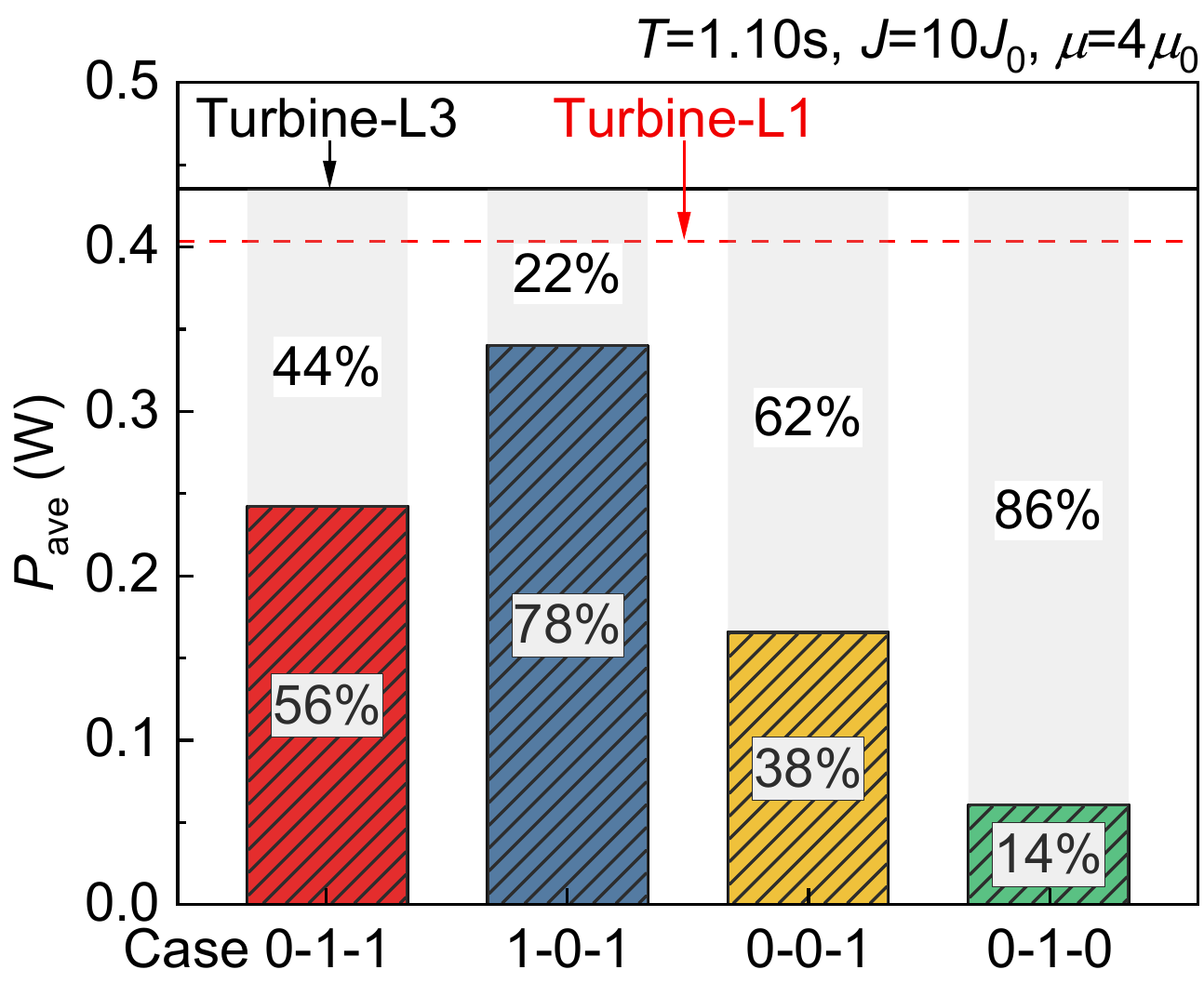}	
\caption{Capability loss of Turbine-L3 in different failure cases}
\label{fig:bad power comparison}
\end{figure}

Fig. \ref{fig:bad power comparison} further compares the averaged power output of these four failure cases. 
It reveals that the failure of Rotor-2 has the least impact on the total power generation. 
Compared with the healthy status, the average power output in case 1-0-1 is only reduced by $22\%$.
This power generation capability is still quite close to that of Turbine-L1.
If either Rotor-1 or Rotor-3 fails to work, the turbine's power generation capability can drop by about $44\%$.
When both Rotor-1 and Rotor-2 are damaged, Turbine-L3 remains at $38\%$ of its rated capability.
In the worst case when only Rotor-2 works, the power output of Turbine-L3 is reduced to $14\%$ of its rated capability.
This suggests that the contribution of the two side rotors (Rotor-1 or Rotor-3) to the power output is greater than that of the middle rotor (Rotor-2).
The aforementioned analyses confirm that Turbine-L3, a MLATS, is far more reliable in extreme conditions than a conditional turbine with a single rotor.

\section{Conclusions}\label{sec:concl}

For a deep understanding of the mechanisms behind pneumatic liquid-tank En-WECs, an integrated numerical model is proposed for the first time.
The coupled hydro-aero-turbo dynamics and power harvesting properties of the liquid-tank system are investigated. 
A scaled prototype of the wave-energy-harvesting (WEH) liquid tank with an impulse air turbine system is created to experimentally validate the numerical model.
Multi - layered impulse air turbine systems (MLATS) are creatively introduced into the liquid-tank system.
Three MLATSs, Turbine-L1, Turbine-L2, and Turbine-L3, are specially designed for analyses.
The inherent mechanisms of the coupled hydro-aero-turbo dynamics of the WEH liquid tank with various turbine properties are investigated systematically.
The MLATS shows improved efficiency and reliability compared to conventional single-rotor turbine designs.

For the integrated numerical model, the rotors of each turbine system rotate freely under the action of air-flow loads, mechanical resistances, and power take-off (PTO) damping.
The Reynolds Averaged Navier-Stokes equations are adopted as the governing equations for both the air domain and the liquid domain.
The fluid domain is divided into a hydro-aero region and $N$ aero-turbo regions. 
The mesh cells in the aero-turbo region are time-invariant relative to the rotor.
A bidirectional transfer of computational quantities is achieved iteratively on the interface between each pair of adjacent regions.

The scaled prototype of the WEH liquid tank has Turbine-L1 installed.
The servo motor of Turbine-L1 serves as the source of the turbine's inherent resistance and measures the rotational speed of the rotor.
Ultrasonic wave gauges are employed to monitor the elevation histories of the free surface in the liquid tank.
Pressure sensors are utilized to measure the air pressure at different positions of the liquid tanks.
The WEH liquid tank model is excited on a six-degree-of-freedom shaking table.
Comparing the numerical and experimental results shows that the current integrated numerical model can accurately simulate the coupled hydro-aero-turbo dynamics (including the rotor speed, liquid motion, and air pressure) of the WEH liquid tank.

The mechanical parameters of the turbine rotor are analysed.
For the considered excitation conditions, the moment of inertia of a rotor does not affect the averaged rotor speed. 
However, it affects the variation range of the rotor speed evidently.
A rotor with a larger moment of inertia tends to have better rotation stability. 
The damping coefficient markedly influences the averaged rotor speed, but has little effect on the variation range of the rotor speed.
The optimal PTO damping for the WEH liquid tank is also identified.
For a WEH liquid tank that is as small as 0.6m long, the maximum averaged power is around 0.4W near the resonance condition. 
The PTO damping does not affect the free-surface oscillation amplitude evidently under any excitation period.

The efficiency performances of three MLATSs are compared.
Each of the three turbines has its maximum averaged power appearing near the resonance condition of the liquid tank. 
For excitation conditions with longer periods, improving Turbine-L1 to Turbine-L2 or Turbine-L3 does not significantly increase the averaged output of the WEH liquid tank. 
Under smaller-period excitation conditions, compared with Turbine-L1, Turbine-L2 and Turbine-L3 can increase the averaged power output by about $25\%$ and about $40\%$, respectively. 
As the number of rotors increases from one to three from Turbine-L1 to Turbine-L3, the free-surface amplitude decreases accordingly, suggesting a larger energy loss of the liquid motion.
In addition, increasing the tank breadth can effectively boost the power output in a nonlinear way.
Under the considered excitation conditions, if the tank breadth is doubled, the maximum averaged power output can be increased by around four times. 

Finally, reliability analyses are conducted on Turbine-L3 through a series of failure tests.
The contribution of the two side rotors (Rotor-1 or Rotor-3) to the power output is greater than that of the middle rotor (Rotor-2).
For conventional single-rotor turbine systems, if the air turbine malfunctions, it means the WEC device has reached the end of its service life.
However, for the present MLATS, if the side rotor or the middle rotor fails to work, the averaged power output is only reduced by $44\%$ and $22\%$, respectively.
The proposed MLATS Turbine-L3 is inherently robust and can adapt to various damage accidents. 
It is much more reliable in extreme conditions than a conventional turbine with a single rotor. 

\section*{Acknowledgements}
This study was funded by National Natural Science Foundation of China (Grant No.~52471271, and U22A20242).
This research was also supported by the advanced computing resources provided by the Supercomputing Center of Dalian University of Technology.

\section*{Data Availability}
The data that support the findings of this study are available from the corresponding author, upon reasonable request.

\printcredits

\section*{Declaration of Generative AI and AI-assisted technologies in the writing process}
During the preparation of this work, generative AI and AI-assisted technologies were not used.

\bibliographystyle{model3-num-names}
\bibliography{Manuscript}

\end{document}